\documentclass[prb,showkeys,twocolumn,floatfix]{revtex4-2}

\usepackage[normalem]{ulem} 
\usepackage{bm}
\usepackage[labelfont=large,position=top]{subfig}
\usepackage{amsfonts}
\usepackage{amsmath}
\usepackage{amssymb}
\usepackage{mathrsfs}
\usepackage{ot1patch}

\usepackage{graphicx}
\usepackage{subfig}

\usepackage{dcolumn}
\usepackage[thinspace,squaren,pstricks,italian,derivedinbase,derived]{SIunits}

\usepackage[normalem]{ulem}

\usepackage[dvipsnames]{xcolor}

\newcommand{\bra}{\langle}
\newcommand{\ket}{\rangle}

\begin{document}

\title{
Validity of perturbation theory in calculations of
magnetocrystalline anisotropy in Co-based layered systems}

\author{M. Cinal}
\email{mcinal@ichf.edu.pl}
\affiliation{Institute of Physical Chemistry, Polish Academy
of Sciences, 01-224 Warsaw, Poland}

\date{\today}

\begin{abstract}

 Validity of second-order perturbation theory (PT) is examined for  magnetocrystalline anisotropy (MCA) energy in Co films with enhanced spin-orbit coupling (SOC) and Co/Pt bilayers.
Comparison with accurate results obtained with the force theorem (FT)  reveals significant discrepancies in the dependence of the MCA energy on the Co thickness.
For systems with strong SOC, the PT fails to
correctly describe the oscillations of the MCA energy,
largely overestimating their amplitude
and even failing (for Co/Pt bilayers) to reproduce their specific periodicity.
These failures specifically concern
the dominating oscillations with the 2-monolayer
period which arise from pairs of quantum well (QW) minority-spin $d$ states
in the Co layer, degenerate at the centre of the Brillouin zone (BZ).
A simplified model of such states  demonstrates
that the large discrepancies between PT and FT predictions arise from
the breakdown  of the PT  in a region
around the BZ centre
where the energy spacing between states
within each pair is small compared to the SOC strength.
 It is also shown that
 the oscillation amplitude of the  MCA energy
calculated with the FT  is limited by
the finite energy spacing
between consecutive QW pairs,
whereas this amplitude grows quadratically with the SOC strength in the PT calculations.
Furthermore, for weak and moderate SOC strengths,
the accuracy of the PT  diminishes with increasing the ratio of the SOC constant
to temperature which defines the broadening of energy levels.
This explains  why the PT overestimates
the amplitude of MCA energy oscillations at low temperatures,
even for the Co film with a relatively weak nominal SOC.
For the Co/Pt bilayer,
the strong temperature dependence of the oscillation amplitude
in the PT approach leads to
the MCA energies which are markedly  different at low and zero temperatures,
of opposite sign and several times larger in magnitude,
compared to the FT results.

\end{abstract}



\maketitle

\pagebreak

\section{INTRODUCTION}
\label{sec-introduction}

Magnetic anisotropy is one of the key properties that determine
the magnetic structure and affect the dynamics of magnetic materials.
It describes the dependence of the system energy on the magnetization direction,
with the energy minimum corresponding to
preferred magnetization orientation, the easy axis.
Two main sources contribute to magnetic anisotropy: the dipole-dipole interaction, which results in the shape anisotropy dependent on the sample geometry, and the spin-orbit coupling (SOC), which modifies the system's electronic structure in a manner dependent on the magnetization direction, leading to the magnetocrystalline anisotropy (MCA) \cite{Gay-Richter86,Gay-Richter87,Li90,Bruno89,MC94}.
 MCA is enhanced in ferromagnet/nonmagnet (FM/NM) layered systems
 due to the reduced symmetry at interfaces and the SOC of NM.
In films with high-symmetry surfaces, the easy axis is oriented
either in-plane or out-of-plane,
while a tilted orientation of the easy axis is observed
for films with stepped surfaces \cite{MCMP16,MCMP19}.
Among layered systems, FM/heavy metal (FM/HM) bilayers
are of particular interest.
In these bilayers,
an in-plane electric current
in a layer of metal with strong SOC, such as Pt or Ta,
generates transverse spin current
that flows into the FM layer enabling magnetization switching
\cite{Buhrman12,Buhrman12b,Lee13},
while the lack of the inversion symmetry 
is essential for emergence of a finite spin-orbit torque acting
on the magnetization.
In the ongoing quest for efficient high-density magnetic storage
with stable magnetization and fast switching characteristics,
materials with high perpendicular magnetic anisotropy and low Gilbert damping
are actively sought and investigated both experimentally
\cite{Suto25,Nakamura25} and theoretically
\cite{Kelly24}.
The renewed interest in FM/HM systems is also driven by the finite Dzyaloshinskii-Moriya interaction (DMI) arising from structural inversion asymmetry, which can stabilize chiral magnetic structures \cite{Ham21,Maziewski21,Maziewski25,Zhang22,Zhu24}).

In theoretical calculations, the MCA energy is  usually determined
using the force theorem (FT) \cite{Weinert95,Wang96-jmmm},
as the difference of occupied band energies
for two different magnetization directions, calculated
in the presence of the SOC,
or the difference of the respective free energies (or grand potentials)
at finite temperature
to improve convergence \cite{MC97,MC22,MC24}.
An alternative approach to the MCA energy,
proposed by Bruno \cite{Bruno89},
is based on the second-order
perturbation theory (PT), with the SOC as the perturbation.
It is relevant for systems with reduced symmetry,
such as magnetic films and multilayers,
as well as bulk crystals with a hexagonal structure
or a cubic structure distorted
by strain.
 For systems with full cubic symmetry,
 however, the fourth-order PT would be required
 but it is not used in practice.

The PT formula for the MCA energy \cite{Bruno89,MC94,MC97,MC22,MC24}
is often applied
to provide insights into the electronic origin of the MCA
in specific systems,
particularly by identifying
contributing quantum states with specific symmetries
\cite{Wang93,Miura13,Qiao18,Miura18-prb, Miura22-jphys-review,Ong16,Sun19,Ke19}.
It can also be used to decompose
the MCA energy
into layer- or atom-resolved terms
\cite{Miura13,Miura18-prb,Okabayashi18},
although this decomposition is not unique \cite{MC22}
because a qualitatively different spatial distribution of this energy
is obtained based on the atom-projected density of states (DOS),
within both the FT \cite{Li13,Li14} and PT \cite{MC97} approaches.
Recently, it has been shown \cite{MC22,MC24} that
the usual PT formula for the MCA energy
\cite{Bruno89,MC94,Wang93,Miura18-prb},
which includes only interband terms (from pairs of electron states),
needs to be amended with intraband terms (from individual states)
for systems without the inversion symmetry,
such as FM/NM  bilayers.
In Co/Pd and Co/Cu bilayers,
the net  interband  and intraband contributions to the MCA energy
are of comparable magnitudes but opposite signs,
thus largely cancelling out.

The MCA energies obtained with the FT and PT are very similar
for layered systems with weak and moderate SOC strengths, such as
the Co film and the Co/Cu bilayer,
while minor differences  are
found for the Co/Pd bilayer \cite{MC22,MC24}.
However,
the discrepancy between the FT and PT results for the MCA energy
becomes significant for the Co/Pt bilayer,
which includes a NM layer with strong SOC.
 Substantial differences between the FT and PT  MCA energies
are also found for CoPt and FePt alloys \cite{Ke19,BlancoRey19}
as well as the thinnest Fe/Pt bilayer
of 2 monolayer (ML) thickness \cite{BlancoRey19}.

The discrepancy between the FT and PT predictions
for the Co/Pt bilayer is particularly pronounced
for the oscillations of the MCA energy, with the oscillation
amplitude a few times larger  in the PT calculations. 
Oscillations of the MCA energy are predicted in numerous theoretical calculations
\cite{MC94,MC97,MC01,MC03,MC22,MC24,Guo99,Sandratskii15,Ong16,Chang17,Qiao18,Werwinski24}
and have also been observed in experiments
for layered systems with varied thicknesses
of Co, Fe, Cu, Pd and Au layers \cite{MP09,MP11,MCMP13,MP16,Slezak20}.
These oscillations are attributed to quantum-well (QW) $d$ states
 in ferromagnetic layers of Co and Fe \cite{MC03,MCMP12,Chang17},
 and  non-magnetic Pd layers \cite{MC98,MC01,MP16},
 as well as to QW $sp$ states in Cu and Au layers \cite{MCMP13,Slezak20}.

This work aims to investigate, in  depth,
inaccuracies of the MCA energy calculated with the PT.
To elucidate the origin of the discrepancies
between the FT and PT results,
the present study  first examines
the  Co film with nominal and enhanced SOC
(Sec. \ref{sec-mca-Co-film}).
Further (Sec. \ref{sec-why-PT-fails}),
the MCA oscillations versus the Co thickness are analyzed
for a model subsystem of QW state pairs
in the central region of  the Brillouin zone (BZ)
which are responsible for the dominating term of these oscillations.
In the second part (Sec \ref{sec-mca-CoNM-bilayers}),
the MCA energies obtained with the FT and PT
are examined for the Co/Pt bilayer, with a focus on the differences
in their oscillations.

\section{THEORY}

In the presence of the SOC,
the energies $\epsilon_m({\bf k})$ of electron states  $|m{\bf k}\ket$
in a ferromagnetic film
or a system with ferromagnetic and nonmagnetic layers
depend on the direction of the magnetization $\bf M$;
for each wave vector ${\bf k}=(k_x,k_y)$,
these states are labelled with a single band index $m$
as the spin is no longer a good quantum number due to the SOC.
With the aid of  the FT,
the MCA energy at zero temperature is then defined
as the difference of band energies
$E_{\text{b}}({\hat{\bf M}})$
[sums of the energies $\epsilon_m({\bf k})$ of the occupied states]
for the out-of-plane and in-plane
magnetization directions, denoted as
${\hat{\bf M}}_{\perp}$ (along the $z$ axis) and
${\hat{\bf M}}_{||}$ (along the $x$ axis), respectively.
If this approach is extended
to finite temperatures the MCA energy $E_{\rm MCA}$ is calculated,
in the canonical ensemble, as the difference
$\Delta F = F({\hat{\bf M}}_{\perp})- F({\hat{\bf M}}_{||})$
of the free energies for the two magnetization directions
\cite{MC22,MC24}.
The shape anisotropy energy $E_{\rm dip}$ due to
the magnetic dipolar interaction \cite{MC94}
must be added to $E_{\rm MCA}$ to obtain the total magnetic anisotropy
energy the sign of which determines the preferable direction of magnetization.

The MCA energy can also be determined
within the grand canonical ensemble
(cf. Ref. \cite{MC97}),
\begin{equation}
\label{eq-mca-omega-FT}
E_{\text{MCA}}=E_{\text{MCA}}^{\text{FT}}=
\Omega({\hat{\bf M}}_{\perp})- \Omega({\hat{\bf M}}_{||}) \, ,
\end{equation}
where the grand potential $\Omega$ at temperature $T$
\begin{equation}
\label{eq-omega-def}
\Omega({\hat{\bf M}})=
\frac{1}{N_{\text{2D}}}
\sum_{m \bf k}g[\epsilon_{m}({\bf k})]
\end{equation}
is defined with the function
$g(\epsilon)=-k_{\text{B}}T
\ln \{1+\exp[(\epsilon_{\text{F}}-\epsilon)/k_{\text{B}}T ] \}$;
here, the $\epsilon_{\text{F}}$ is the Fermi energy
(or, more precisely, the chemical potential)
and $k_{\text{B}}$ denotes the Boltzmann constant.
With
$N_{\text{2D}}$ equal to
the number  $\bf k$ points in
the two-dimensional Brillouin zone (BZ),
the MCA energy is defined per primitive unit surface cell,
or per surface atom for systems with one atom
per a cell.
The value of $\epsilon_{\text{F}}$
is fixed and corresponds to the number of electrons $N$
for one of the magnetization directions,
e.g., in-plane,
though taking  $\epsilon_{\text{F}}$ 
for the other direction gives nearly identical results
for $E_{\rm MCA}^{\rm FT}$.
The temperature $T$  is originally introduced to improve the convergence of the integration over the BZ by smearing the Fermi level.
However, this also has a physical effect by reducing the
 amplitude of the MCA energy oscillations with increasing the film thickness
 as predicted theoretically \cite{MC97} and
 confirmed experimentally \cite{MP09,MCMP12} for Fe and Co films.

The applied definition of $E_{\rm MCA}$ using the grand potential
[Eq. (\ref{eq-mca-omega-FT})]
is convenient
since it assumes
the common Fermi level for  both  magnetization directions.
In particular,  it is suitable
for the model subsystem of QW states investigated
in Sec. \ref{sec-model-QW-states}  where
the change of the Fermi energy
with the magnetization direction,
$\Delta \epsilon_{\text{F}}=
\epsilon_{\text{F}}({\hat{\bf M}}_{\perp})-
\epsilon_{\text{F}}({\hat{\bf M}}_{||})$,
would be needed as an extra external parameter
(determined for the whole system)
if the MCA energy was defined  as the free energy difference $\Delta F$.
A similar approach, also based on the grand potential,
though with
$\Omega=E_{\text{b}}-\epsilon_{\text{F}}N$ at zero temperature,
was successfully used to calculate the MCA energy
of Co and Fe films in Refs. \cite{Li13,Li14}.
Nevertheless, as noted therein, the  MCA energies
obtained with the FT in the canonical and grand canonical ensembles
are (almost) the same as confirmed by the present results
in Sec. \ref{sec-mca-Co-film}.

Another method of calculating the MCA energy
is provided by  the PT with
the spin-orbit interaction $H_{\text{so}}$
treated as a perturbation in the system Hamiltonian
$H+H_{\text{so}}$
where $H$ is the unperturbed Hamiltonian
describing  the system in the absence of the SOC.
For each electron state, its energy
 $\epsilon_{m}=\epsilon_{n\sigma}^{\text{per}} =\epsilon_{n\sigma}  + \epsilon_{n\sigma}^{(1)} + \epsilon_{n\sigma}^{(2)}$
 is then represented with the first-order and second-order
corrections to the energy
 $\epsilon_{n\sigma}=\epsilon_{n\sigma}({\bf k})$
 of the unperturbed state
 $|n\sigma{\bf k}\rangle$ labelled
 with band index  $n$  for each spin $\sigma$ ($\uparrow$ or $\downarrow$).
In the resulting expansion of the grand potential
$\Omega=\Omega_0 + \Omega^{(1)} + \Omega^{(2)}$,
the first-order correction $\Omega^{(1)}$ vanishes.
Indeed,
for systems with the inversion symmetry,
we have $\epsilon_{n\sigma}^{(1)}=0$ at each  $\bf k$,
while, in the absence of this symmetry,
although the corrections $\epsilon_{n\sigma}^{(1)}({\bf k})$
can be finite,
the contributions to $\Omega^{(1)}$
from  $\bf k$ and $- \bf k$ cancel out
due to the relation
$\epsilon_{n\sigma}^{(1)}(-{\bf k})=
- \epsilon_{n\sigma}^{(1)}({\bf k})$ \cite{MC22}.
As a result, the MCA energy for layered system
is given by the following second-order PT expression
\cite{MC97,MC22,MC24}
\begin{equation}
\label{eq-mca-omega-PT}
E_{\text{MCA}}=E_{\text{MCA}}^{\text{PT}}=
\Omega^{(2)} ({\hat{\bf M}}_{\perp})-
\Omega^{(2)} ({\hat{\bf M}}_{||})
\end{equation}
where 
\begin{eqnarray}
\label{E2-corr}
\Omega^{(2)}({\hat{\bf M}}) & = &
\frac{1}{2} \frac{1}{N_{\text{2D}}} \sum_{{\bf k} } 
\sum_{\sigma, \sigma'}
\sum_{n,n'}
\frac{f_0(\epsilon_{n\sigma}({\bf k}))-f_0(\epsilon_{n'\sigma'}({\bf k}))}%
{\epsilon_{n\sigma}({\bf k})-\epsilon_{n'\sigma'}({\bf k})}
\nonumber   \\
&& \times
\left|
\langle  n'{\bf k}\sigma'|H_{\text{so}}| n{\bf k}\sigma\rangle
\right|^2
\end{eqnarray}
is calculated with  the occupation factor
$f_0(\epsilon)=f(\epsilon; \epsilon_{\text{F}}=\epsilon_{\text{F0}} )$
corresponding to the unperturbed system with the Fermi energy
$\epsilon_{\text{F0}}$.
The identical corrections are found for the free energy,
$F^{(1)}=\Omega^{(1)}=0$ and $F^{(2)}=\Omega^{(2)}$
(see Appendix in Ref. \onlinecite{MC22})
so that the same PT formula is obtained  for the MCA energy
defined as the difference of the free energies.
This formula can be also represented
(see, e.g., Ref. \cite{BlancoRey19}) in a fully equivalent form
where the ratio
$\frac{1}{2}(f_{n\sigma}-f_{n'\sigma'})/%
(\epsilon_{n\sigma}-\epsilon_{n'\sigma'})$
[with  $f_{n\sigma}$ denoting $f_0(\epsilon_{n\sigma})$]
is replaced with the non-symmetrical expression
$f_{n\sigma}(1-f_{n'\sigma'})/%
(\epsilon_{n\sigma}-\epsilon_{n'\sigma'})$.
That alternative formula reproduces Eq. (\ref{E2-corr})
once the contributions from the pairs
of states $(n\sigma,n'\sigma')$ and $(n'\sigma',n\sigma)$
are summed.

The MCA energy defined with
Eqs. (\ref{eq-mca-omega-PT}) and (\ref{E2-corr})
is given by
the sum of contributions from pairs of electrons states,
with same and opposite spins,
\begin{equation}
\label{eq-mca-PT-ns_n_np}
E_{\text{MCA}}^{\text{PT}}=
\frac{1}{2} \frac{1}{N_{\text{2D}}} \sum_{{\bf k} }
\sum_{n\sigma, n'\sigma'} E_{\text{MCA}}^{n\sigma, n'\sigma'}({\bf k}) \,
\end{equation}
which results in the representation of the MCA energy
as the sum of spin-pair terms
\begin{equation}
\label{eq-mca-PT-spin-pairs}
E_{\text{MCA}}^{\text{PT}}=
\sum_{\sigma, \sigma'} E_{\text{MCA}}^{\sigma \sigma'} \, .
\end{equation}
The terms with different spins,
$E_{\text{MCA}}^{\downarrow\uparrow}$
and $E_{\text{MCA}}^{\uparrow\downarrow}$,
comprise only interband contributions
[$(n'\sigma')\neq(n\sigma)$] while
the terms with the same spins include both interband and interband
[$(n'\sigma)=(n\sigma)$]
contributions,
\begin{equation}
\label{eq-mca-PT-spin-pairs-intra-inter}
E_{\text{MCA}}^{\sigma \sigma}=
E_{\text{MCA,inter}}^{\sigma \sigma}
+ E_{\text{MCA,intra}}^{\sigma \sigma} \,
\end{equation}
($\sigma=\downarrow$ and $\sigma=\uparrow$);
see Refs. \onlinecite{MC22,MC24}.

While the second-order interband terms come from
electron energy corrections $\epsilon_{n\sigma}^{(2)}$
of the same order,
finite second-order intraband  terms
$f'_0(\epsilon_{n\sigma})[\epsilon_{n\sigma}^{(1)}]^2$
originate  from first-order corrections
$\epsilon_{n\sigma}^{(1)}=
\langle  n{\bf k}\sigma|H_{\text{so}}| n{\bf k}\sigma\rangle$
in the absence of the inversion symmetry.
These terms arise from the power series expansion of the
$g(\epsilon_{m}=\epsilon_{n\sigma}^{\text{per}})$
terms of $\Omega$ [Eq.~(\ref{eq-omega-def})]
around $\epsilon=\epsilon_{n\sigma}$ \cite{MC22,MC24}
and 
reflect small shifts
of the Fermi surface sheets
(or rather the Fermi contour lines in two-dimensional $\bf k$ space)
 due to the SOC \cite{MC24}.
The intraband terms are not included in the original PT formula \cite{Bruno89}
 where the changes in the occupations
 of the electron states are neglected.
 However, the inclusion of such terms in the MCA energy
 is shown to be crucial for systems
 without the inversion symmetry, like Co/Cu and Co/Pd bilayers \cite{MC24}.
It is found that the total intraband and interband terms of this energy,
\begin{equation}
\label{eq-mca-PT-intra}
E_{\text{MCA,intra}}= E_{\text{MCA,intra}}^{\downarrow \downarrow }
                      + E_{\text{MCA,intra}}^{\uparrow \uparrow }
\end{equation}
and
\begin{equation}
\label{eq-mca-PT-inter}
E_{\text{MCA,inter}}=   E_{\text{MCA,inter}}^{\downarrow \downarrow }
                      + E_{\text{MCA,inter}}^{\uparrow \uparrow }
                      + E_{\text{MCA}}^{\downarrow \uparrow }
                      + E_{\text{MCA}}^{\downarrow \uparrow } \, ,
\end{equation}
respectively, are of comparable magnitude, though of opposite signs,
  so that they largely cancel out in the sum
\begin{equation}
\label{eq-mca-PT-sum-intra-inter}
E_{\text{MCA}}^{\text{PT}}=
E_{\text{MCA,intra}} + E_{\text{MCA,inter}} \,
\end{equation}
which  gives the total MCA energy in the PT approach.

The intraband terms of the MCA energy differ
from  the unconventional second-order terms identified in
Refs. \onlinecite{Hubner96,Hubner97}.
Although they both  come from changes of electron energies
linear in the SOC strength and are localized along lines
in the BZ, the  latter terms do not originate from
the first-order PT corrections $\epsilon^{(1)}$.
Furthermore, unlike the intraband terms, they are not related to
the lack of the inversion symmetry.
Instead, they arise under specific conditions:
when two energy bands are degenerate
near the Fermi level along a line
in the $\bf k$ space
with no or very weak dispersion along that line.
Then, if this degeneracy is  lifted by the SOC
for only one of the two magnetization directions,
it results in a second-order contribution to the MCA energy
from a narrow stripe-shaped region around the line.
In contrast, the intraband contributions to the MCA energy arise from each band
that intersects the Fermi level.
This occurs  at  $\bf k$ points which form a line,
the respective line of the Fermi contour  in the two-dimensional BZ.
Formation of the Fermi contour in thin films,
represented by a set of discrete lines,
can also be illustrated by discretization
of the bulk Fermi surface into multiple
circular lines within a free-electron model,
as described in Ref. \onlinecite{Balcerzak06}.

The spin-orbit interaction is given
by the following sum over sites $j$ 
in atomic layers $l$ of constituent transition metals
with the SOC constants $\xi_{l}$,
\begin{equation}
\label{eq-hso}
H_{\text{so}}=\sum_{lj} \xi_{l}
{\bf L}({\bf r}-{\bf R}_{lj})\cdot {\bf S} \, .
\end{equation}
It includes the operator of electron spin $\bf S$
and the operators of orbital angular momentum ${\bf L}({\bf r}-{\bf R}_{lj})$
which depend on the electron position  $\bf r$
with respect to the consecutive atomic locations ${\bf R}_{lj}$.
Although the SOC operator itself does not depend on the direction of the magnetization, such dependence arises \cite{Abate-Asdente65,Li90,MC22}
 when this operator is applied to the   electron states
  $\psi({\bf r})
  =\psi^{\uparrow}({\bf r})|\uparrow\ket +
  \psi^{\downarrow}({\bf r})|\downarrow\ket $
  represented with two spatial wave functions
  $\psi^{\uparrow}({\bf r})$  and $\psi^{\downarrow}({\bf r})$
  in  the spin basis $|\sigma\ket$  ($\sigma=\uparrow$, $\downarrow$)
  corresponding to the eigenstates of the operator $S_{\zeta}$ of spin component
  along the magnetization direction ${\hat{\bf M}}$.
The spin operator is then represented as ${\bf S}=(S_{\xi},S_{\eta},S_{\zeta})$
in the rotated frame of reference $O\xi\eta\zeta$
with the $\xi$ and $\eta$ axes
perpendicular to the spin quantization axis $\zeta $
while the fixed frame of reference $Oxyz$
is used for ${\bf L}=(L_x,L_y,L_z)$
and linked to the system geometry with the $z$ axis
oriented perpendicular to the
film surface and the $x$ and $y$ axes parallel to it.
For the considered (001) fcc films, the $x$, $y$ and $z$ axes
are along (100), (010), and (001) crystallographic directions, respectively.

\begin{figure}
        \includegraphics*[width=7.5cm]{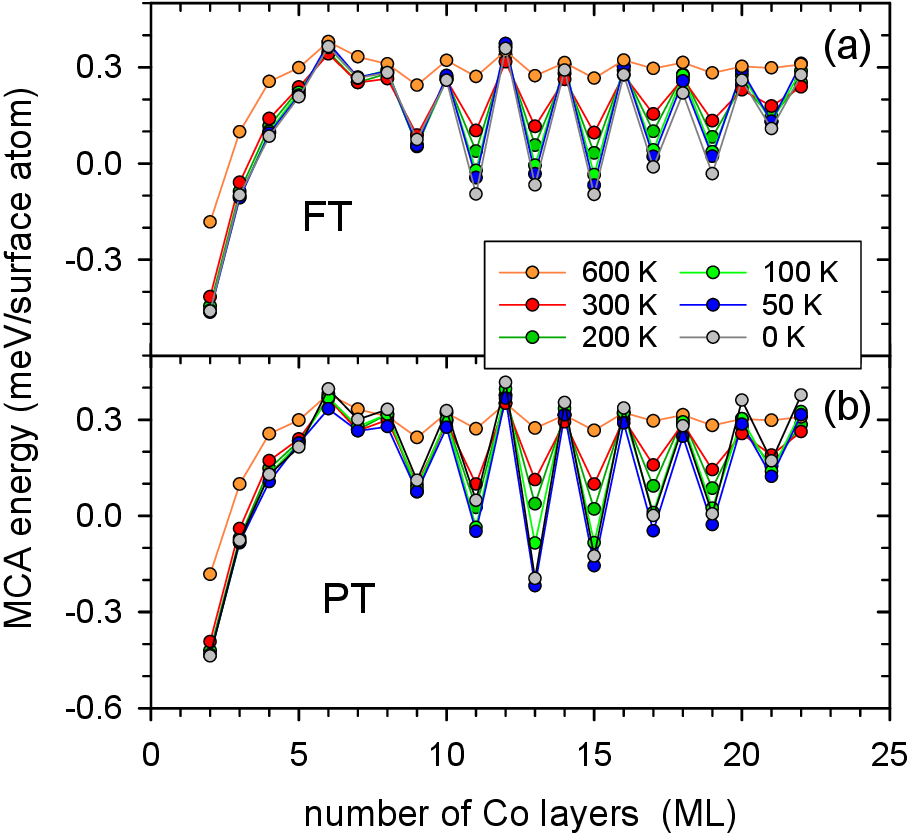}
         \caption{
           MCA energy of the (001) fcc  Co film with the nominal SOC strength ($\xi=\xi_{\text{Co}}$) calculated  with (a) the FT and (b) the PT at  various temperatures.
    }
  \label{fig-Co_film_eso_per_multi_temp}
\end{figure}

\begin{figure}
        \includegraphics*[width=7.5cm]{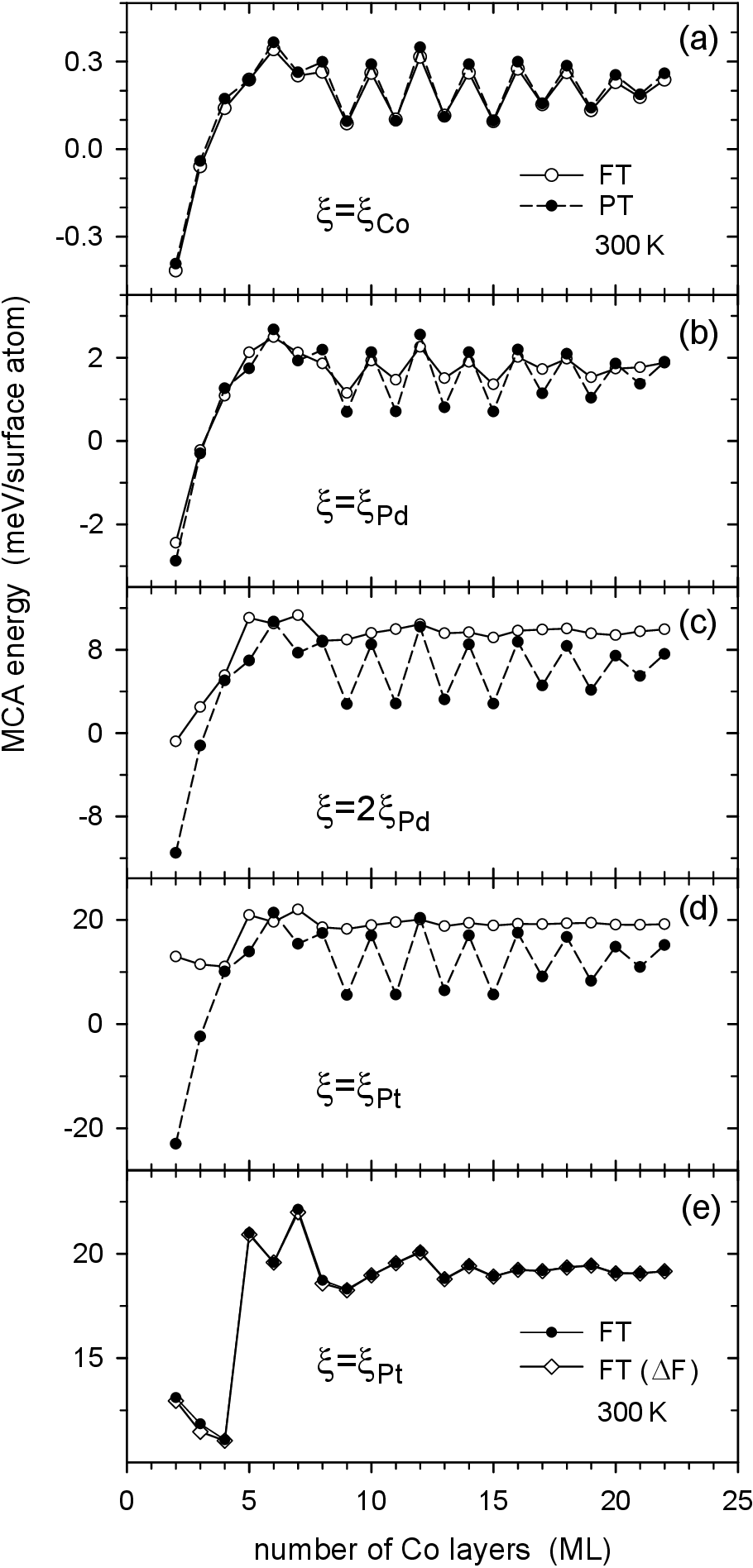}
      \caption{
      MCA energy of the (001) fcc  Co film calculated
      with (a-e) the FT [Eq. (\ref{eq-mca-omega-FT})]
      and (a-d) the PT [Eq. (\ref{eq-mca-omega-PT})]
      for various strengths $\xi$ of the SOC coupling
      at  $T=300$~K. For comparison, the results of the alternative FT formula
     $\Delta F= F({\bf M}_{\perp})-F({\bf M}_{||})$
     for $\xi=\xi_{\text{Pt}}$ at the same $T$ are also shown in panel (e). }
  \label{fig-Co_film_eso_per_var_xi_300K}
\end{figure}

\begin{figure}
        \includegraphics*[width=6cm]{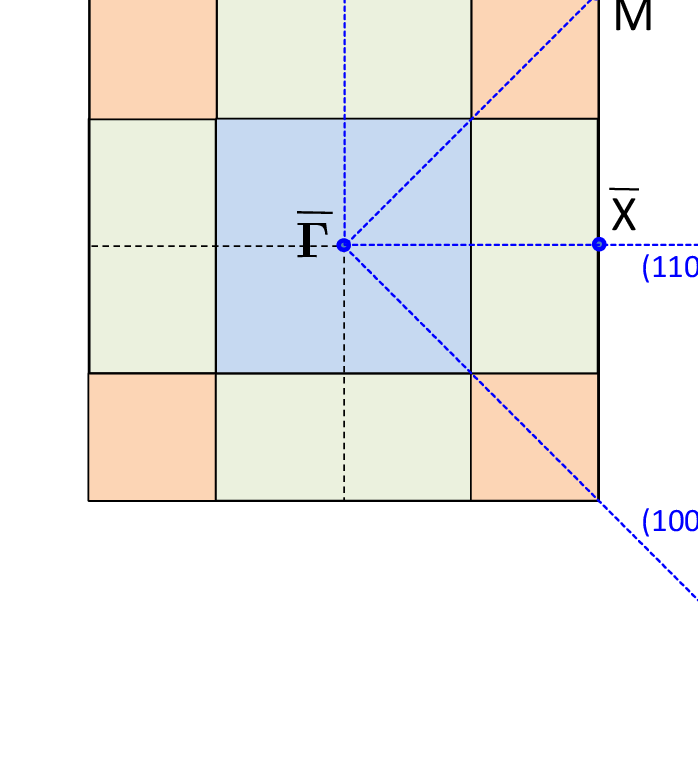}
      \caption{Division of the BZ into
      the $\overline{\Gamma}$ region (blue),
       the $\overline{M}$ region (orange),
       and the $\overline{X}$ region (green),
       centred around the respective high-symmetry and
       equivalent $\bf k$-points.
    }
  \label{fig-Brillouin_zone}
\end{figure}

\begin{figure}[t]
   \includegraphics*[width=7.5cm]{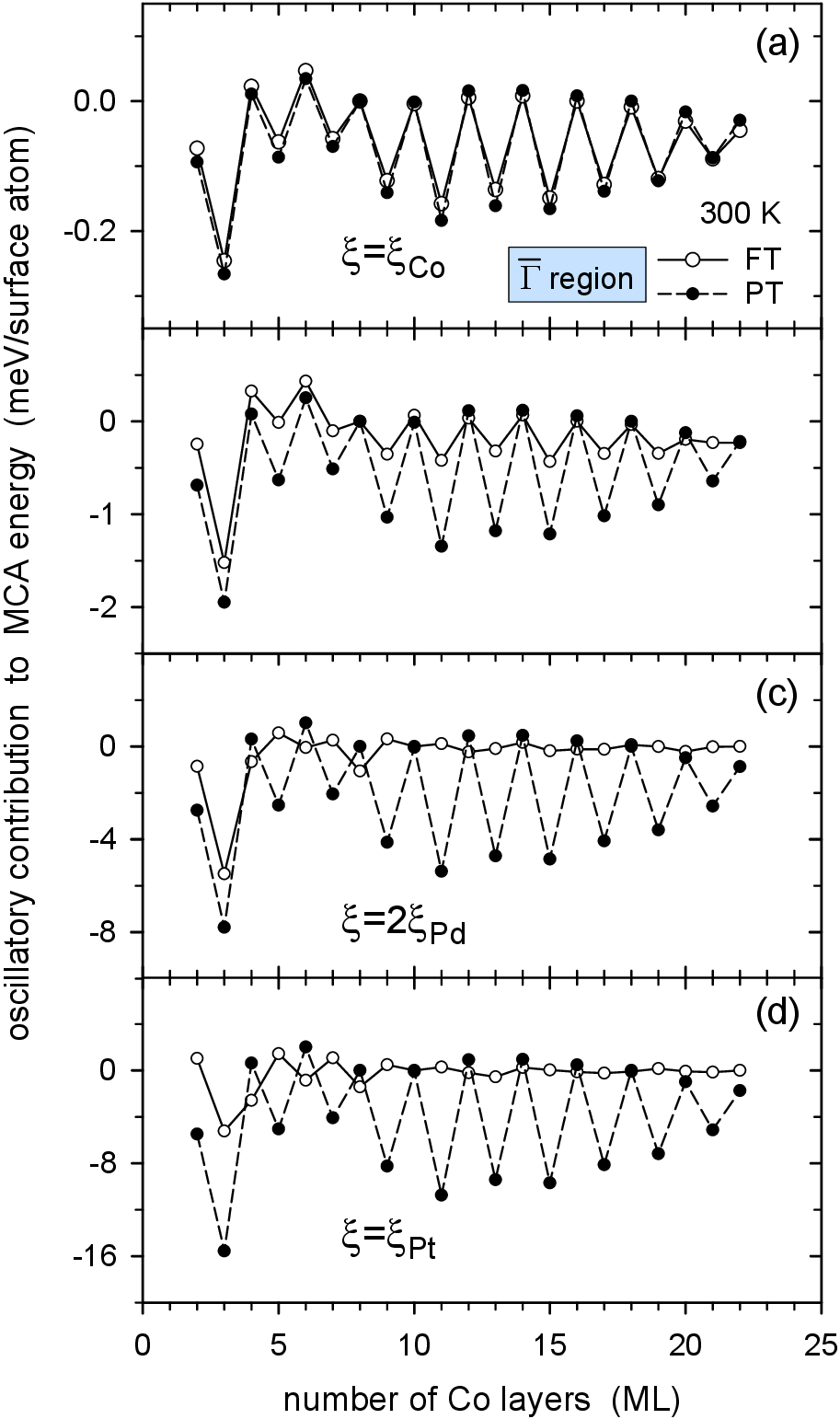}
      \caption{
      (a-d) Oscillatory component of the contribution to the MCA energy from the $\overline{\Gamma}$ region in the BZ
      (Fig.~\ref{fig-Brillouin_zone}), calculated  with the FT and the PT
      for the (001) fcc  Co film with various SOC strengths $\xi$
      at $T=300$~K.
    }
  \label{fig-Co_film_osc_eso_per_var_xi_gam05_300K}
\end{figure}

\begin{figure}
   \includegraphics*[width=7.5cm]{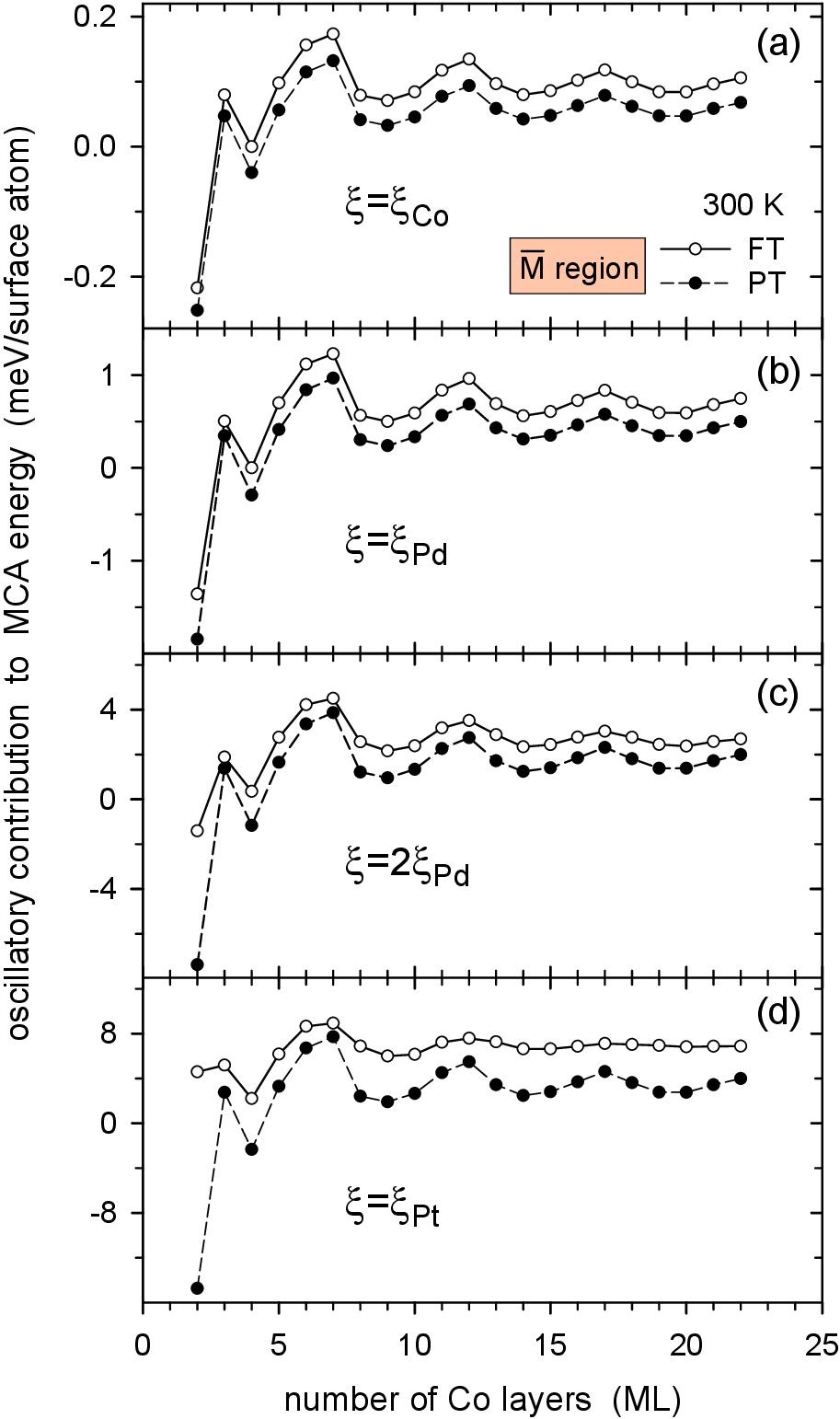}      %
      \caption{
      (a-d) Oscillatory component of the contribution to the MCA energy from the $\overline{M}$ region in the BZ  (Fig.~\ref{fig-Brillouin_zone}),
      calculated  with the FT and the PT for the (001) fcc  Co film
      with various SOC strengths $\xi$ at $T=300$~K. The FT and PT plots
       are vertically shifted with respect to each other for easier comparison.
      In each panel, the energy scale is the same as in  Fig.~\ref{fig-Co_film_osc_eso_per_var_xi_gam05_300K}.
    }
  \label{fig-Co_film_osc_eso_per_var_xi_m05_300K}
\end{figure}

\begin{figure}
   \includegraphics*[width=7.5cm]{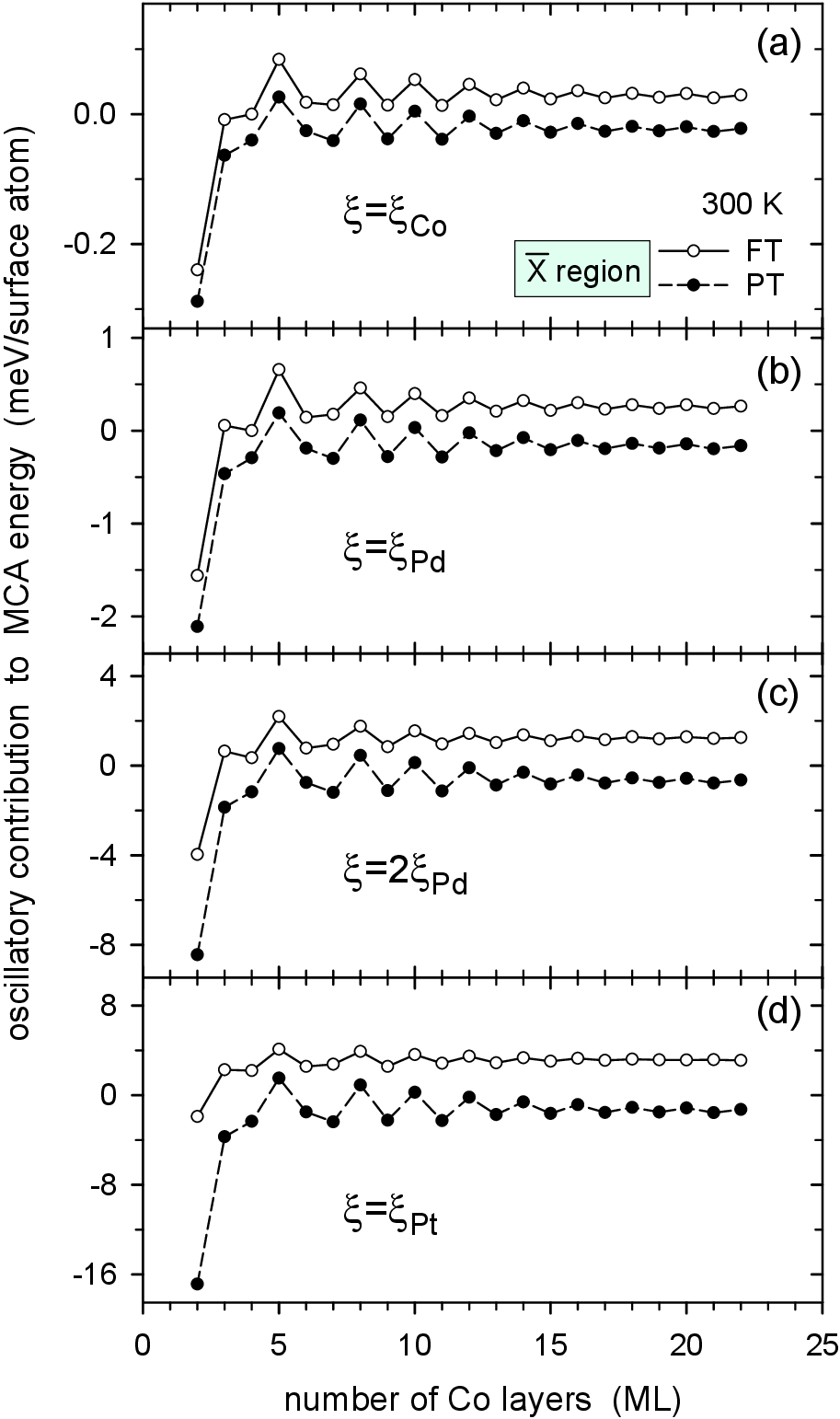}      %
      \caption{
      (a-d) Oscillatory component of the contribution to the MCA energy from the $\overline{X}$ region in the BZ (Fig.~\ref{fig-Brillouin_zone}),
      calculated  with the FT and the PT for the (001) fcc  Co film with various SOC strengths $\xi$ at $T=300$~K. The FT and PT plots
       are vertically shifted with respect to each other for easier comparison.
       In each panel, the energy scale  is the same as in  Fig.~\ref{fig-Co_film_osc_eso_per_var_xi_gam05_300K}.
    }
  \label{fig-Co_film_osc_eso_per_var_xi_x05_300K}
\end{figure}

The electronic structure is described using
a realistic TB model,
with nine basis orbitals (per atom) of $s$, $p$ and $d$ symmetries
for each spin $\sigma$ and hopping parameters fitted
to ab initio bulk bands \cite{Papaconstantopoulos86}.
The applied model also includes shifts of on-site orbital energies,
to make each atomic layer electrically neutral and account for
different orientations of orbitals at surfaces and interfaces,
as well as local exchange splittings self-consistently adjusted
to layer magnetic moments.
The details of this TB model are described in
Refs. \onlinecite{MC01,MC03, MC22}.

The use of a TB model in the custom-built code
allows for efficient calculations of the MCA energy
in thick layered systems in both the FT and PT approaches
and targeted analysis of the obtained results,
with modest computer resources.
The 0.01 meV accuracy
of the MCA energy is reached at $T=300$~K
with a moderate number
$N_{\text{2D}}=(2N_k +1)\times (2N_k +1)\approx 3700$
of $\bf k$ points using a square mesh with $N_k=30$,
while larger numbers of $\bf k$ points are needed for
convergence at lower temperatures,
up to 
$N_{\text{2D}}\approx 40000$ ($N_k=100$)
for $T=50$~K (see Ref. \onlinecite{MC24}).
The calculations of the MCA energy  are also done
for zero temperature,
using the triangle method based on division of the BZ into a triangular mesh and  linearization of electron energies within each triangle.
The triangle formula for  interband contributions
in the PT approach is derived in Ref. \onlinecite{MC94}.
The intraband term of the MCA energy at $T=0$ is given
by a sum of the Dirac delta functions
$\lim_{T\rightarrow 0} f'_0(\epsilon_{n\sigma})= -\delta(\epsilon_{n\sigma}-\epsilon_{\text{F0}})$
 multiplied by the squares of
 $\langle  n{\bf k}\sigma|H_{\text{so}}| n{\bf k}\sigma\rangle$.
Accordingly, the triangle method for the projected DOS
\cite{Freeman79} can be readily adopted
for numerical integrations of such intraband contributions
\cite{MC24}.
In the FT approach, the grand potential
$\Omega=E_{\text{b}}-\epsilon_{\text{F}}N$
 is calculated
with the analytical formula for the integral
$\Omega=\int_{-\infty}^{\epsilon_{\text{F}}}
(\epsilon -\epsilon_{\text{F}}) n(\epsilon) d \epsilon$
where the DOS $n(\epsilon)$
is also determined with the triangle method \cite{Jepsen71}.
The square mesh of $\bf k$ points with $N_k=100$ and each square divided
into two triangles is found sufficient for convergence in both the FT and PT calculations at zero temperature.

\section{RESULTS AND DISCUSSION}

\subsection{Magnetocrystalline anisotropy of Co film
with enhanced spin-orbit coupling. Effect of temperature}
\label{sec-mca-Co-film}

Validity of the PT in calculations of the MCA energy
and possible inaccuracies of this method
are conveniently investigated for a simple ferromagnetic system of
the (001) fcc Co film with a modified strength $\xi$ of the SOC.
The effect of temperature on
the deviations of the PT predictions
from the exact results of the FT calculations
is also studied.
For a fcc Co film with the nominal SOC constant ($\xi=\xi_{\text{Co}}$),
 it is found (Fig. \ref{fig-Co_film_eso_per_multi_temp})
 that the MCA energies obtained
 with the FT and PT are almost identical at $T=300$~K but differ significantly
 when the temperature becomes low enough.
At $T=50$ and $100$~K,
the oscillations of 
the MCA energy with increasing the Co film thickness $N_{\text{Co}}$
have significantly larger amplitude in the PT approach than in the FT calculation.

\begin{figure*}[t]
   \includegraphics*[width=16cm]{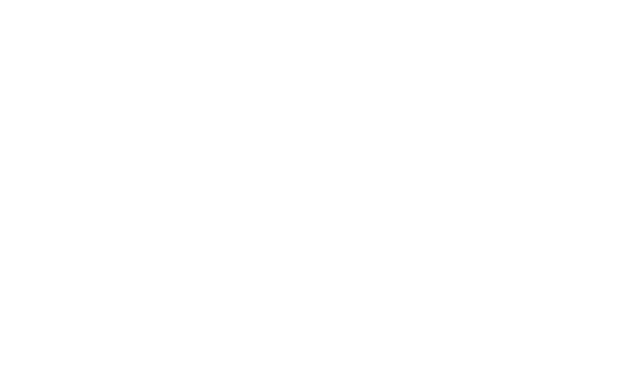}
      \caption{
     Distribution of MCA energy in the BZ for
     the fcc (001)(a,d) Co(11 ML)
     and (b,e) Co(12 ML) films
     with the nominal SOC strength ($\xi=\xi_{\text{Co}}$)
     at  $T=300$~K, obtained with the FT (top row) and the PT (bottom row).
      (c,f)~Difference between the 12 ML and 11 ML distributions.
      The plots show the symmetrized distributions
       in one quarter of the BZ (see Appendix  \ref{app-symmetrization}). }
  \label{fig-ma-bz-Co-film-FT-PT-xiCo}
\end{figure*}

\begin{figure*}
   \includegraphics*[width=16cm]{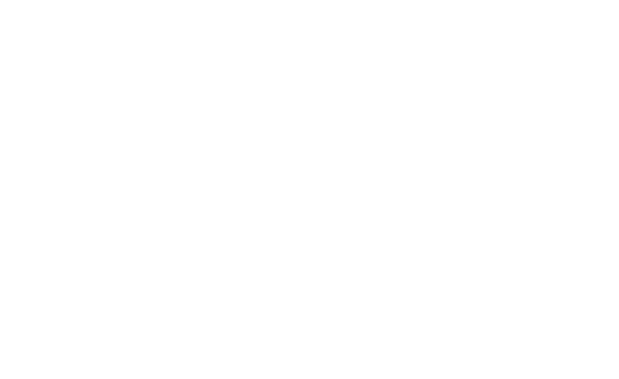}
      \caption{
     Distribution of MCA energy in the BZ for
     the fcc (001) (a,d) Co(11 ML)
     and (b,e) Co(12 ML)  films
     with the  SOC strength of Pt ($\xi=\xi_{\text{Pt}}$)
     at  $T=300$~K,obtained with the FT (top row) and the PT (bottom row).
     (c,f)~Difference between the 12 ML and 11 ML distributions.
     The plots show the symmetrized distributions
 (see Appendix  \ref{app-symmetrization}).}
  \label{fig-ma-bz-Co-film-FT-PT-xiPt}
\end{figure*}

To test the effect of the SOC strength
we consider Co films with the nominal  SOC constant
$\xi_{\text{Co}}=0.085\text{ eV}$, as well as
that of Pd $\xi=\xi_{\text{Pd}}=0.23\text{ eV}$,
 its double $\xi=2\xi_{\text{Pd}}=0.46\text{ eV}$,
 and finally the SOC constant of Pt $\xi=\xi_{\text{Pt}}=0.65\text{ eV}$
 \cite{EBMC14,MC22}.
 The MCA energies for such systems are determined with both the FT and the PT
at $T=300$~K.
It is found (Fig. \ref{fig-Co_film_eso_per_var_xi_300K}) that
 the PT reproduces the FT result for $E_{\text{MCA}}$ very well only for $\xi=\xi_{\text{Co}}$
  while   significant discrepancies between the PT and FT  predictions
  for the MCA energy
 are present for the films with the stronger SOC.
 This particularly concerns
 the oscillatory variation of $E_{\text{MCA}}$ with increasing
 the Co film thickness $N_{\text{Co}}$.
 For $\xi=\xi_{\text{Pd}}$,
 the amplitude of the dominating oscillations with
 the 2 ML period is over two times larger
in the PT calculations than when using the FT.
Further, for the two largest SOC constants,
$\xi=2\xi_{\text{Pd}}$ and $\xi=\xi_{\text{Pt}}$,
 the oscillations of the MCA energy calculated
 with the PT are very large compared to those in the FT approach.
This implies that the oscillation amplitude obtained with the FT
for strong SOC does not follow
the $\xi^2$ dependence predicted by the PT for the MCA energy.

However, if we focus on the upper envelope of the oscillatory variation,
the magnitudes of the corresponding MCA energies obtained with
the FT and the PT show good agreement even for strong SOC
(Fig.  \ref{fig-Co_film_eso_per_var_xi_300K}).
Thus, the total MCA energy (its upper envelope) in  the FT approach
scales as $\xi^2$, in contrast to its oscillations at large $\xi$.
The choice of the upper envelope is dictated by the fact that,
within the PT calculations,
the 2 ML period oscillations of the MCA energy
$E_{\text{MCA}}^{\text{PT}} =
\Omega^{(2)}({\bf M}_{\perp})-\Omega^{(2)}({\bf M}_{||})$
in the Co film
originate from the {\em negative} contributions  to
$\Omega^{(2)}$ for ${\bf M}={\bf M}_{\perp}$
which arise from QW states
crossing the Fermi level \cite{MC03,MCMP12}.
These contributions, which determine the position of  the lower envelope
in the $E_{\text{MCA}}^{\text{PT}}(N_{\text{Co}})$ variation,
become smaller with increasing temperature. Consequently,
the corresponding MCA oscillations become smaller
while the upper envelope of $E_{\text{MCA}}^{\text{PT}}(N_{\text{Co}})$
remains almost unchanged; see Fig. \ref{fig-Co_film_eso_per_multi_temp}.

Note also that the MCA energies obtained
with the FT using the grand potential
$\Omega$ [Eq. (\ref{eq-mca-omega-FT})]
are almost identical to the respective MCA energies defined,
also within the FT approach,
as the free energy difference $\Delta F$.
This is illustrated for the Co film with the SOC of Pt in
Fig. \ref{fig-Co_film_eso_per_var_xi_300K}(e),
where the difference between the results of the two types of the FT calculations
is hardly noticeable while for weaker SOC
these  two MCA energies
are indistinguishable  in the scales of Fig. \ref{fig-Co_film_eso_per_var_xi_300K}(a)-(c).

To further analyze the  described discrepancies between the FT and PT results
at different strengths of the SOC
we examine the contributions to  the MCA energies from three different regions of the two-dimensional BZ which are centred (if viewed in the
extended BZ picture) around the high-symmetry points, $\overline{\Gamma}$, $\overline{M}$ and  $\overline{X}$, correspondingly
(Figs. \ref{fig-Brillouin_zone},
\ref{fig-Co_film_osc_eso_per_var_xi_gam05_300K},
\ref{fig-Co_film_osc_eso_per_var_xi_m05_300K},
\ref{fig-Co_film_osc_eso_per_var_xi_x05_300K}).
As previously  established \cite{MC03},
the oscillations of the 2 ML period which dominate in the variation of the MCA energy at the nominal SOC strength of Co both within the PT and FT approaches
come from the central region  of the BZ around
the  $\overline{\Gamma}$ point.
In the PT approach,
oscillations from this $\overline{\Gamma}$ region are also predominant at stronger SOC, since they scale as $\xi^2$,
but they are much smaller in the FT calculations
and no longer dominate.
Accordingly,
all three parts of the BZ then contribute significantly to
the oscillatory pattern of $E_{\text{MCA}}^{\text{FT}}(N_{\text{Co}})$,
with terms  of different periods and similar amplitudes.
The region around the $\overline{M}$  point ($\overline{M}$ region)
contributes to the thickness variation of the MCA energy
with an oscillatory term of 5.15 ML period
which comes from the QW states derived from the bulk band
of the $Z_3$ symmetry \cite{MC03}.
This term is very similar in the PT and FT approaches
not only for $\xi=\xi_{\text{Co}}$
but also  for $\xi=\xi_{\text{Pd}}$,
while its oscillation amplitude is significantly smaller in
the FT approach only for $\xi=\xi_{\text{Pt}}$.
Lastly, similar conclusions hold for the  contributions from the
region around the $\overline{X}$  point, the $\overline{X}$ region.
Thus, it is found that the discrepancy between
the MCA energies obtained with the PT and FT
originates mainly from the $\overline{\Gamma}$ region.

These findings are visualized with the plots of
the MCA energy distributions within the BZ
(Figs. \ref{fig-ma-bz-Co-film-FT-PT-xiCo}
and \ref{fig-ma-bz-Co-film-FT-PT-xiPt}).
For Co films with the nominal SOC ($\xi=\xi_{\text{Co}}$),
the distributions calculated with the FT and PT
are nearly identical.
However,
a significant difference emerges at strong SOC.
The peaky features  in the
PT distribution  of $E_{\text{MCA}}$
in some regions of the BZ
are largely smoothed out in the FT distribution
for the SOC of Pt.
Specifically, the PT predicts,
for any SOC strength, regular changes
in the MCA energy contributions around
the $\overline{\Gamma}$ point
with increasing the Co thickness
(from $N_{\text{Co}}=11$ to 12~ML in
Figs. \ref{fig-ma-bz-Co-film-FT-PT-xiCo}
and \ref{fig-ma-bz-Co-film-FT-PT-xiPt}),
leading to 2~ML period oscillations.
These changes
are barely present in the FT distribution at
strong SOC.
This is particularly evident in
the difference between the MCA energy distributions
in the Co(12 ML) and Co(11 ML) films
which represents the distribution
of the oscillation amplitude; see
Figs.  \ref{fig-ma-bz-Co-film-FT-PT-xiCo} (c), (f)
and \ref{fig-ma-bz-Co-film-FT-PT-xiPt}) (c), (f).
The amplitude distribution is nearly flat in the FT calculations
for the SOC of Pt,
while being almost identical to the respective PT distribution
for the SOC of Co.
Nevertheless,
if the rapid variations of MCA contributions within the BZ
are disregarded,
the overall shapes of the MCA energy distributions are  similar
in the FT and PT approaches for all considered SOC strengths.

\begin{figure}
   \includegraphics*[width=8cm]{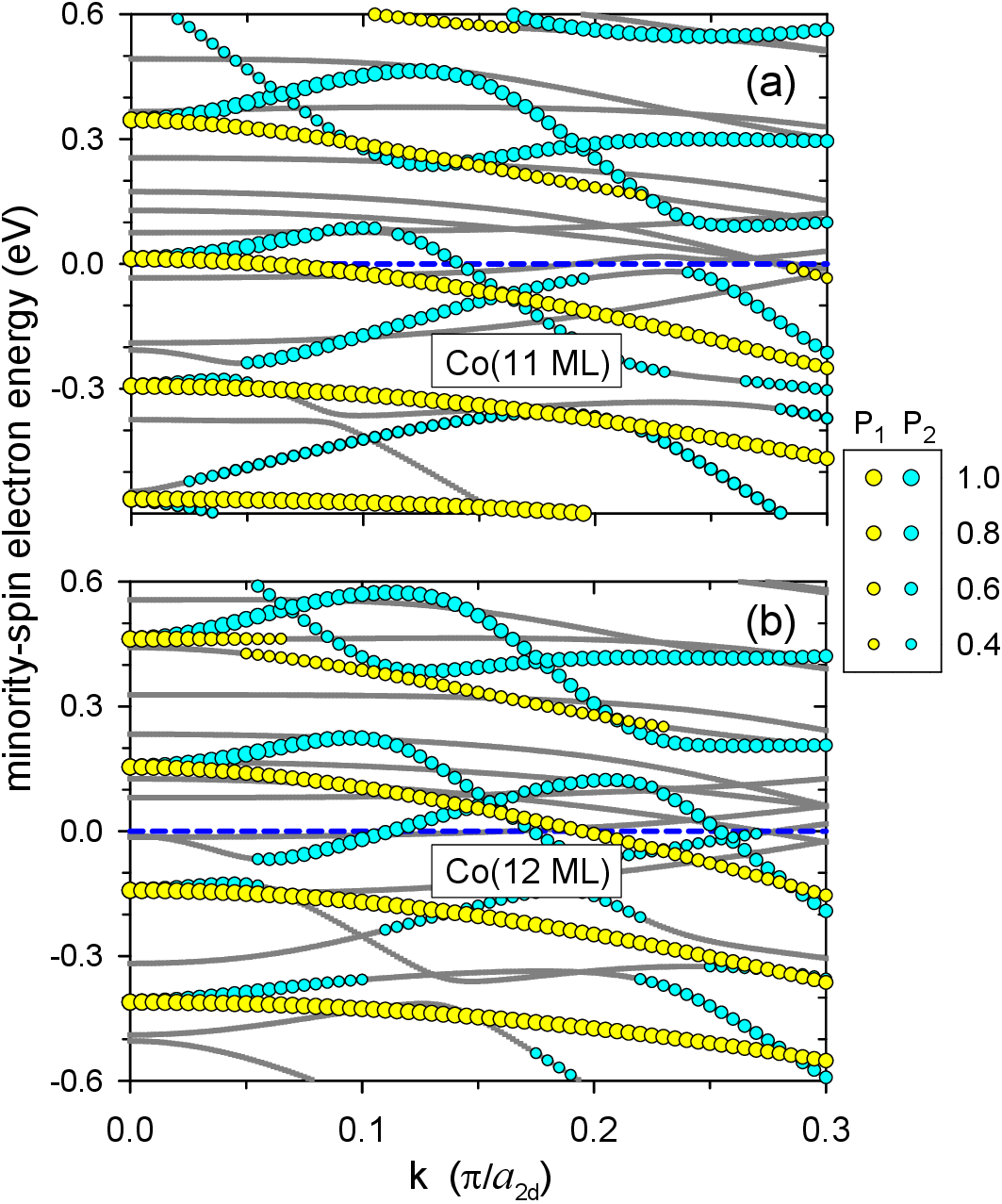}
      \caption{
      Minority-spin electron energies along the
      $\overline{\Gamma}-\overline{X}$ line near the BZ centre
      for the (001) fcc Co(11 ML) and Co(12 ML) films without the SOC.
      The colour-marked bands correspond to the states with projections
      $P_1\ge 0.4$ onto the $|\phi_1\ket=(|yz\ket -|zx\ket)/\sqrt{2}$ orbitals
      (yellow circles),
       and
       $P_2\ge 0.4 $ onto the $|\phi_2\ket=(|yz\ket +|zx\ket)/\sqrt{2}$ orbitals
       (light blue circles); $P_1$~and~$P_2$ are equal to 1 at $k=0$.
       The Fermi energy is  at $\epsilon_{\text{F0}}=0$
       (horizontal dashed line).
      }
  \label{fig-Co_film-bands-zero-soc}
\end{figure}

\begin{figure}
   \includegraphics*[width=7cm]{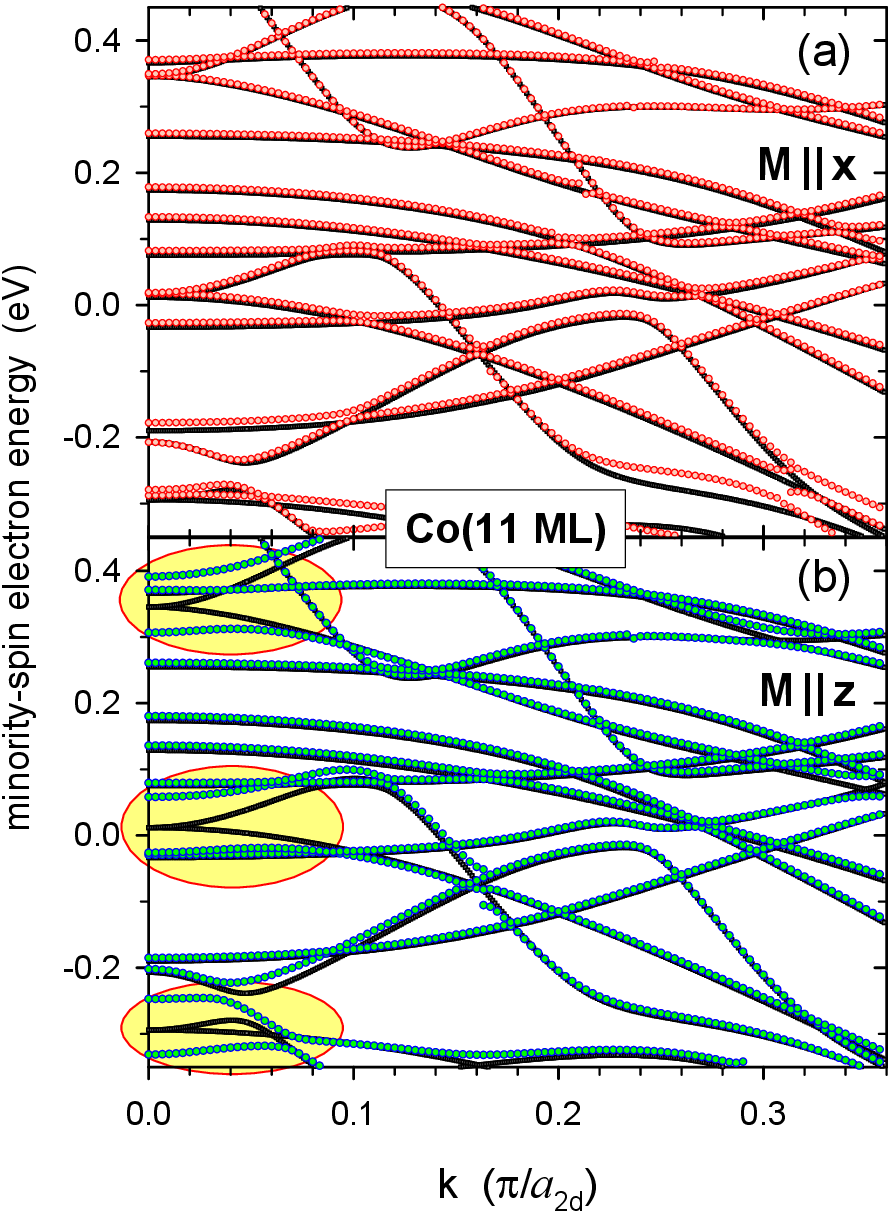}
      \caption{(a,b)~Electron energies along the
      $\overline{\Gamma}-\overline{X}$ line near the BZ centre
      for the (001) fcc Co(11 ML) film:
      minority-spin energies in the film without the SOC (black lines)
      and the energies of states with a dominating
       minority-spin  component (with a probability larger than 0.5)
      calculated with the nominal SOC ($\xi=\xi_{\text{Co}}$)
       for the (a)~in-plane magnetization ${\bf M}\,||\,x$ (red circles)
       and (b)~out-of-plane magnetization ${\bf M}\,||\,z$ (green circles).
       The yellow ovals highlight pairs of states
       which are built predominantly of the $yz$ and $zx$  orbitals and are
        split by the SOC 
        for ${\bf M}\,||\,z$. }
  \label{fig-Co_film-bands-soc}
\end{figure}

\begin{figure}
   \includegraphics*[width=8cm]{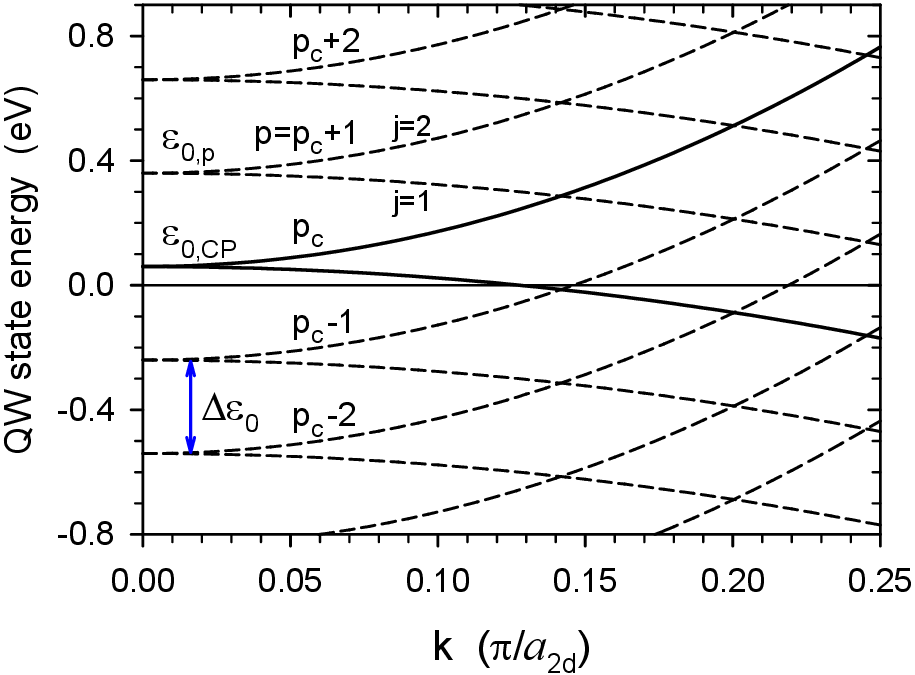}
      \caption{
      Energies $\epsilon_{j,p}$ 
      ($j=1$, $2$ and
      $p=p_{\text{c}}$, $p_{\text{c}} \pm 1$, $p_{\text{c}} \pm 2,\,\ldots$)
      of model QW state pairs in the Co film in the absence of the SOC
      [Eqs. (\ref{eq-eps1}) and (\ref{eq-eps2})]
      versus $k=|{\bf k}|$ along the $\overline{\Gamma}-\overline{X}$  line.
      The index $p=p_{\text{c}}$ denotes the central QW state pair
       with the energy
       $\epsilon_{\text{0,CP}}$ at $k=0$ (here 0.06~eV).
        The Fermi energy is at $\epsilon_{\text{F0}}=0$ (horizontal line).
       }
  \label{fig-model-two-state-series}
\end{figure}

\begin{figure}
   \includegraphics*[width=8cm]{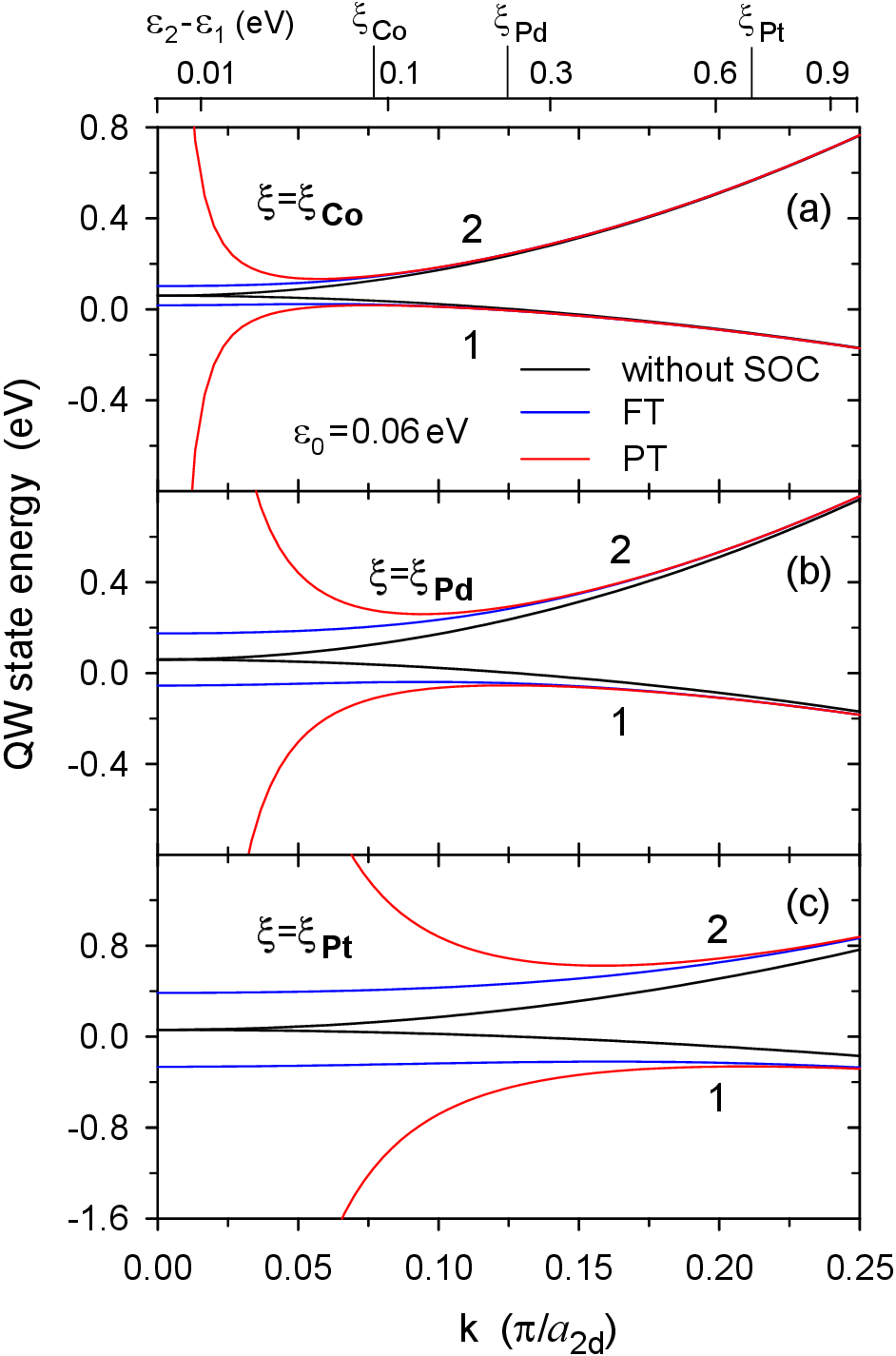}
      \caption{
      Energies $\epsilon_{1,p}$ and $\epsilon_{2,p}$
      of a model QW state pair
      [Eqs. (\ref{eq-eps1}) and (\ref{eq-eps2})]
      with $\epsilon_{0,p}=0.06$~eV
      in the Co film
      without the SOC (black lines)
       versus $k=|{\bf k}|$ along the $\overline{\Gamma}-\overline{X}$ line;
       $\epsilon_{\text{F0}}=0$.
       Corresponding energies
      $\epsilon_{1,p}^{\text{FT},\perp}$ and
      $\epsilon_{2,p}^{\text{FT},\perp}$
      (blue lines),
       and $\epsilon_{1,p}^{\text{PT},\perp}$ and
           $\epsilon_{2,p}^{\text{PT},\perp}$
      (red lines) for the out-of-plane direction of magnetization,
      obtained using the FT and the PT, respectively,
      with the SOC constants of (a) Co, (b) Pd, and (c) Pt;
      see Eqs. (\ref{eq-eps-FT}) and (\ref{eq-eps-PT}).
      The top scale shows the energy separation
       $\Delta\epsilon=\epsilon_{2,p}-\epsilon_{1,p}$ in each pair
       of QW states, with the marked values of the SOC constants. }
  \label{fig-model-two-state-exact_PT}
\end{figure}

\subsection{Why perturbation theory fails
for magnetocrystalline energy oscillations
at strong spin-orbit coupling}
\label{sec-why-PT-fails}

\subsubsection{Model subsystem of quantum-well states that come in pairs}
\label{sec-model-QW-states}

The 2 ML period oscillations of the MCA energy
in the (001) fcc Co film
come from pairs of  QW states which, in the absence of the SOC, are degenerate
at the very centre of the BZ,
the $\overline{\Gamma}$ point [${\bf k}=(0,0)$],
and originate from the minority-spin  band
of the $\Delta_5$ symmetry in bulk fcc Co \cite{MC03}.
Such pairs of QW states can be identified in the unperturbed band
structure of the (001) fcc Co film by projecting the quantum states
with $\bf k$ close to $\overline{\Gamma}$  onto the $(yz,zx)$ orbital
subspace; see Fig. \ref{fig-Co_film-bands-zero-soc}.
The energies of these states shift as the Co thickness increases.
As a result, the Fermi energy is close to the energies of a QW state pair
for some thicknesses, such as $N_{\text{Co}}=11$, 13, and 15~ML,
but lies between the energies of two neighboring pairs for other thicknesses,
such as $N_{\text{Co}}=10$, 12, and 14~ML.
Such regular energy shifts give rise to an oscillatory contribution
to the MCA energy
from the $\overline{\Gamma}$  region,
as predicted by the PT formula,
Eq. (\ref{eq-mca-omega-PT}).
Within the FT approach, the contribution to the  MCA energy
from the QW states in the $\overline{\Gamma}$ region
arises from differences in the electron energy bands
for the two magnetization directions
and this contribution can oscillate as the energies of
the perturbed QW states shift with increasing the film thickness.
For the in-plane magnetization,
the energy bands around the $\overline{\Gamma}$ point
are only slightly modified  by the SOC,
as illustrated in  Fig. \ref{fig-Co_film-bands-soc}.
In contrast, significant
energy splittings of electron occur for
the QW states at ${\bf k}=(0,0)$ and near this point
for the out-of-plane magnetization.

To further investigate the possibly different predictions of the FT and PT
for the MCA energy oscillations,
a simple model can be used which approximates
the MCA contributions from the specified pairs of minority-spin QW states.
In the absence of the SOC, the two states in each pair
(labelled with index $p$)
are degenerate at ${\bf k}=(0,0)$, both having the energy  $\epsilon_{0,p}$,
and their energies change quadratically with the
distance $k=|{\bf k}|=\sqrt{k_x^2 + k_y^2}$ from the centre of the BZ:
\begin{eqnarray}
 \epsilon_{1,p}(k) &=& \epsilon_{0,p} + b_1 k^2 \, ,
 \label{eq-eps1}\\
 \epsilon_{2,p}(k) &=& \epsilon_{0,p} + b_2 k^2 \, .
 \label{eq-eps2}
\end{eqnarray}
The dispersion parameters
$b_1$ and $b_2$ ($b_2>b_1$) depend on the direction of ${\bf k}$,
i.e., the angle $\phi=\arctan (k_y / k_x)$
so they are constant along each line starting
at $\overline{\Gamma}$. Deviations from
the quadratic dependence are neglected because,
as shown in Fig. \ref{fig-Co_film-bands-zero-soc},
they are not significant
in the interval $0\leq k \lesssim 0.1\pi/a_{\text{2d}}$
where the MCA oscillatory contributions
with the 2 ML period arise
(Figs. \ref{fig-ma-bz-Co-film-FT-PT-xiCo}
and \ref{fig-ma-bz-Co-film-FT-PT-xiPt}).
Here, the lattice constant $a_{\text{2d}}=a/\sqrt{2}$
corresponds to the two-dimensional square lattice of
the (001) fcc surface ($a$ is the fcc lattice constant).

The considered states
are built of $yz$ and $zx$ orbitals,
\begin{eqnarray}
 \psi_{1,p}({\bf k}) &=&
 c|\,p\,{\bf k}\,yz\downarrow\ket -d|\,p\,{\bf k}\,zx\downarrow\ket
  \label{eq-psi1} \\
 \psi_{2,p}({\bf k}) &=&
 d|\,p\,{\bf k}\,yz\downarrow\ket +c|\,p\,{\bf k}\,zx\downarrow\ket \, ,
 \label{eq-psi12}
\end{eqnarray}
they are orthogonal and normalized ($c^2+d^2=1$, $c$ and $d$
can be chosen real \cite{MC98}).
Their components
$|\,p\,{\bf k}\mu\sigma \ket = {\cal N}
\sum_{l} \sin(k_{z,p} z_l) |{\bf k} l\mu\sigma\ket$
are the two-dimensional Bloch states
built of the orbitals $\mu$ ($yz$ and $xz$) with
spin $\sigma=\downarrow$, located at position $z_l=la/2$
in each layer $l$
and normalized with the factor ${\cal N}$.
The actual composition of these states
depends on the
direction of the ${\bf k}$ vector,
with $c=d=1/\sqrt{2}$ for $k_x=k_y$
(along the $\overline{\Gamma}-\overline{X}$ line)
and  $c=0$, $d=1$ for $k_x=0$
(along the $\overline{\Gamma}-\overline{M}$ line).
For simplicity, it is  assumed
that the coefficients $c$ and $d$ depend so slowly on
the distance $k=|{\bf k}|$ from the centre of the BZ
that this dependence can be neglected altogether.
Contributions from other orbitals are also neglected,
in agreement with the near-unity projections ($P_1\approx 1$ and $P_2\approx 1$)
of the QW states in the Co film onto the corresponding orthogonal
combinations
 \begin{eqnarray}
 |\phi_1\ket &=& c|yz\ket - d|zx\ket \, ,\\
 |\phi_2 \ket &=& d|yz\ket + c|zx\ket
 \label{eq-phi12}
\end{eqnarray}
of the $yz$ and $zx$ orbitals; see Fig. \ref{fig-Co_film-bands-zero-soc}.

The different pairs of the QW states $\psi_{j,p}$
with energies  $\epsilon_{j,p}$ ($j=1$, $2$)
correspond to
different energies $\epsilon_{0,p}$
and discrete values of
$k_{z}=k_{z,p}\approx\pi p/[(N_{\text{Co}}+1)d]$
where $p=1,\ldots,N_{\text{Co}}$ and $d=a/2$.
This arises from the quantization of the $z$-component
of the three-dimensional wave vector $({\bf k},k_z)$
in the QW of the film.
The energy $\epsilon_{0,p}$
of the \mbox{$p$-th} pair QW states at ${\bf k}=(0,0)$
can be approximated by the energy
$\epsilon_{n\downarrow}^{\text{bulk}}({\bf k},k_z)$
of the minority-spin $\Delta_5$  bulk band
at the $(0,0,k_z=k_{z,p})$ point.
For the QW states
with $\epsilon_{0,p}$ close to
the Fermi energy,
the values of $k_{z,p}$ are close to
the extremal radius
of the Fermi surface sheet
of this bulk band,
which determines the oscillation period of 2.12 ML
associated with the considered QW states \cite{MC03}.

The model system comprises several QW state pairs
($p=1,2,\ldots,P$), with energies  $\epsilon_{0,p}$ at ${\bf k}=(0,0)$
separated by $\Delta\epsilon_0$ from each other,
as illustrated in Fig. \ref{fig-model-two-state-series}.
The assumed values of $\Delta\epsilon_0=0.3$~eV
and the dispersion parameters $b_1$ and $b_2$,
are fitted to the energy bands near  the Fermi energy
in the Co(12 ML) film
(Fig.~\ref{fig-Co_film-bands-zero-soc}).
For thicker films,
the spacing $\Delta\epsilon_0$ between the neighboring pairs
decreases due to finer quantization of $k_z=k_{z,p}$.
The considered subsystem can  be limited to a few QW state pairs
(even $P=7$ in the present calculations)
since  pairs with energies
far from the Fermi level
give negligible contributions to the MCA energy.
This is because
the respective terms in  $\Omega$
are either very small or
finite but nearly  identical for the in-plane and out-of-plane magnetizations
so they cancel out.

A key parameter that affects the MCA energy  of this subsystem is
 the energy $\epsilon_{\text{0,CP}}=\epsilon_{0,p}$
of the QW state pair  at ${\bf k}=(0,0)$
which is closest to the Fermi level.
This pair is hereafter referred to as the {\em central pair} of QW states
with index $p=p_{\text{c}}$.
Accordingly, the energies $\epsilon_{0,p}$ of other pairs at $k=0$
(where $\epsilon_{1,p}=\epsilon_{2,p}$)
are shifted with respect to $\epsilon_{\text{0,CP}}$
by the $\Delta\epsilon_0$ spacing or its multiples,
i.e., they are equal to
$\epsilon_{0,p}=\epsilon_{\text{0,CP}}+
(p-p_{\text{c}})\Delta\epsilon_0$
where $p=p_{\text{c}}\pm 1, p_{\text{c}}\pm 2, \ldots$.
The zero energy level  of $\epsilon_{\text{0,CP}}=0$
corresponds to the Fermi energy $\epsilon_{\text{F0}}=0$.
For simplicity,
the same value $\epsilon_{\text{F}}=0$ of the Fermi energy
is assumed in the model of QW states for the perturbed system.
The effects of a possible shift of the Fermi energy due to
the SOC perturbation are discussed in the last part of
Sec. \ref{sec-ma-QW-state-FT-PT}.

In each pair, both QW states
$\psi_{1,p}$ and $\psi_{2,p}$
are of the same spin ($\sigma=\downarrow$) so
the matrix elements of the SOC operator
$\bra\psi_{i,p}|H_{\text{SO}}|\psi_{j,p}\ket $  ($i,j=1,2$)
 reduce to
 \begin{equation}
 \bra\psi_{i,p}|H_{\text{SO}}|\psi_{j,p}\ket=
 -\frac{1}{2}\xi \bra\phi_i|L_{\zeta}|\phi_j\ket
 \label{eq-mat_element-Hso}
 \end{equation}
 where
$\zeta$ denotes the direction of the film magnetization (the net spin).
For $\phi_1$ and $\phi_2$ defined in Eqs. (\ref{eq-phi12}),
 all four matrix elements (\ref{eq-mat_element-Hso}) vanish
for the in-plane magnetization ($\zeta=x$ or $y$).
This can be immediately shown when
$L_x$ and $L_y$ are represented as linear combinations of
the ladder operators $L_+$ and $L_-$,
and the orbitals
$|yz\ket=(i/\sqrt{2})(|2,-1\ket + |2,1\ket)$
and
$|zx\ket=(1/\sqrt{2})(|2,-1\ket - |2,1\ket)$
are expressed in terms of the eigenstates $|L,m\ket$
 of ${\bf L}^2$ and  $L_{z}$
(where $L$ is the orbital quantum number and
$m$ is the magnetic  quantum number).
 The diagonal elements
 $\bra\psi_{j,p}|H_{\text{SO}}|\psi_{j,p}\ket$ ($j=1$, $2$)
 of the SOC operator also vanish for
 the out-of-plane orientation of magnetization,
 with $L_{\zeta}=L_z$.
 However, the  off-diagonal element
$\bra\psi_{1,p}|H_{\text{SO}}|\psi_{2,p}\ket$
 is finite for $\zeta=z$ and equals to $-\frac{1}{2}i\xi$,
 independently of the direction of the  $\bf k$ vector
as it is  proportional to $c^2+d^2=1$.

 The matrix of the full Hamiltonian $H=H_0+H_{\text{SO}}$
 is block diagonal, with 2x2 blocks corresponding to individual
QW state pairs.
 This is so because   the matrix elements of both $H_0$ and $H_{\text{SO}}$
 between the QW states $\psi_{j,p}$ and $\psi_{j',p'}$
 from different pairs ($p\neq p'$) vanish since such states are orthogonal
 due to different $k_{z,p}$ and $k_{z,p'}$
 \cite{MC98,MC03}.
For each pair,
the Hamiltonian matrix  $H_{ij}=\bra\psi_{i,p}|H|\psi_{j,p}\ket$ ($i,j=1,2$)
takes the following form
 \begin{equation}
\begin{bmatrix}
H_{11} & H_{12}  \\
H_{21} & H_{22} \\
\end{bmatrix}
=
\begin{bmatrix}
\epsilon_{1,p} &  V  \\
V^{*}  & \epsilon_{2,p} \\
\end{bmatrix} \, .
\label{eq-ham-matrix}
 \end{equation}
with $V=(H_{\text{SO}})_{12}=
\bra\psi_{1,p}|H_{\text{SO}}|\psi_{2,p}\ket$.
  For the in-plane magnetization,
  corresponding to $V=0$,
  the QW states are not perturbed by the SOC and
  their energies remain unchanged,
   \begin{equation}
 \epsilon_{j,p}^{\text{FT},||} =\epsilon_{j,p} \;\;\; (j=1,2).
 \label{e-eps-F-in-plane}
 \end{equation}
 However, this is not the case
for the energies  of the QW states
at the out-of-plane magnetization,
 \begin{equation}
 \epsilon_{j,p}^{\text{FT},\perp} =
 \frac{1}{2}(\epsilon_{1,p}+\epsilon_{2,p})
 \mp   \frac{1}{2} \sqrt{(\Delta\epsilon)^2 +\xi^2}
 \label{eq-eps-FT}
 \end{equation}
(where $\Delta\epsilon=\epsilon_{2,p}-\epsilon_{1,p}$),
which are obtained by the diagonalizing the pair Hamiltonian
 with the finite term $V=-\frac{1}{2}i\xi$.
Here and in the following discussion, the plus and minus signs
 correspond to the lower state ($j=1$) and the upper ($j=2$) state in each pair,
 respectively.
This exact solution, obtained through the diagonalization,
is used in the FT approach for the simplified model of the QW states.
It reproduces, with
good accuracy, the splittings of the QW state energies around
the centre of the BZ
that arise due to the SOC in the band
structure of the full Co film system;
see  Figs. \ref{fig-Co_film-bands-soc} and
\ref{fig-model-two-state-exact_PT}.
In particular, the model predicts that
the energies
for the out-of-plane magnetization
are split as follows at ${\bf k}=(0,0)$,
 \begin{equation}
 \epsilon_{j,p}^{\text{FT},\perp}(k=0) =
 \epsilon_{0,p} \mp   \frac{1}{2}\xi \, ,
  \label{eq-eps-FT-k=0}
 \end{equation}
 and they remain almost constant in a central region of the BZ
 where $\Delta\epsilon$  is smaller than $\xi/2$.

 It is also possible to find an approximate solution
for the energies of the perturbed  QW states,
 \begin{equation}
  \epsilon_{j,p}^{\text{PT},\perp} =
 \epsilon_{j,p}
 \mp \frac{\xi^2}{4\Delta\epsilon}  \, ,
 \label{eq-eps-PT}
 \end{equation}
 where the second-order correction is
 obtained using the PT in a similar way as
in calculations of  the total MCA energy with
Eqs. (\ref{eq-mca-omega-PT}) and (\ref{E2-corr}).
This expression can  also be derived  by expanding
$\epsilon_{j,p}^{\text{FT},\perp}$ in powers of $\xi^2$,
up to the linear term.
The PT solution well reproduces the exact (FT) energies for
 $\Delta\epsilon \geq \xi$ but becomes largely inaccurate
 for $\Delta\epsilon$ smaller than $\xi/2$,
 with the difference
 $\epsilon_{j,p}^{\text{PT},\perp}-\epsilon_{j,p}^{\text{FT},\perp}$
 ($j=1,2$)
 equal to $\mp 0.0429\xi$ at $\Delta\epsilon=\xi$,
 $\mp 0.1910\xi$ at $\Delta\epsilon=\xi/2$,
 and $\mp 0.6096\xi$ at $\Delta\epsilon=\xi/4$.
 This is illustrated
 in Fig. \ref{fig-model-two-state-exact_PT}
  which shows the  energies of a pair of QW states calculated
  using the FT and the PT for various SOC strengths
  and plotted as functions of $k$.
  The values of the respective SOC
constants are marked on the top scale that indicates
 the energy separation
 $\Delta\epsilon=\epsilon_{2,p}-\epsilon_{1,p}=(b_2-b_1)k^2$
 between the upper and lower states within each pair.

 Accordingly,
most significant  deviations of the PT energies
from the FT energies
occur close to the $\overline{\Gamma}$ point
and the size of this region in the BZ  scales as $\sqrt{\xi}$.
In particular,
the energies $\epsilon_{j,p}^{\text{PT},\perp}$
diverge at $k=0$ (where $\Delta\epsilon=0$)
while the exact energies $\epsilon_{j,p}^{\text{FT},\perp}$
remain finite [cf. Eq. (\ref{eq-eps-FT-k=0}].
This large discrepancy between the FT and PT energies
at small $k=|{\bf k}|$ arises because the standard
PT formula [Eq. (\ref{eq-eps-PT})]  fails
if the difference
$\Delta\epsilon$ between the
unperturbed energies
$\epsilon_{2,p}$ and $\epsilon_{1,p}$
 is substantially smaller than the module $|V|=\frac{1}{2}\xi$
 of the perturbation matrix element.
In this case, the PT formulation
for degenerate
states should be applied,
with  the $2\times 2$ Hamiltonian matrix being initially diagonalized.
This, in fact, would reproduce the exact (FT) result because
all perturbation terms vanish
as the SOC does not  couple
states from different pairs of QW states.
Effective application of the degenerate PT
to the entire system of the Co film
is also not possible due to the presence
of  numerous closely spaced quantum states,
as seen in Figs. \ref{fig-Co_film-bands-zero-soc} and
            \ref{fig-Co_film-bands-soc}.
This would require prior diagonalization of the full Hamiltonian matrix,
as in the FT approach.
In consequence, the pragmatic PT approach to the calculations of the MCA energy
involves the use of the second-order non-degenerate PT.
This method has demonstrated good accuracy
at moderate strengths of the SOC
\cite{Bruno89,MC94,MC97,MC22,MC24,Ke19,BlancoRey19},
despite  its limitations for individual electron energies.
However, as discussed in Sec. \ref{sec-mca-Co-film},
this standard PT approach can lead to significantly inaccurate results for the MCA energy, particularly its oscillatory variation, at strong SOC.

\subsubsection{Contributions to magnetocrystalline energy from quantum-well states. Force theorem versus perturbation theory}
\label{sec-ma-QW-state-FT-PT}

The discrepancy  between the  FT and PT results
is now investigated by examining the contribution to the MCA energy
from the model subsystem of
the QW state pairs at various SOC strengths.
The contribution at a specific wavevector $\bf k$
is then defined in the FT approach as follows
    \begin{equation}
 E_{\text{MCA,QW}}^{\text{FT}} ({\bf k}) =
 \Omega_{\text{QW}}^{\text{FT},\perp}({\bf k})-
 \Omega_{\text{QW}}^{\text{FT},||}({\bf k})
  \label{eq-ma-qw-FT}
 \end{equation}
 where the grand potential
    \begin{equation}
 \Omega_{\text{QW}}^{\text{FT},D}({\bf k}) =
 \sum_p [g(\epsilon_{1,p}^{\text{FT},D} )
        + g(\epsilon_{2,p}^{\text{FT},D}) ]
  \label{eq-ma-qw-Omega-FT-dir}
 \end{equation}
 is calculated  with the energies
 $\epsilon_{j,p}^{\text{FT},D}$
 of the perturbed QW state pairs
 for the two considered magnetization directions,
 out-of-plane ($D=\perp$) and in-plane ($D=||$);
 see Eqs. (\ref{eq-eps-FT}) and (\ref{e-eps-F-in-plane}).
 In the PT approach,
 the respective contribution to the MCA energy,
    \begin{equation}
 E_{\text{MCA,QW}}^{\text{PT}} ({\bf k})=
  \frac{\xi^2}{4}
  \sum_{p}
 \frac{f_0(\epsilon_{2,p})-f_0(\epsilon_{1,p})}{\epsilon_{2,p}-\epsilon_{1,p}} \, .
  \label{eq-ma-qw-PT}
 \end{equation}
is given in terms of
the  energies $\epsilon_{j,p}$ and occupation factors
 $f_0(\epsilon_{j,p})$ in the unperturbed system.

\begin{figure}
   \includegraphics*[width=8cm]{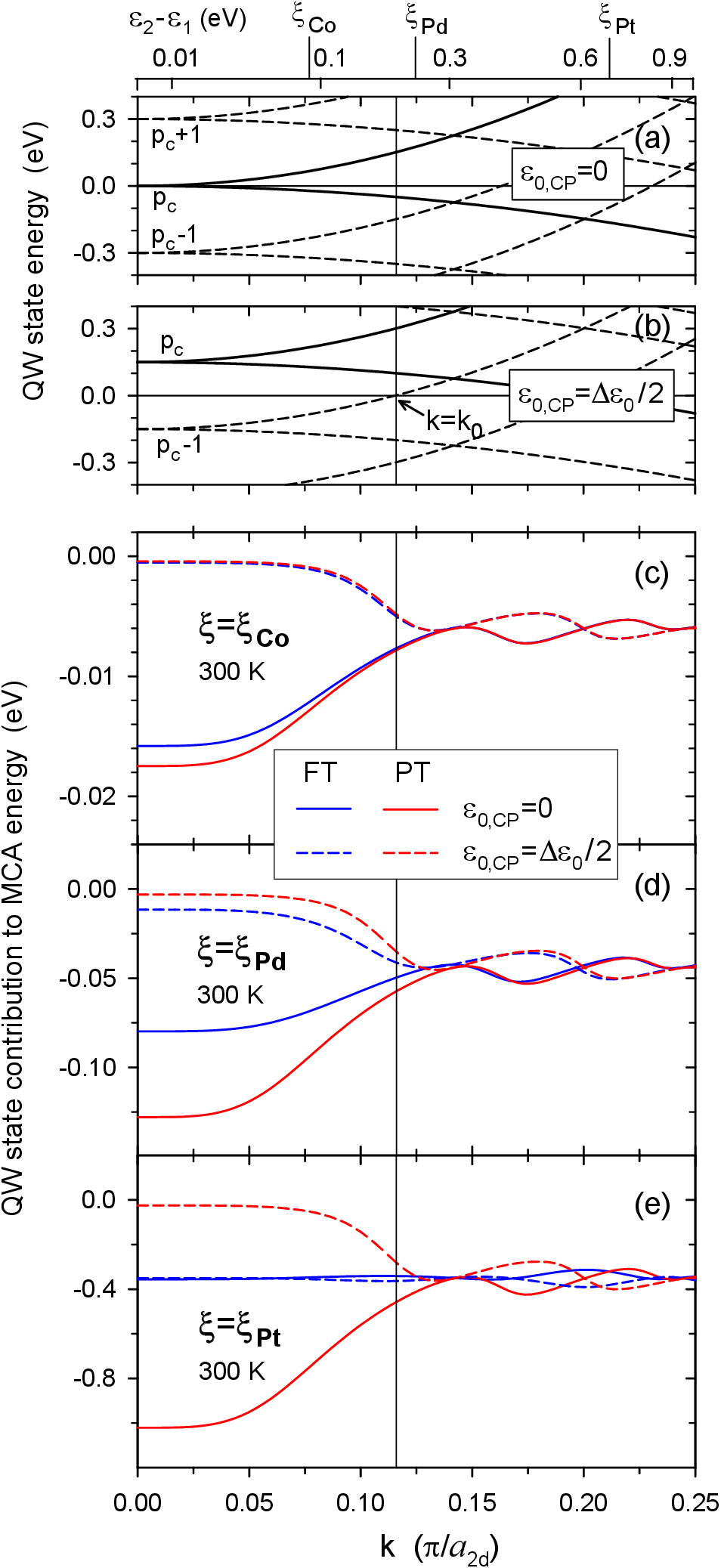}
      \caption{
      (a,b) Energies of the model QW states
      in the Co film along the $\overline{\Gamma}-\overline{X}$ line
       near the BZ centre
      for (a)~$\epsilon_{\text{0,CP}}=0$ and
       (b)~$\epsilon_{\text{0,CP}}=\Delta\epsilon_0/2$.
      (c-e)~Contribution to the MCA energy
      from these states  calculated using  the FT  (blue lines)
      and  the PT (red lines)
      with the SOC constants  (c)~$\xi=\xi_{\text{Co}}$,
      (d)~$\xi=\xi_{\text{Pd}}$,  and (e)~$\xi=\xi_{\text{Pt}}$
        for
       $\epsilon_{\text{0,CP}}=0$ (solid lines) and
       $\epsilon_{\text{0,CP}}=\Delta\epsilon_0/2$ (dashed lines)
       at $T=$300~K,
       with $\Delta\epsilon_0=0.3$~eV and
        $\epsilon_{\text{F}}=\epsilon_{\text{F0}}=0$.
        The top scale shows the energy separation
       $\Delta\epsilon=\epsilon_{2,p}-\epsilon_{1,p}$ in each pair
       of QW states, with the marked values of the SOC constants.
       }
  \label{fig-model_ma_vs_k_xiCo_xiPd_xiPt_FT_PT_300K}
\end{figure}

\begin{figure}
   \includegraphics*[width=8cm]{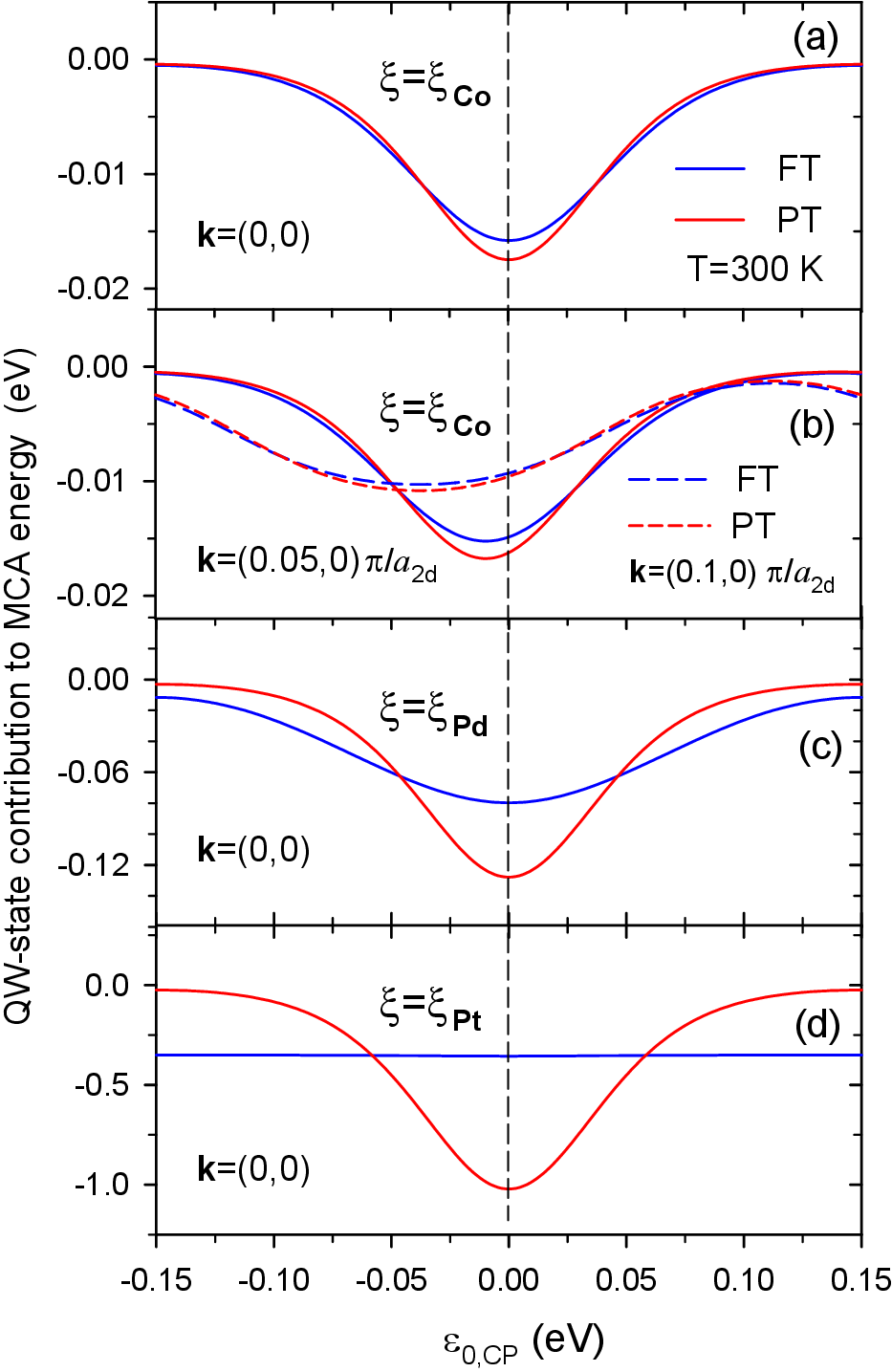}
      \caption{
      (a-c) Contribution to the MCA energy
      from the model QW states
      at (a,~c,~d)~${\bf k}=(0,0)$,
      (b)~${\bf k}=(0.05,0)\pi/a_{\text{2d}}$ and
      (b)~${\bf k}=(0.1,0)\pi/a_{\text{2d}}$
      in the Co film,
      calculated  using the FT (blue line) and  the PT (red line)
      versus the central QW state energy $\epsilon_{\text{0,CP}}$
       (see Fig.~\ref{fig-model-two-state-series}).
       The results are obtained for the SOC constants of
       (a, b)~$\xi=\xi_{\text{Co}}$, (b)~$\xi=\xi_{\text{Pd}}$,
       and (c)~$\xi=\xi_{\text{Pt}}$,
       with $\Delta\epsilon_0=0.3$~eV and
       $\epsilon_{\text{F}}=\epsilon_{\text{F0}}=0$.
         }
  \label{fig-model-ma-vs-e0-k=0-var-xi-300K}
\end{figure}

The contributions to the MCA energy from  the QW states
around the $\overline{\Gamma}$ point are plotted along the $\overline{\Gamma}-\overline{X}$ line in
Fig.~\ref{fig-model_ma_vs_k_xiCo_xiPd_xiPt_FT_PT_300K}.
The $k$ profiles of these contributions vary
with the position $\epsilon_{0,\text{CP}}$
of the central pair of QW states at $k=0$ relative to
the Fermi level ($\epsilon_{\text{F0}}=0$),
This dependence is significant in the PT calculations,
exhibiting a clear trend for any SOC, and
for the SOC of Co and Pd (but not Pt) using the FT.
The lowest profile is found
for $\epsilon_{0,\text{CP}}=0$, when the central pair contributes
across the entire $k$ range (down to $k=0$).
The highest profile occurs for
$\epsilon_{0,\text{CP}}=\Delta\epsilon_0/2$
(or, equivalently, $\epsilon_{0,\text{CP}}=-\Delta\epsilon_0/2$),
corresponding to the  Fermi energy
at the middle position $(\epsilon_{0,p-1}+\epsilon_{0,p})/2 $
between the two neighbouring QW state pairs,
as shown in Fig. \ref{fig-Co_film-bands-zero-soc}(b).
In this case,
no pair  contributes significantly in the interval of
$k$ near $k=0$ where both states (lower and upper) of each pair lie either
below or above the Fermi level.
This interval constitutes a small region around the centre of the BZ
($0 \leq k <k_0$, e.g.,
$k\lesssim 0.12\pi/a_{\text{2d}}$ for $\Delta\epsilon_0=0.3$~eV,
 along  the $\overline{\Gamma}-\overline{X}$ line).
  Within this region,
 the difference between the two profiles is most significant,
 being  the largest at the $\overline{\Gamma}$  point ($k=0$).
 The boundary of this region is located near the $k=k_0$ point where,
 for a subsystem of QW states with the central pair ($p=p_{\text{c}}$)
at $\epsilon_{0,\text{CP}}=\Delta\epsilon_0/2$, 
 the upper state from the neighbouring pair  ($p=p_{\text{c}}-1$,
 $\epsilon_{0,p}=-\Delta\epsilon_0/2$) crosses the Fermi level;
 see Fig. \ref{fig-model_ma_vs_k_xiCo_xiPd_xiPt_FT_PT_300K} (a,b).
 Indeed, for $k>k_0$,
 this latter pair  gives a very similar contribution
 to the MCA energy
as the central pair in a system with $\epsilon_{0,\text{CP}}=0$.
 Accordingly, the region size $k_0$
is determined not by the SOC strength but by the spacing $\Delta\epsilon_0$
between the pairs of QW states and their $k$ dispersion.

The full $\epsilon_{\text{0,CP}}$ dependencies of
the QW state contributions
at specific $\bf k$ points are shown
in Fig. \ref{fig-model-ma-vs-e0-k=0-var-xi-300K}.
In both  the FT and PT
approaches,
the contributions calculated at ${\bf k}=(0,0)$
for the SOC strengths of Co and Pd,
reach a negative minimum value
for $\epsilon_{\text{0,CP}}=0$ (at the Fermi level)
and a maximum close to 0
for the energy $\epsilon_{\text{0,CP}}=\Delta\epsilon_0/2$.
For non-zero ${\bf k}$ near the $\overline{\Gamma}$ point,
the positions of these minimum and maximum are slightly shifted to
leftwards [to smaller $\epsilon_{\text{0,CP}}$;
see Fig. \ref{fig-model-ma-vs-e0-k=0-var-xi-300K}(b)] which results from
the asymmetric dispersion
of the lower and upper bands in each QW state pair
[$b_1 \neq -b_2$ in Eqs. (\ref{eq-eps1}) and (\ref{eq-eps2})].
Note that
the QW state contribution is the same for
$\epsilon_{\text{0,CP}}=-\Delta\epsilon_0/2$
and $\epsilon_{\text{0,CP}}=\Delta\epsilon_0/2$
since the systems with such positions of
 $\epsilon_{\text{0,CP}}$  are equivalent.

These findings show that the simple model of QW states well captures
the predictions of the FT and the PT 
for the MCA energy oscillations found for the full film system
(Figs. \ref{fig-Co_film_eso_per_var_xi_300K},
\ref{fig-ma-bz-Co-film-FT-PT-xiCo},
and \ref{fig-ma-bz-Co-film-FT-PT-xiPt}).
The position of the central pair energy $\epsilon_{0,\text{CP}}$
changes by about $\Delta\epsilon_0/2$ when the Co thickness
increases from $N_{\text{Co}}$ to $N_{\text{Co}}+1$ as illustrated
for the films of Co(11~ML) and Co(12~ML)
in  Fig. \ref{fig-Co_film-bands-zero-soc}.
Thus, the  $\epsilon_{0,\text{CP}}$ dependence of the QW state contribution explains the 2~ML period  oscillations of the MCA energy
(Figs. \ref{fig-Co_film_eso_per_multi_temp}
and \ref{fig-Co_film_osc_eso_per_var_xi_gam05_300K}), which arise
in the $\overline{\Gamma}$ region and have minima
at $N_{\text{Co}}=9$, 11, 13 ML and other thicknesses
for which $\epsilon_{0,\text{CP}}$ is close to 0.
The profiles of the model QW state contributions around the BZ centre,
calculated using the FT and the PT
for $\epsilon_{\text{0,CP}}=0$ and $\epsilon_{\text{0,CP}}=\Delta\epsilon_0/2$,
closely resemble the MCA energy distribution profiles
obtained with the FT and PT
for full film systems with similar positions of the central pair energy,
such as the Co(11 ML) and Co(12 ML) films, respectively.
The similarity between the full and QW state  profiles
of the MCA energy distributions holds not only for the moderate SOC  of Co
but also for the SOC of Pt;
see Figs. \ref{fig-ma-bz-Co-film-FT-PT-xiPt} and \ref{fig-model_ma_vs_k_xiCo_xiPd_xiPt_FT_PT_300K}(e).

The obtained $k$ profiles of the QW state contribution
also  show that
the discrepancy  between the FT and PT
distributions of the MCA energy close to the $\overline{\Gamma}$ point
is clearly related to the inaccuracies of the PT energies of the QW states.
The difference  between the FT and PT profiles
is most substantial in a $k$ interval where these energies become
inaccurate because
the energy separation $\Delta\epsilon=(b_2-b_1)k^2$
between the states in each pair
is considerably smaller than the SOC constant $\xi$.
The width of this interval is much smaller than
$k_0\approx 0.12\pi/a_{\text{2d}}$
for the nominal SOC of Co
and grows with the SOC strength
(as $\sqrt{\xi}$).
It reaches $k_0$ for $\xi\sim\xi_{\text{Pd}}$ and exceeds $k_0$
for $\xi=\xi_{\text{Pt}}$.
This is illustrated  in
 Fig.~\ref{fig-model_ma_vs_k_xiCo_xiPd_xiPt_FT_PT_300K},
 where the top scale shows $\Delta\epsilon$  and identifies the $k$ points
 where $\Delta\epsilon$ is equal to  $\xi_{\text{Co}}$, $\xi_{\text{Pd}}$,
 and $\xi_{\text{Pt}}$.

The relative difference between the PT and FT contributions
also grows with the SOC strength.
While the QW state contribution
is still well described (even down to $k=0$) by the PT
for $\xi=\xi_{\text{Co}}$ at $T=300$~K,
the FT and  PT profiles of this contribution
largely differ for the SOC of Pd at the same temperature,
though they still maintain similar shapes.
For $\xi=\xi_{\text{Pt}}$, the  PT  completely fails to describe this
contribution.
Its $k$ profiles and $\epsilon_{\text{0,CP}}$
dependence have  the same shapes as for the weaker SOC
(due to the $\xi^2$ scaling),
which  dramatically differs
from the nearly flat $k$ profiles and
the extremely weak $\epsilon_{\text{0,CP}}$  dependence
of  the exact FT contribution; see
Figs.~\ref{fig-model_ma_vs_k_xiCo_xiPd_xiPt_FT_PT_300K}(e)
and~\ref{fig-model-ma-vs-e0-k=0-var-xi-300K}(c).
This explains why  the  FT distribution of the full MCA energy remains
almost unchanged in the central part of the BZ
when the film thickness is increased by 1 ML;
see Fig. \ref{fig-ma-bz-Co-film-FT-PT-xiPt}(a) and (c).
Consequently,
the oscillations of $E_{\text{MCA}}^{\text{FT}}(N_{\text{Co}})$
are many times smaller compared to
the pronounced 2 ML oscillations predicted
by the PT calculations [Fig. \ref{fig-ma-bz-Co-film-FT-PT-xiPt}(b) and (d)].

\begin{figure}
   \includegraphics*[width=8cm]{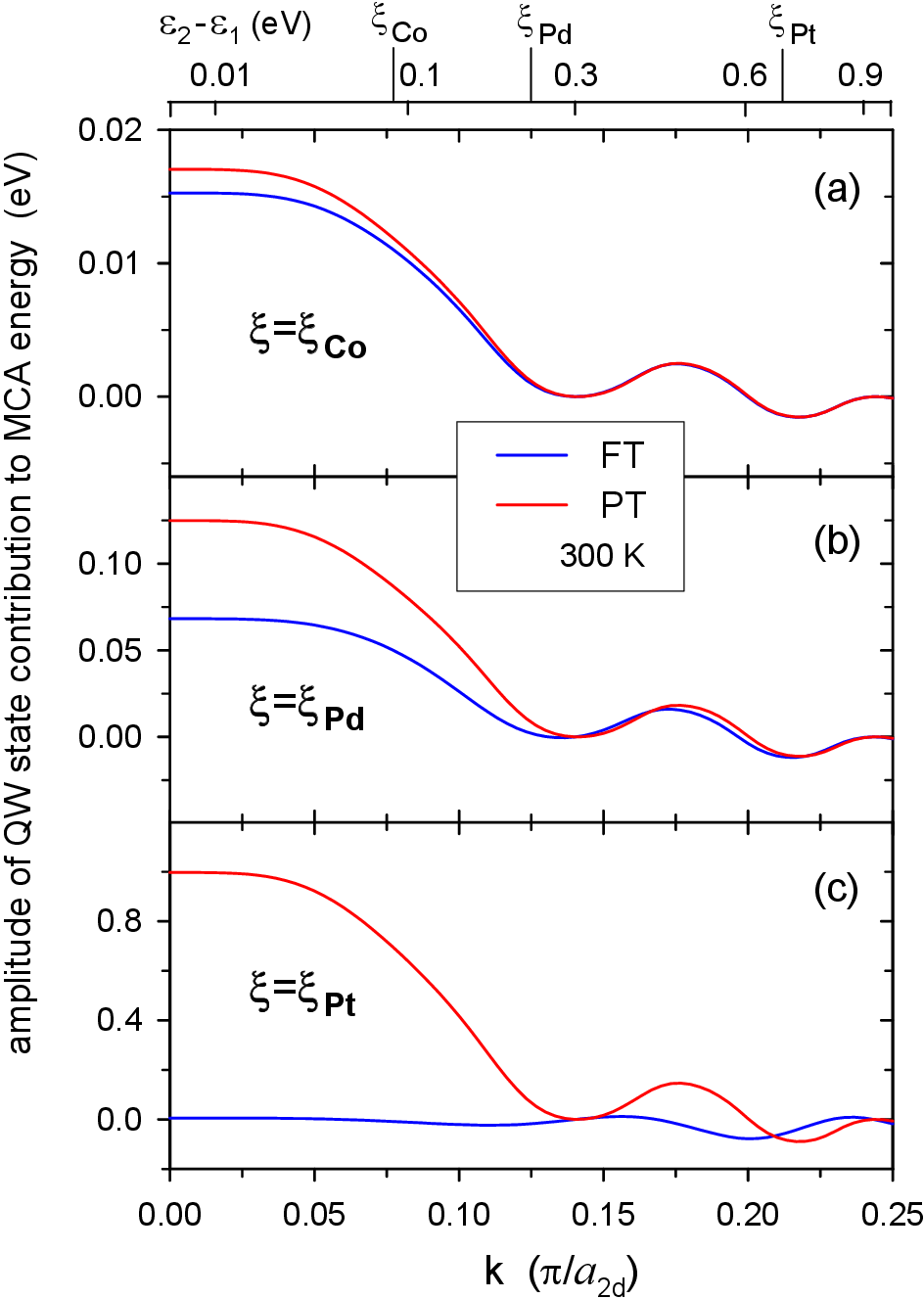}
      \caption{
      Oscillation amplitude
      $A_{\text{MCA,QW}}^{\text{Y}} ({\bf k})$ (Y=FT, PT)
      [Eq. (\ref{eq-model-ma-amplitude-qw-states}) with
      $\epsilon_{\text{0,CP}}=0$]
      of the  MCA energy contribution
       from the model QW states in the Co film
       along the $\overline{\Gamma}-\overline{X}$ line
      near the BZ centre.
      The amplitude is calculated using  the FT  (blue line)
      and  the PT (red line)
      for (a)~$\xi=\xi_{\text{Co}}$,
      (b)~$\xi=\xi_{\text{Pd}}$,  and (c)~$\xi=\xi_{\text{Pt}}$,
      at $T=$300 K, with $\Delta\epsilon_0=0.3$~eV.
       The top scale shows the energy separation
       $\Delta\epsilon=\epsilon_{2,p}-\epsilon_{1,p}$ in each pair
       of states, with the marked values of the SOC constants.
       }
  \label{fig-model_ma_amp_vs_k_xiCo_xiPd_xiPt_FT_PT_300K}
\end{figure}

In the model system,
the change of the QW state contribution to the MCA energy
upon increasing the Co thickness by 1 ML
is given by the difference
\begin{equation}
  A_{\text{MCA,QW}}^{\text{Y}} ({\bf k},\epsilon_{\text{0,CP}}) =
  \left. E_{\text{MCA,QW}}^{\text{Y}} ({\bf k}) \right|_{\epsilon'_{\text{0,CP}}}
  -  \left. E_{\text{MCA,QW}}^{\text{Y}} ({\bf k})
  \right|_{\epsilon_{\text{0,CP}}}
   \label{eq-model-ma-amplitude-qw-states}
\end{equation}
(Y=FT, PT)
between the values of this  contribution
for two positions of the central pair energy, $\epsilon_{\text{0,CP}}$ and $\epsilon'_{\text{0,CP}}=\epsilon_{\text{0,CP}}\pm\Delta\epsilon_0/2$,
corresponding to Co thicknesses  $N_{\text{Co}}$ and $N_{\text{Co}}+1$, respectively.
This quantity represents the local amplitude of the 2 ML period oscillations
of the MCA energy due to QW states in the $\overline{\Gamma}$ region
and can be either positive or negative.
In the definition of $\epsilon'_{\text{0,CP}}$,
the plus and minus signs are used for $\epsilon_{\text{0,CP}} <0$ and
$\epsilon_{\text{0,CP}} \geq 0 $, respectively,
so that both  energies fall in the energy interval
$[-\Delta\epsilon_0/2,\Delta\epsilon_0/2]$ relevant to central pairs.
(alternatively, the plus sign could solely be used if this condition is relaxed
and the periodicity in $\epsilon_{\text{0,CP}}$ is accounted for).

At the $\overline{\Gamma}$ point, the amplitudes $A_{\text{MCA,QW}}^{\text{FT}}$
and $A_{\text{MCA,QW}}^{\text{PT}}$  attain their maximum values for
$\epsilon_{\text{0,CP}}=0$ (with corresponding
$\epsilon'_{\text{0,CP}}=\Delta\epsilon_0/2$)
and  vanish for $\epsilon_{\text{0,CP}}=-\Delta\epsilon_0/4$
(with $\epsilon'_{\text{0,CP}}=\Delta\epsilon_0/4$).
This occurs
because the contributions $E_{\text{MCA,QW}}^{\text{Y}}$ (Y=FT, PT)
are symmetric functions of $\epsilon_{\text{0,CP}}$
at ${\bf k}=(0,0)$;
see Fig. \ref{fig-model-ma-vs-e0-k=0-var-xi-300K}.
For ${\bf k}$ points near the $\overline{\Gamma}$ point,
 the amplitude $A_{\text{MCA,QW}}^{\text{Y}} ({\bf k},\epsilon_{\text{0,CP}})$
 attains  its maximum
 for a central pair energy slightly  off the Fermi level ($\epsilon_{\text{0,CP}}\neq  0$).
 This behaviour results from
 the asymmetric  $\epsilon_{\text{0,CP}}$ dependence of
 the $E_{\text{MCA,QW}}^{\text{Y}}$ (Y=FT, PT) contribution for $k >0$
 [see Fig. \ref{fig-model-ma-vs-e0-k=0-var-xi-300K}(b)],
 which is a direct consequence of  the asymmetric dispersions
 ($b_1 \neq -b_2$) of the lower and upper bands in each QW state pair.
Note that
the central pair energy
can be at various positions within the interval
$-\Delta\epsilon_0/2 \leq \epsilon_{\text{0,CP}} \leq \Delta\epsilon_0/2$,
depending on  the Co film thickness.
This is because the shifts of this energy
with increasing $N_{\text{Co}}$ are not exactly equal to $\pm\Delta\epsilon_0/2$
since the actual oscillation period of 2.12 ML of the MCA energy
slightly differs  from 2 ML \cite{MC03}.
 This leads to oscillations of $E_{\text{MCA}}$
 with a modulated amplitude, similar to a beating pattern, with a modulation period of approximately 17  ML ($\approx 2/0.12$~ML),
 such as seen in Figs. \ref{fig-Co_film_eso_per_multi_temp} and \ref{fig-Co_film_osc_eso_per_var_xi_gam05_300K}.

\begin{figure}
   \includegraphics*[width=8cm]{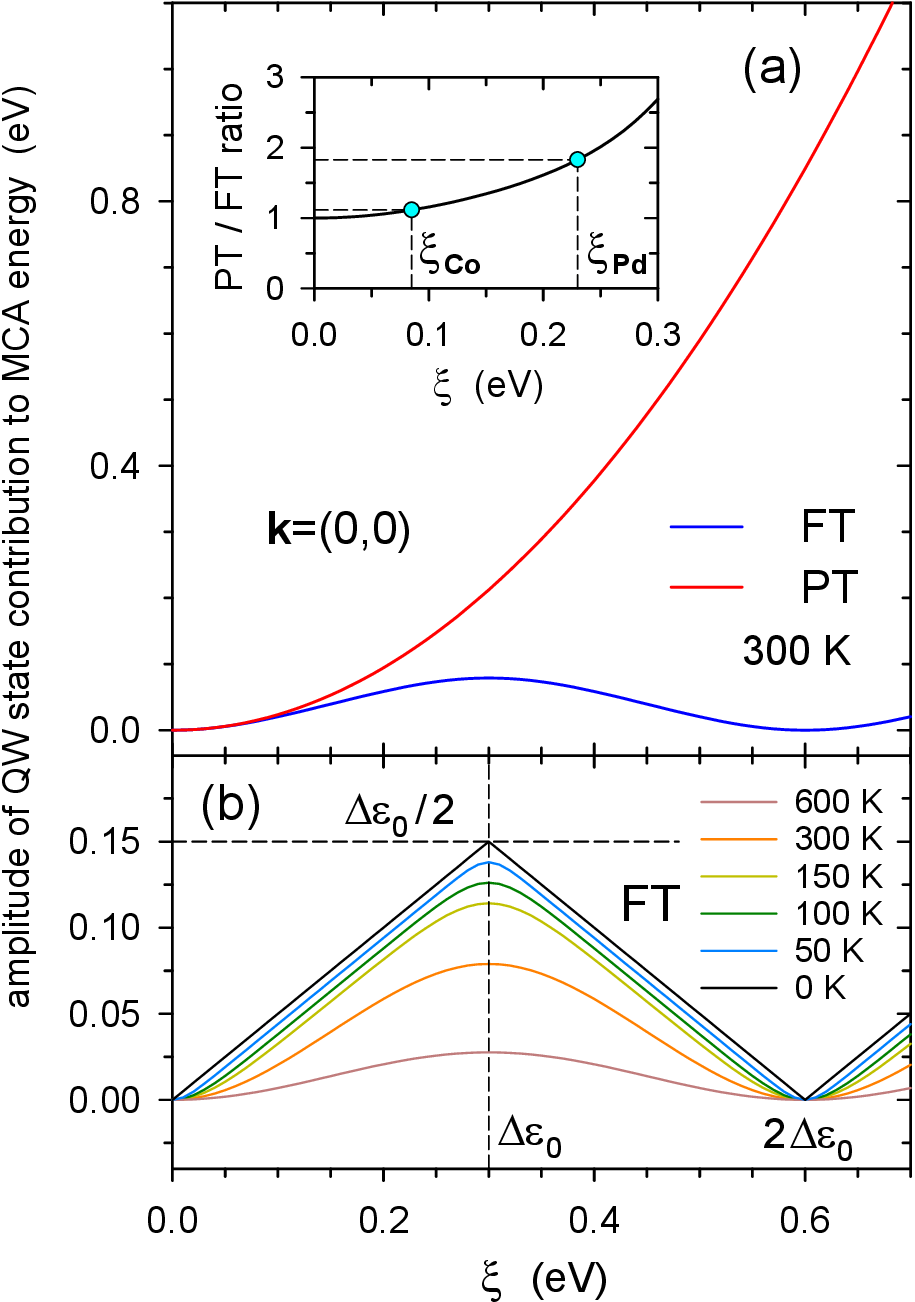}
      \caption{
      (a,b) Oscillation amplitude
      $A_{\text{MCA,QW}}^{\text{Y}} ({\bf k})$ (Y=FT, PT)
      [Eq. (\ref{eq-model-ma-amplitude-qw-states}) with
      $\epsilon_{\text{0,CP}}=0$]
      of the MCA energy contribution from the model QW states
      at ${\bf k}=(0,0)$  in the Co film versus the SOC constant $\xi$.
      The amplitude is
      calculated with (a) the FT (blue line)  and  the PT (red line)
      at $T=300$~K, and using (b) the FT at various temperatures,
      for the spacing of $\Delta\epsilon_0=0.3$~eV between QW state pairs
      (see Fig. \ref{fig-model-two-state-series}).
      (a) Insert: PT to FT amplitude ratio versus $\xi$. }
  \label{fig-model_amp_ma_vs_xi_FT_PT_k=0}
\end{figure}

The oscillation amplitudes
$A_{\text{MCA,QW}}^{\text{FT}} ({\bf k})$
(for $\xi=\xi_{\text{Co}}$ and $\xi_{\text{Pd}}$)
and $A_{\text{MCA,QW}}^{\text{PT}} ({\bf k})$
(for all $\xi$), calculated with $\epsilon_{\text{0,CP}}=0$,
are sizable  in the interval
$0\leq k\lesssim k_0\approx 0.12\pi/a_{\text{2d}}$
(for $\Delta\epsilon_0=0.3$~eV)
and largest at the $\overline{\Gamma}$ point; see Fig. \ref{fig-model_ma_amp_vs_k_xiCo_xiPd_xiPt_FT_PT_300K}.
 For $\xi$ less than 0.1 eV,
the amplitudes
obtained with the FT and the  PT
are very similar, differing  less than 16\%
at ${\bf k}=(0,0)$;
see Fig. \ref{fig-model_amp_ma_vs_xi_FT_PT_k=0}.
However, the difference between these amplitudes  increases rapidly
with increasing $\xi$.
For the SOC of Pd ($\xi=0.23$~eV),
the PT amplitude at ${\bf k}=(0,0)$  is nearly twice as large as the FT amplitude.
For stronger SOC,
the discrepancy between the FT and PT predictions
becomes even more pronounced,
with the PT oscillation amplitude
being several, or even tens, times larger than
the corresponding amplitude obtained with the FT.
This explains the presence of similar discrepancies
between the amplitudes of the 2 ML period oscillations
of the MCA energy obtained with the FT and PT
for the full Co film with enhanced  SOC
(Fig. \ref{fig-Co_film_eso_per_var_xi_300K}).

\begin{figure}
   \includegraphics*[width=8cm]{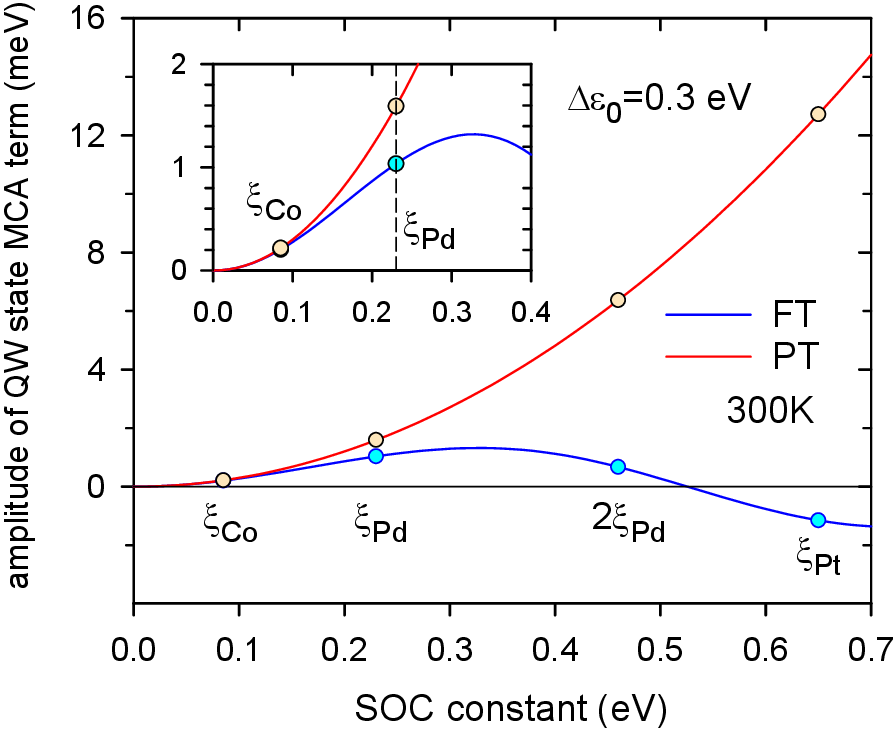}
      \caption{
       Oscillation amplitude
      ${\cal A}_{\text{MCA,QW}}^{\text{Y}} ({\bf k})$ (Y=FT, PT)
      [Eqs. (\ref{eq-ma-model-amp-region})
      and  (\ref{eq-model-ma-amplitude-qw-states}) with
      $\epsilon_{\text{0,CP}}=0$]
     of the  MCA energy term
      from the model QW states with $k \leq 0.5\pi/a_{\text{2d}}$
      in the Co film versus the SOC constant $\xi$.
      The amplitude is calculated using the FT (blue line)
      and  the PT (red line) at $T=300$~K,
      with  $\Delta \epsilon_0=0.3$~eV.
      Insert: Close-up of the main plot.
                                         }
  \label{fig-model_amp_ma_vs_xi_FT_PT_region}
\end{figure}

\begin{figure}
   \includegraphics*[width=8cm]{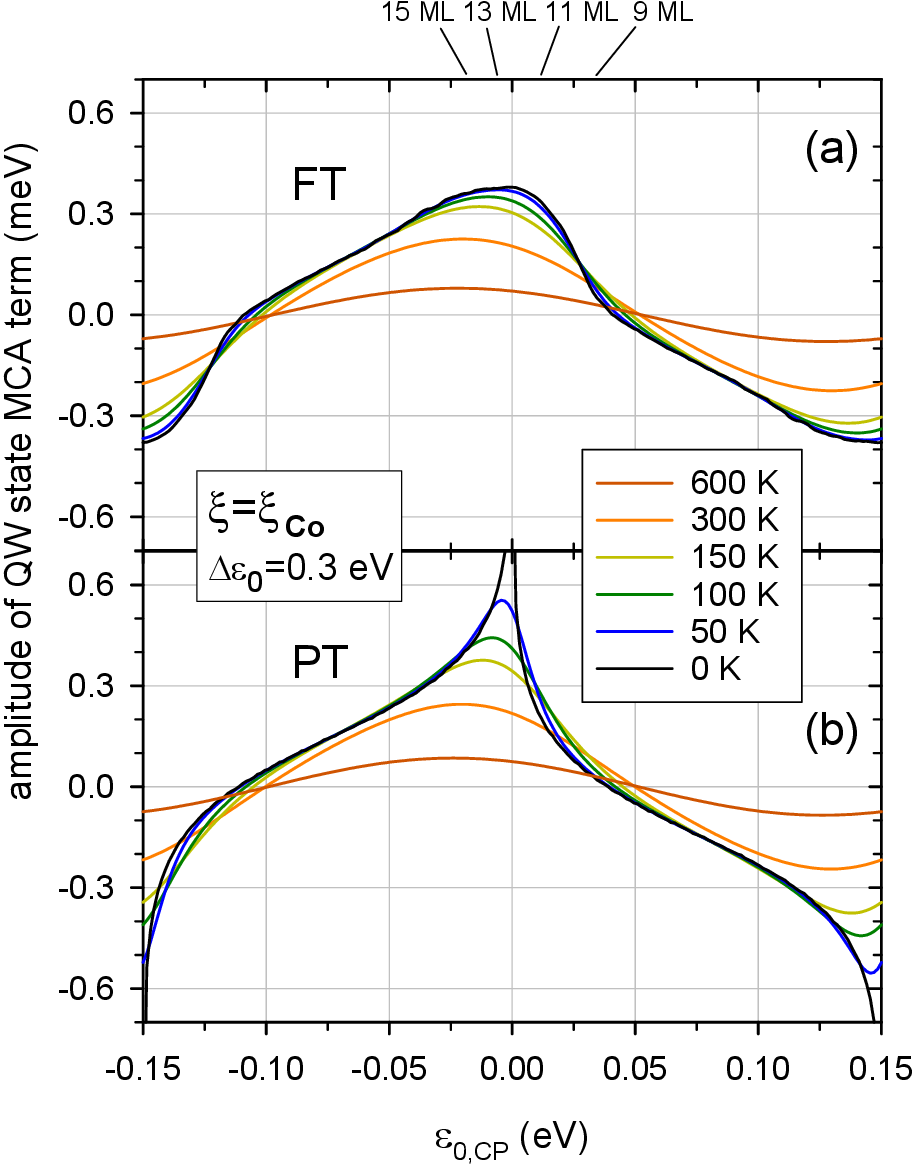}
      \caption{
       Oscillation amplitude
       ${\cal A}_{\text{MCA,QW}}^{\text{Y}}$ (Y=FT, PT)
      of the MCA energy term
      [Eq. (\ref{eq-ma-model-amp-region})]
      from the model QW states with $k \leq 0.5 \pi/a_{\text{2d}}$
      in the Co film with the nominal SOC ($\xi=\xi_{\text{Co}}$)
      versus the central pair energy  $\epsilon_{\text{0,CP}}$.
      The amplitude is calculated
      using (a) the FT and  (b) the PT  at various temperatures,
      with $\Delta \epsilon_0=0.3$~eV. The values of $\epsilon_{\text{0,CP}}$ for selected Co films are marked on the top scale.}
  \label{fig-model_amp_ma_vs_e0_temp_FT_PT_region}
\end{figure}

The (maximum) amplitude of the oscillatory term in
the total MCA energy  due to
the considered QW states can be estimated
by integrating $A_{\text{MCA,QW}}^{\text{Y}}({\bf k})$
(Y=FT and PT) for $\epsilon_{\text{0,CP}}=0 $ over of the BZ region
with $0\leq k \leq k_{\text{L}}$ where $k_{\text{L}}=0.5 \pi/a_{\text{2d}}$,
which covers most of the $\overline{\Gamma}$ region
(very similar results are obtained for a smaller region
with \mbox{$k_{\text{L}}=k_0=0.12\pi/a_{\text{2d}}$}
where the integrated amplitude is most substantial).
For the sake of simplicity,
the parameters $b_1$ and $b_2$ which define the dispersion of the QW states
[Eqs. (\ref{eq-eps1}) and (\ref{eq-eps2})] are assumed to be constant,
by taking their respective averages over different $\bf k$ directions.
Then,
the integral that  defines the size of the oscillatory MCA term
is approximated as follows
 \begin{equation}
  \label{eq-ma-model-amp-region}
 {\cal A}_{\text{MCA,QW}}^{\text{Y}} =
 \frac{2\pi}{\Omega_{\text{BZ}}}\int_0^{k_L}
 A_{\text{MCA,QW}}^{\text{Y}}(k)k dk
 \end{equation}
(Y=FT and PT), with  the normalized factor  given by the inverse
of the BZ volume $\Omega_{\text{BZ}}$; cf. Eq. (\ref{E2-corr}).
The so-estimated amplitude of the MCA energy oscillations at $T=300$~K
(Fig. \ref{fig-model_amp_ma_vs_xi_FT_PT_region})
agrees well with the exact results for the MCA energy of the Co film
with enhanced SOC (Fig. \ref{fig-Co_film_eso_per_var_xi_300K}).
In particular, the oscillation amplitudes  have similar magnitudes
in the calculations for the full film and the QW state model,
in both FT and PT approaches.
Notably, the PT amplitudes
are much larger than the FT amplitudes,
except for the film with the nominal Co SOC strength
where the two amplitudes are nearly equal at $T=300$~K.
The model also correctly reproduces the temperature dependence
(Fig.~\ref{fig-model_amp_ma_vs_e0_temp_FT_PT_region}),
with the PT amplitudes substantially increasing  with lowering the temperature,
in contrast to much weaker  temperature dependence of the FT amplitudes
(Fig.~\ref{fig-Co_film_eso_per_multi_temp}).

Figure \ref{fig-model_amp_ma_vs_e0_temp_FT_PT_region} also shows how
the integrated amplitudes
${\cal A}_{\text{MCA,QW}}^{\text{FT}}$ and
${\cal A}_{\text{MCA,QW}}^{\text{PT}}$
change with
the central pair energy $\epsilon_{\text{0,CP}}$ used to define
$A_{\text{MCA,QW}}^{\text{Y}}({\bf k})$ (Y=FT, PT) in Eq. (\ref{eq-model-ma-amplitude-qw-states}).
The maxima of these amplitudes
are slightly shifted from the Fermi level
($\epsilon_{\text{F0}}=0$, $\epsilon_{\text{F}}=0$),
towards negative $\epsilon_{\text{0,CP}}$,  with smaller shifts
for lower $T$. These shifts result from the aforementioned fact that the amplitude $A_{\text{MCA,QW}}^{\text{Y}}({\bf k})$ attains its maximum at $\epsilon_{\text{0,CP}}=0$ only for ${\bf k}=(0,0)$ while the position
of this maximum is shifted to finite $\epsilon_{\text{0,CP}}$  for
${\bf k}\neq(0,0)$ as
the dispersions of the QW states are not symmetric
($b_1\neq -b_2$).
With $\epsilon_{\text{0,CP}}$ values of
0.0335~eV for  Co(9~ML), 0.0118~eV
 for Co(11~ML), -0.0053~eV for Co(13~ML), and -0.0180~eV for Co(15~ML)
 (marked on Fig. \ref{fig-model_amp_ma_vs_e0_temp_FT_PT_region}),
 the model predicts the largest oscillation amplitude
 ${\cal A}_{\text{MCA,QW}}^{\text{PT}}$, with the strongest temperature dependence,
 for the Co(13) film.
This is in  perfect agreement
with the PT calculations for the full Co film system
 (Fig. \ref{fig-Co_film_eso_per_multi_temp}).
 For larger values of $\epsilon_{\text{0,CP}}$, both negative and positive,
  the oscillation amplitude of the QW state term
 decreases and finally reverses its sign
 which corresponds to a change of the oscillation phase
 in the thickness dependence.

The  model calculations show that
at weak and moderate SOC
(such as that of Co) and  sufficiently high temperatures,
the PT reproduces surprisingly well
the predictions of the FT for the QW state contributions
to the MCA energy
even at very  small $k$, where
it fails for the energies of the perturbed QW states.
To explain this apparent contradiction
the QW state contribution at $k=0$ (the $\overline{\Gamma}$ point)
 is expanded in a power series of $\xi$.
This expansion is done by noting that the energies
$\epsilon_{1,p}^{\text{FT},D}$ and  $\epsilon_{2,p}^{\text{FT},D}$
at this point are equal to
$\epsilon_{0,p} -   \frac{1}{2}\xi$ and
$\epsilon_{0,p} +  \frac{1}{2}\xi$,
respectively, for the out-of-plane magnetization ($D=\perp$)
and  $\epsilon_{0,p}$ for its in-plane direction ($D=||$).
The zeroth and odd-order terms in
$E_{\text{MCA,QW}}^{\text{FT}}(k=0)$, Eq. (\ref{eq-ma-qw-FT}),
cancel out, while the second-order term, given by the sum of
$\frac{1}{4}f'(\epsilon_{0,p}) \xi^2$ over $p$,
 reproduces the PT result [Eq. (\ref{eq-ma-qw-PT})]
 at  $k=0$ where the energies
  $\epsilon_{1,p}$ and $\epsilon_{2,p}$ are equal to~$\epsilon_{0,p}$.
  The dominating term $\frac{1}{4}f'(\epsilon_{\text{0,CP}}) \xi^2$
  in this expansion comes from the central QW state pair,
  the one with $\epsilon_{0,p}$ closest to the Fermi energy.
  The fourth-order term
  $\frac{1}{96} f'''(\epsilon_{\text{0,CP}}) \xi^4$,
  associated with this pair,
  can serve as an estimate of the error
  in the second-order PT expression
   for $E_{\text{MCA,QW}}^{\text{PT}}(k=0)$,
   resulting from the neglect of higher-order terms.
For $\epsilon_{\text{0,CP}}=0$ at the Fermi energy,
the  fourth-order term is positive
and is the following
fraction of the second-order term
\begin{equation}
R_{4/2}=-\frac{1}{96}\left ( \frac{\xi}{k_{\text{B}}T}\right)^2
\approx -\left ( \frac{\xi}{10k_{\text{B}}T}\right)^2  \, ,
\label{eq-mmodel-ratio-4th-to-2nd-terms}
\end{equation}
which depends on the ratio
of the SOC constant $\xi$ to $k_{\text{B}}T$.
In particular,
 for $\xi=0.085$eV and $T=300$~K ($k_{\text{B}}T=0.02587$~eV),
 we obtain $R_{4/2}=-0.1125$
 which elucidates why the PT is sufficiently accurate for
 the Co film with the nominal SOC at this temperature
 [see Fig.~\ref{fig-model_ma_vs_k_xiCo_xiPd_xiPt_FT_PT_300K}(c)].
 However, for the SOC of Pd ($\xi=0.23$eV),
the ratio $R_{4/2}$ reaches the value of $-0.8234$ at $T=300$~K.
This indicates, in agreement with the numerical results
(Figs. \ref{fig-model_ma_amp_vs_k_xiCo_xiPd_xiPt_FT_PT_300K}
and  \ref{fig-model_amp_ma_vs_xi_FT_PT_k=0}),
that the second-order PT provides only a rough
approximation to $E_{\text{MCA,QW}}^{\text{PT}}(k=0)$
for this SOC  strength, and
higher-order terms would be needed for an accurate description of
the QW state contribution to the MCA energy.
Note that the negative second-order term predicted
by the PT is partially compensated
by the positive fourth-order term,
which results in  a reduction of  the oscillation amplitude.
This is consistent with the numerical results which show
 a smaller amplitude  of the MCA energy
 when calculated with the FT
 compared to the second-order PT,
for both the QW state subsystem
(Fig. \ref{fig-model_ma_amp_vs_k_xiCo_xiPd_xiPt_FT_PT_300K})
and  the entire film (Fig. \ref{fig-Co_film_eso_per_var_xi_300K}),
with  $\xi=\xi_{\text{Co}}$ and $\xi=\xi_{\text{Pd}}$.
For the strongest considered SOC, $\xi=\xi_{\text{Pt}}=0.65$~eV,
the ratio of $R_{4/2}$ takes a huge value of -6.5760
which demonstrates
that the PT formula is completely inadequate
to describe the considered QW state contribution to the MCA energy.
This explains why the PT so inaccurately predicts
the thickness dependence of the MCA energy
in the Co film with the SOC of Pt
[Fig. \ref{fig-Co_film_eso_per_var_xi_300K}(d)].

The ratio $R_{4/2}$
increases with decreasing  temperature
which indicates that,  in the PT approach,
contributions from QW state pairs
with energies very close to the Fermi level
at ${\bf k}=(0,0)$
become  less accurate at lower $T$.
This effect of decreasing temperature is reflected
in the growing difference between the FT and PT results
for the amplitude
${\cal A}_{\text{MCA,QW}}^{\text{Y}}$ (Y=FT, PT)
of oscillations arising from the QW state subsystem with
$\epsilon_{\text{0,CP}}\approx 0$
(Fig.~\ref{fig-model_amp_ma_vs_e0_temp_FT_PT_region}).
Such a trend is also found for
the MCA energy of the full Co film system,
in particular,
at thicknesses of 13 and 15 ML (Fig.~\ref{fig-Co_film_eso_per_multi_temp}).
For $\epsilon_{\text{0,CP}}=0$,
the contribution $E_{\text{MCA,QW}}^{\text{PT}}(k=0)$
diverges as $1/T$,
leading to a divergent oscillation amplitude
${\cal A}_{\text{MCA,QW}}^{\text{PT}}$
at zero temperature.
However, the  oscillation amplitude
is  finite even at $T=0$
because, for the actual Co films,
the energy $\epsilon_{\text{0,CP}}$
never lies exactly at the Fermi level.
Instead, it takes small,
non-zero values, as reported above.

In the FT approach,
the oscillation amplitude of the contribution
$E_{\text{MCA,QW}}^{\text{FT}}({\bf k})$
also increases with decreasing $T$ for any~$\xi$,
but, unlike in the PT calculations,remains  finite even in the $T=0$ limit.
This is illustrated for $k=0$
in Fig. \ref{fig-model_amp_ma_vs_xi_FT_PT_k=0} (b).
Consequently, the temperature dependence
 of the MCA energy (especially its oscillations)
 calculated with the FT is substantially weaker than
 that in the PT calculations,
 as seen  for the Co film with the nominal SOC
in Fig. \ref{fig-Co_film_eso_per_multi_temp}
and previously reported for the Co/Pd bilayer in
Ref. \onlinecite{MC24} (Fig. 10 therein).
The amplitude
$A_{\text{MCA,QW}}^{\text{FT}} (k=0)$
reaches the maximum at $\xi=\Delta\epsilon_{0}$
and vanishes for  $\xi=0$ and $\xi=2 \Delta\epsilon_{0}$,
which is reflected by a triangular $\xi$ dependence at $T=0$.
Thus, the oscillation amplitude
decreases with $\xi$ for $\xi>\Delta\epsilon_{0}$
which further explains the large discrepancy between the
PT and FT predictions for stronger SOC.
Note that the value of $\xi=2 \Delta\epsilon_{0}$  is equal to $0.6$~eV
for the assumed interpair spacing $\Delta\epsilon_{0}=0.3$~eV
(corresponding to $N_{\text{Co}}\approx 12$~ML)
and it is close to the SOC of Pt ($\xi=0.65$~eV).

To understand such a non-monotonic $\xi$ dependence of the FT amplitude
let us note that the energies of the $p$-th pair at $k=0$
for the SOC constant $\xi$,
\begin{eqnarray}
\epsilon_{1,p}^{\text{FT},\perp} & = & \epsilon_{0,p}-\xi/2 \, ,  \\
\epsilon_{2,p}^{\text{FT},\perp} & = & \epsilon_{0,p}+\xi/2 \, ,
\end{eqnarray}
are equal to the energies of the neighbouring pairs,
\begin{eqnarray}
\epsilon_{2,p-1}^{\text{FT},\perp}=\epsilon_{0,p-1}+\xi'/2 \, , \\
\epsilon_{1,p+1}^{\text{FT},\perp}=\epsilon_{0,p+1}-\xi'/2 \, ,
\end{eqnarray}
respectively, calculated with the SOC constant of $\xi'=2 \Delta\epsilon_{0}-\xi$.
Thus, when the contributions from all the QW states
are summed up in Eqs. (\ref{eq-ma-qw-FT}), (\ref{eq-ma-qw-Omega-FT-dir}),
the oscillation amplitude of the resultant MCA energy contribution
 $E_{\text{MCA,QW}}^{\text{FT}}(k=0)$
 is the same for
 the SOC strengths of $\xi$  and $\xi'=2 \Delta\epsilon_{0}-\xi$,
 if  very tiny contributions to this amplitude from the end QW state pairs,
 which lie well below ($p=1$) or  well above ($p=P$) the Fermi level,
 are neglected.
 Then, the amplitude $A_{\text{MCA,QW}}^{\text{FT}}(k=0)$,
as a function of $\xi$,
 is symmetric  about the point $\xi=\Delta\epsilon_{0}/2$, as shown in
Fig.~\ref{fig-model_amp_ma_vs_xi_FT_PT_k=0}(b).

The FT results for the QW state contribution
$E_{\text{MCA,QW}}^{\text{FT}}$
are obtained using
Eq.~(\ref{eq-ma-qw-Omega-FT-dir}),
which incorporates the Fermi energy $\epsilon_{\text{F}}$
in the presence of the SOC via the $g(\epsilon)$ function.
So far, for simplicity,
the same Fermi energy,
$\epsilon_{\text{F}}=\epsilon_{\text{F0}}=0$, has been assumed
in the FT and PT calculations for the QW state subsystem.
However, in an actual Co film with a finite SOC,  the Fermi energy is
shifted slightly with respect to $\epsilon_{\text{F0}}$,
such that
$\epsilon_{\text{F}}=\epsilon_{\text{F0}}+\Delta\epsilon_{\text{F}}$.
The shift $\Delta\epsilon_{\text{F}}$ results from
the changes of all electron energies
within the film due to the SOC,
so it  is treated as an external parameter
in the model of the QW state subsystem.
In the FT calculations,
the MCA energy contribution
  [Eq. (\ref{eq-ma-qw-FT})],
  expressed with the  grand potential of this subsystem
 [Eq. (\ref{eq-ma-qw-Omega-FT-dir})],
effectively depends on the position of
the central pair with respect to the Fermi level,
$\epsilon_{\text{0,CP}}-\epsilon_{\text{F}}$
(alongside the interpair energy spacing $\Delta\epsilon_0$,
the  $k$ dispersion parameters $b_1$, $b_2$
and the SOC coupling constant~$\xi$).
Thus, the change of the Fermi energy by
$\Delta\epsilon_{\text{F}}$
is equivalent to calculating the QW state contribution
at $\epsilon_{\text{0,CP}}-\Delta\epsilon_{\text{F}}$
(instead of $\epsilon_{\text{0,CP}}$)
while  keeping $\epsilon_{\text{F}}$ fixed at $\epsilon_{\text{F0}}=0$.
As a direct result, the profile
of $E_{\text{MCA,QW}}^{\text{FT}}(k=0)$
 determined as a function of $\epsilon_{\text{0,CP}}$
 for $\epsilon_{\text{F}}=0$
(Fig. \ref{fig-model-ma-vs-e0-k=0-var-xi-300K})
is shifted rightwards by $\Delta\epsilon_{\text{F}}$
when the Fermi energy is $\epsilon_{\text{F}}=\epsilon_{\text{F0}}+\Delta\epsilon_{\text{F}}$.
This shift, in turn, moves the profile minimum from
$\epsilon_{\text{0,CP}}=0$ to
$\epsilon_{\text{0,CP}}=\Delta\epsilon_{\text{F}}$.

For a Co film with $N_{\text{Co}}=12$~ML,
the Fermi energy shift
$\Delta\epsilon_{\text{F}}=\epsilon_{\text{F}}-\epsilon_{\text{F0}}$
is small for the nominal SOC of Co($0.0041$~eV)
but increases significantly with stronger SOC,
reaching $0.0271$~eV for $\xi=\xi_{\text{Pd}}$,
$0.0980$~eV for $\xi=2\xi_{\text{Pd}}$,
and $0.180$~eV for $\xi=\xi_{\text{Pt}}$.
These shifts change only slightly
for thicker Co films.
Notably,
the shifts $\Delta\epsilon_{\text{F}}$
are well approximated by the $\xi^2$ dependence predicted
by the PT  \cite{MC24}
across the whole considered SOC range.
This indicates that the Fermi energy
is less sensitive than the MCA energy
to inaccuracies of the PT description at strong SOC perturbations.

However, the shift of the Fermi energy is irrelevant
if we are primarily interested in
the oscillation amplitude
of the QW state contribution
to the MCA energy, and
comparing it between the FT and the PT.
The amplitude
$A_{\text{MCA,QW}}^{\text{FT}} ({\bf k})$
is invariant under such a shift
since it corresponds to shifting
the $\epsilon_{\text{0,CP}}$ profile of this contribution.
Consequently,
the oscillation amplitude
can be equivalently calculated
by assuming  $\epsilon_{\text{F}}=\epsilon_{\text{F0}}=0$
and using the same two values of the central pair energy,
 $\epsilon_{\text{0,CP}}$ and
 $\epsilon'_{\text{0,CP}}=\epsilon_{\text{0,CP}}\pm\Delta\epsilon_0/2$,
as in the PT approach.
In particular,  the energies
$\epsilon_{\text{0,CP}}=0$ and
$\epsilon'_{\text{0,CP}}=\Delta\epsilon_0/2$
can be used to determine the maximum amplitude
at ${\bf k}=(0,0)$.
These assumptions were made in the FT calculations,
the results of which are shown in
Figs. \ref{fig-model_ma_vs_k_xiCo_xiPd_xiPt_FT_PT_300K}-%
\ref{fig-model_amp_ma_vs_e0_temp_FT_PT_region}.

\subsection{ Magnetocrystalline anisotropy in C\lowercase{o}/Pt bilayer
calculated with force theorem and  perturbation theory}
\label{sec-mca-CoNM-bilayers}

\begin{figure}
   \includegraphics*[width=7cm]{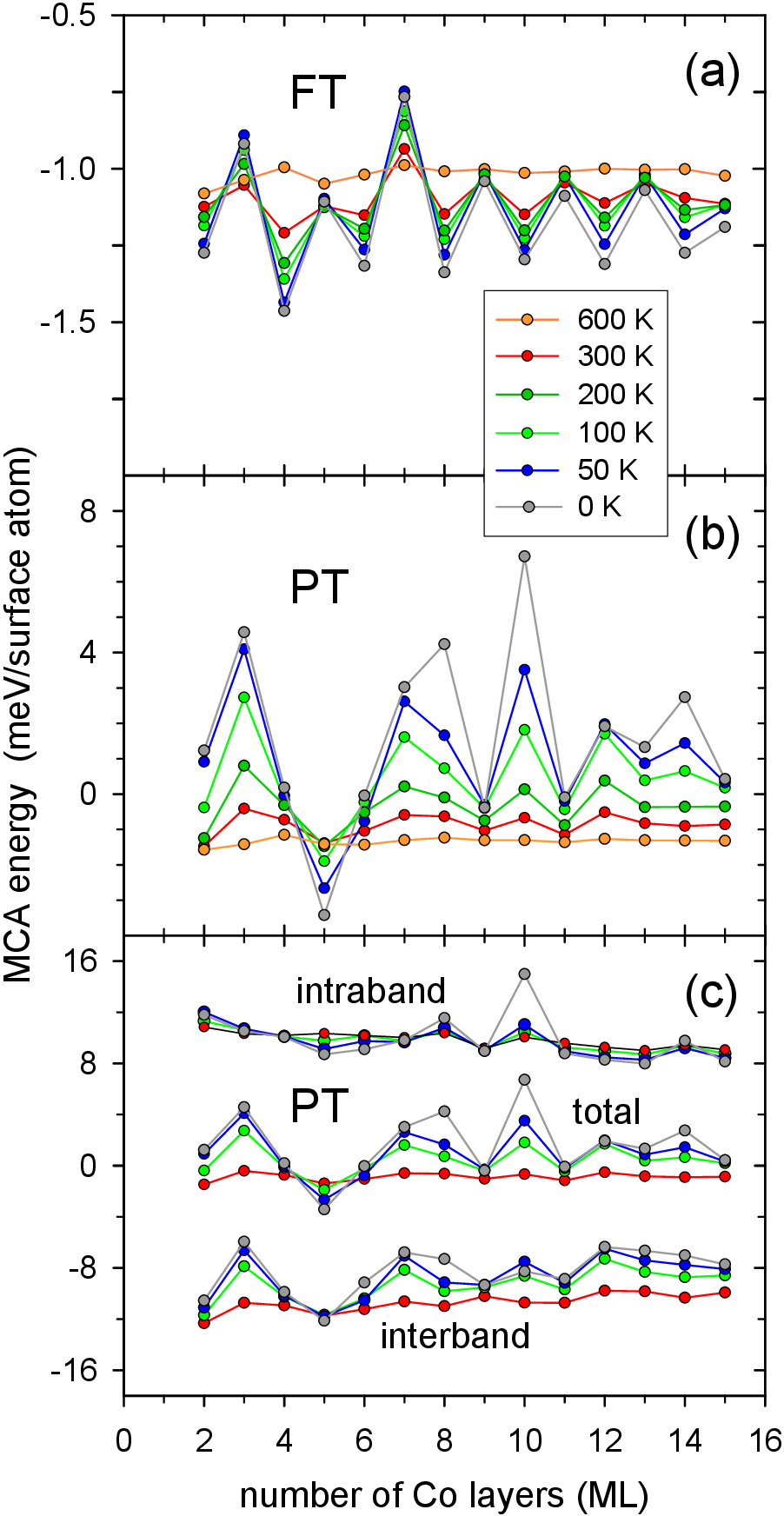}
      \caption{
      MCA energy  of the (001) fcc  Co/Pt(8 ML) bilayer
      calculated  with the (a) FT  and (b,c) the PT
      at various temperatures.
      (c) Intraband and interband terms of $E_{\text{MCA}}^{\text{PT}}$.
    }
  \label{fig-ma_FT_PT_CoPt8_var_temp}
\end{figure}

\begin{figure}
   \includegraphics*[width=8cm]{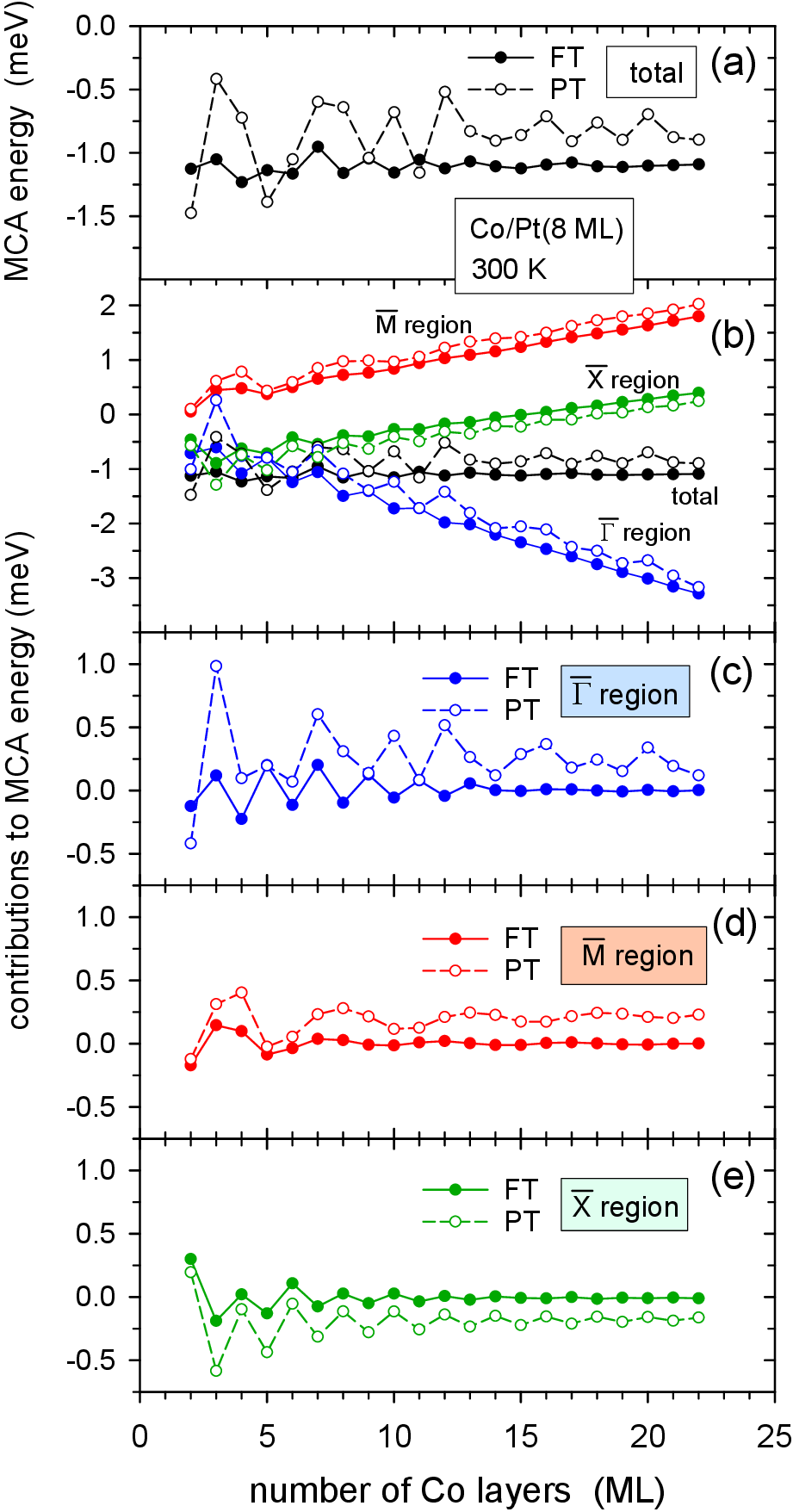}
      \caption{(a) MCA energy calculated  with the FT (solid circles)
      and the PT (open circles)
      for the (001) fcc  Co/Pt(8 ML) bilayer at $T=300$~K.
    (b)~Contributions to this energy from  different
      regions in the BZ (Fig.~\ref{fig-Brillouin_zone}).
      (c-e)~Oscillatory components of the MCA energy contributions
      from the (c) $\overline{\Gamma}$,
      (d) $\overline{M}$ and (e) $\overline{X}$ regions.
      Note: the same energy scale  is used  in panels (a,c-e).
    }
  \label{fig-CoPt-ma-FT-PT-regions}
\end{figure}

\begin{figure*}
   \includegraphics*[width=16cm]{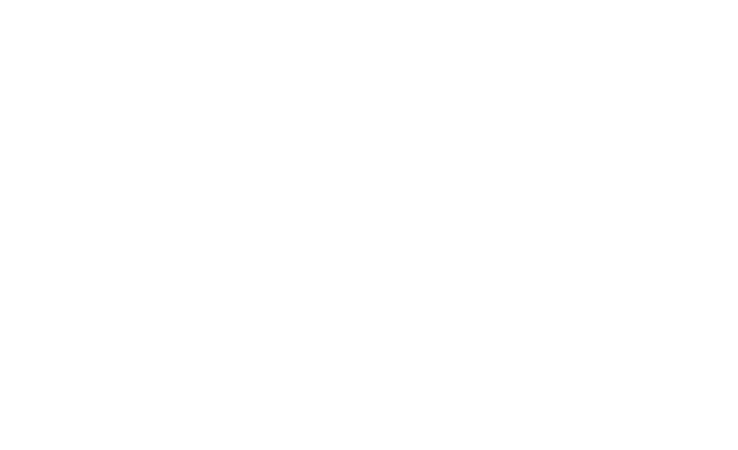}
      \caption{Distribution of the MCA energy in  the BZ
      for the (001) fcc (a,d) Co(11 ML)/Pd(8 ML) and (b,e) Co(12 ML)/Pd(8 ML) bilayers  with $N=11$ at  $T=300$~K,
      obtained with the FT (top row) and the PT (bottom row).
       (c,f)~Difference between the 12 ML and 11 ML distributions.
       The plots show  the symmetrized distributions
       (see Appendix~\ref{app-symmetrization}) in one quarter of the BZ.
       }
  \label{fig-ma_bz_per_CoPt8-FT-PT}
\end{figure*}

The PT also fails to accurately reproduce the MCA energy obtained with the FT
for the Co/Pt bilayer, as reported in Ref. \onlinecite{MC22} for T = 300 K.
The present results for other temperatures
(Fig. \ref{fig-ma_FT_PT_CoPt8_var_temp}) confirm this discrepancy, showing that the MCA energy oscillations in this system have a different pattern and larger amplitude in the PT calculations compared to the exact FT results.
This difference becomes more pronounced at low temperatures.
 The discrepancy between the FT and PT results
is different in different regions of the BZ
(Fig. \ref{fig-CoPt-ma-FT-PT-regions})
and most evident in  the contribution to the MCA energy
from the $\overline{\Gamma}$ region,
as for the Co film with an enhanced SOC constant
(Figs.
\ref{fig-Co_film_osc_eso_per_var_xi_gam05_300K}-%
\ref{fig-Co_film_osc_eso_per_var_xi_x05_300K}).
These findings are visualized by the respective
distributions of the MCA energy in the BZ, with
the PT contributions largest around the $\overline{\Gamma}$ point
and the FT distribution strongly smoothed out;
see Fig. \ref{fig-ma_bz_per_CoPt8-FT-PT}.
In both approaches,
the oscillation amplitude is much larger in
the contribution to
$E_{\text{MCA}}=E_{\text{MCA}}^{\text{Y}}$ (Y=FT and PT)
from the $\overline{\Gamma}$ region
than the  $\overline{M}$  region
while the $\overline{X}$ region contributes the least
to the MCA energy oscillations.
The oscillatory term from the $\overline{M}$ region
exhibits a clear period of 5~ML in both the FT and PT approaches,
while oscillations of the 2~ML period dominate in
the contribution from the $\overline{\Gamma}$ region,
as for the Co film.

\begin{figure}
       \includegraphics*[width=8cm]{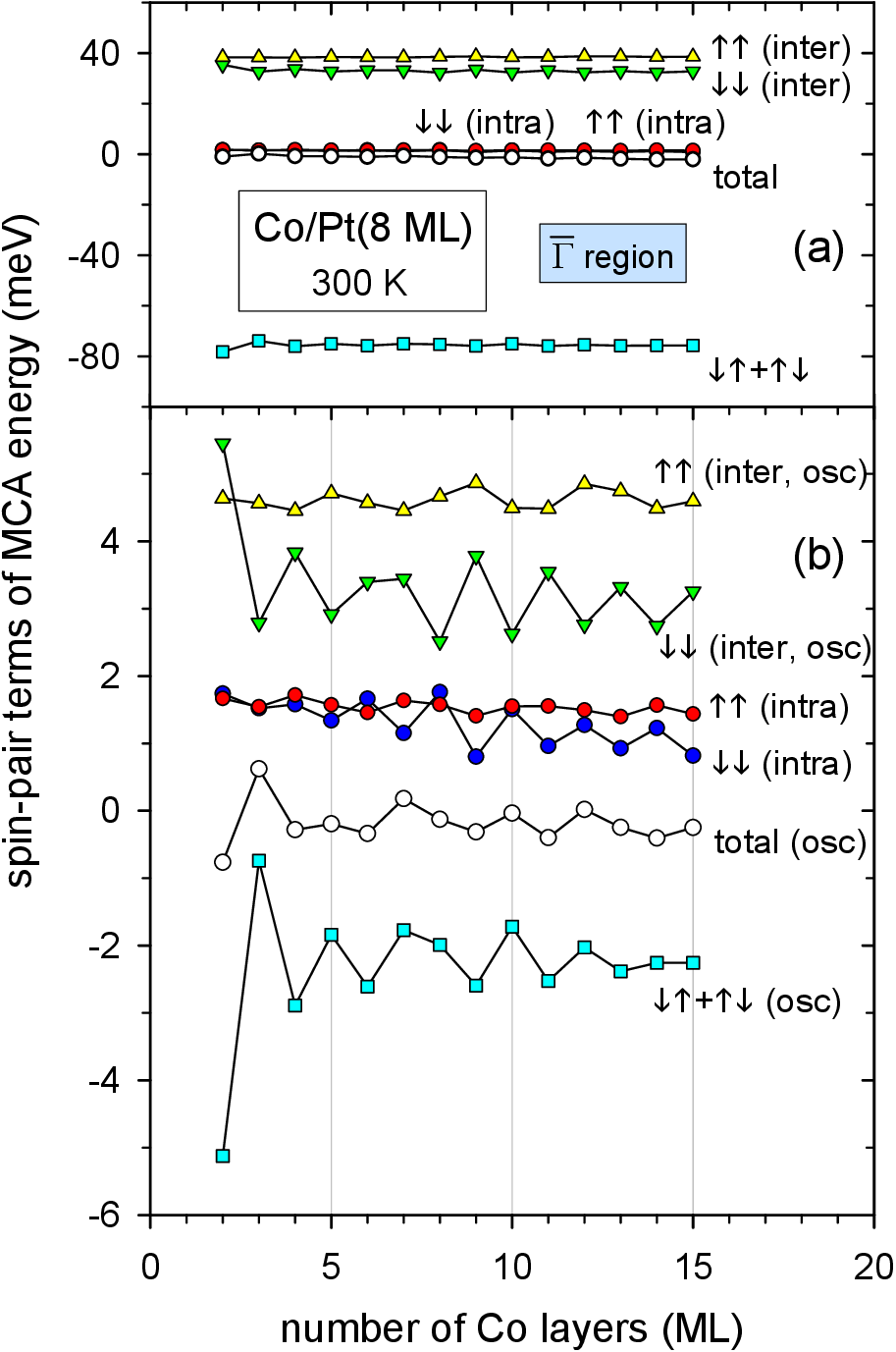}
      \caption{
      (a) Spin-pair terms of the contribution to
      the MCA energy  from the $\overline{\Gamma}$ region in the BZ
       calculated with the PT for the (001) fcc  Co/Pt(8 ML) bilayer
       at $T=300$~K. (b) Oscillatory parts of the spin-pair terms and the total
        $\overline{\Gamma}$ region contribution. The $\downarrow\downarrow$ and $\uparrow\uparrow$ terms are separated into their respective interband and intraband parts; see Eq. (\ref{eq-mca-PT-spin-pairs-intra-inter}).
    }
  \label{fig-CoPt-ma-spin-pair-Gamma-region}
\end{figure}

\begin{figure}
     \includegraphics*[width=8cm]{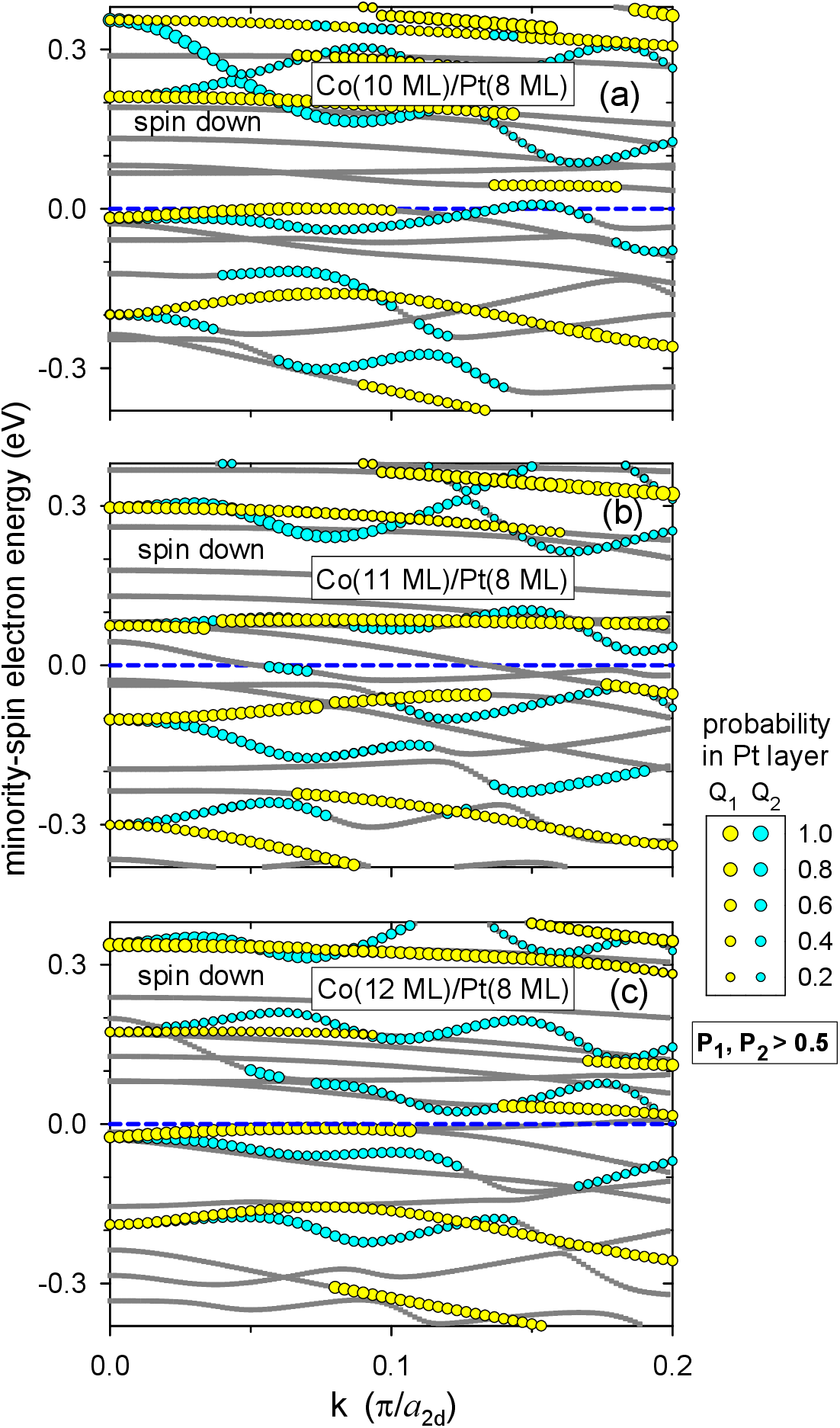}
      \caption{
      (a,b,c) Minority-spin
      electron energies along the
      $\overline{\Gamma}-\overline{X}$ line for
      the (001) fcc (a)~Co(10 ML)/Pt(8 ML), (b)~Co(11 ML)/Pt(8 ML)  and
      (c)~Co(12 ML)/Pt(8 ML) bilayers without the SOC.
      The colour-marked bands indicate states with projections
      $P_1\ge 0.5$
      onto the $|\phi_1\ket=(|yz\ket -|zx\ket)/\sqrt{2}$ orbital
      (yellow circles)
       and
       $P_2\ge 0.5 $ onto the $|\phi_2\ket=(|yz\ket +|zx\ket)/\sqrt{2}$ orbital
       (light blue circles); $P_1$~and~$P_2$ are equal to 1 at $k=0$.
       The size of colour symbols is proportional to the
       total probabilities of these states in the Pt layer,
       $Q_1$ and $Q_2$, respectively, according to the side scale.
        The Fermi energy is at $\epsilon_{\text{F0}}=0$ (horizontal dashed line).
      }
  \label{fig-CoPt-bilayer-energy-bands-zero-soc}
\end{figure}

\begin{figure}
     \includegraphics*[width=8cm]{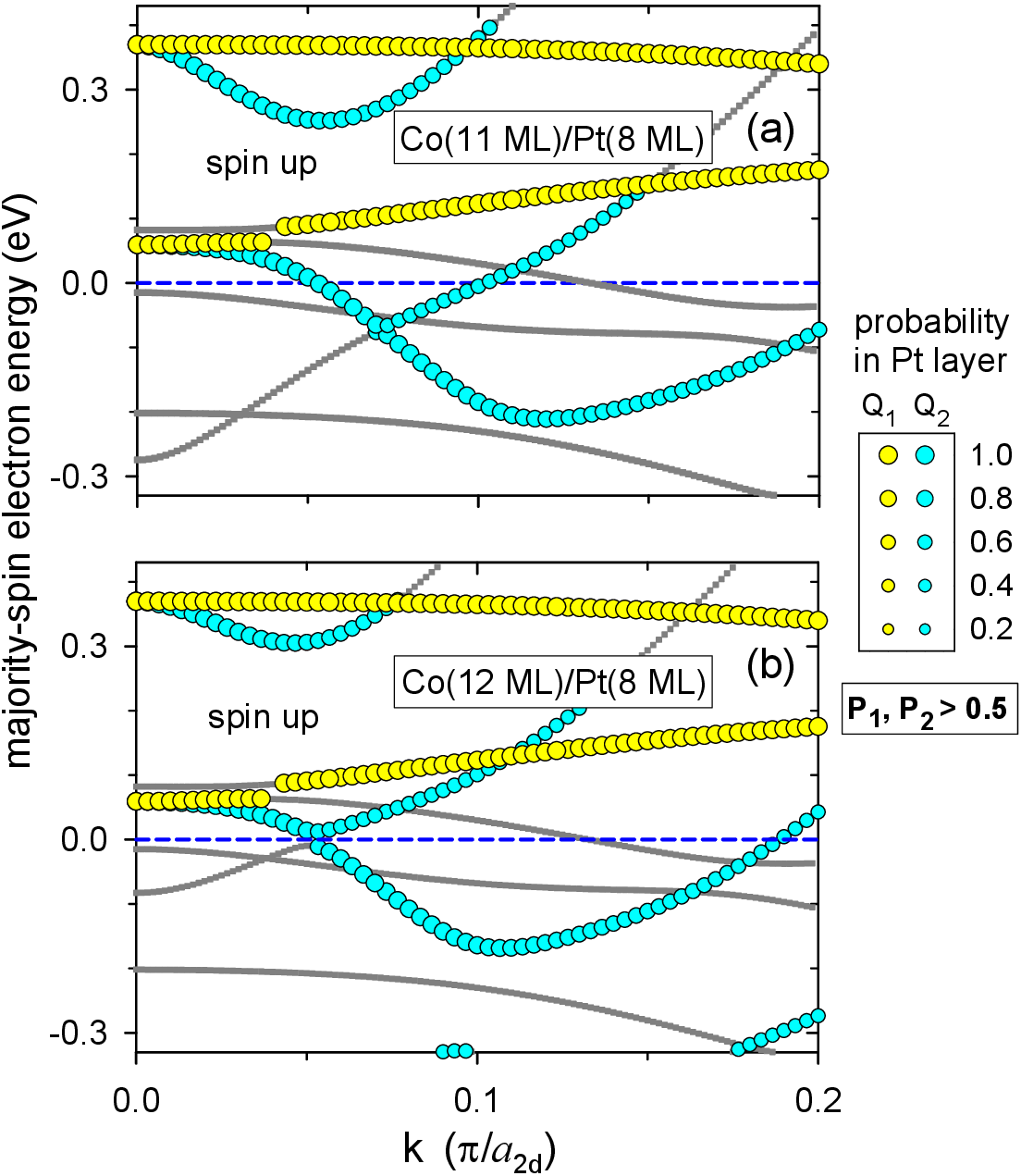}
      \caption{
      (a,b) Majority-spin electron energies along the
      $\overline{\Gamma}-\overline{X}$ line for
      the (001) fcc (a) Co(11 ML)/Pt(8 ML) and
      (b) Co(12 ML)/Pt(8 ML) bilayers without the SOC.
      The colour-marked bands (yellow and light blue circles)
      are defined  as in
      Fig.  \ref{fig-CoPt-bilayer-energy-bands-zero-soc}.
      }
  \label{fig-CoPt-bilayer-energy-bands-zero-soc-majority-spin}
\end{figure}

\begin{figure}
      \includegraphics*[width=8cm]{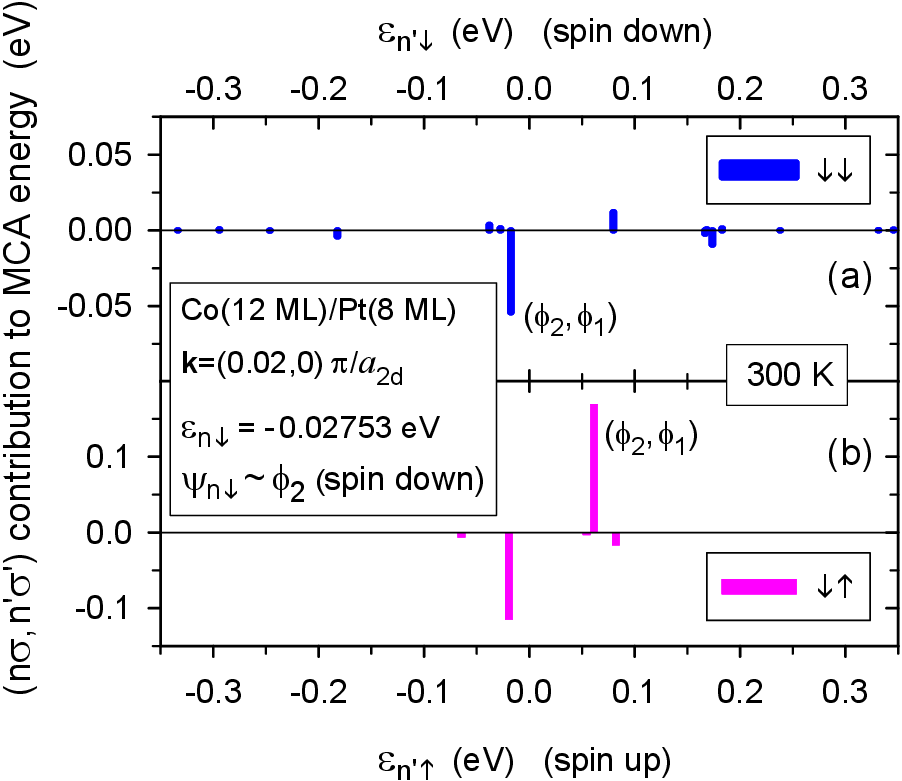}
      \caption{
      (a,b) MCA energy contributions $E_{\text{MCA}}^{n\sigma, n'\sigma'}$
      from  selected pairs of electron states
      at ${\bf k}=(k'_x,k'_y)=(0.02,0)\pi/a_{\text{2d}}$
      in the Co(12~ML)/Pt(8~ML) bilayer at $T=300$~K.
      The minority-spin state $|n{\bf k}\sigma\ket$ ($\sigma=\downarrow$)
      (predominant $|\phi_2\ket=(|yz\ket +|zx\ket)/\sqrt{2}$
      orbital composition, $\epsilon_{n\downarrow}=-0.02753$~eV)
      is coupled to  other states $|n'{\bf k}\sigma'\ket$ of
      (a) minority ($\sigma'=\downarrow$)
      and (b) majority ($\sigma'=\uparrow$) spins.
      The two largest contributions from coupling to states
      with the predominant $|\phi_1\ket=(|yz\ket -|zx\ket)/\sqrt{2}$
      composition are marked.
      }
  \label{fig-CoPt-bilayer-ma-n-np}
\end{figure}

In the PT calculations, the oscillations come
from all spin-pair terms $E_{\text{MCA}}^{\sigma\sigma'}$,
with  same and opposite spins $\sigma$ and $\sigma'$
(Fig. \ref{fig-CoPt-ma-spin-pair-Gamma-region}).
Two largest oscillatory contributions
from the $\overline{\Gamma}$ region
 originate from the spin-pair terms involving the minority-spin states,
 $E_{\text{MCA}}^{\downarrow\downarrow}$  and
 $E_{\text{MCA}}^{\downarrow\uparrow}+
 E_{\text{MCA}}^{\uparrow\downarrow}=
 2 E_{\text{MCA}}^{\downarrow\uparrow}$,
with the clear 2 ML period and similar amplitudes.
However, these contributions
have opposite phases
so that largely cancel out when they are summed.
 As a result, the net amplitude of the 2 ML period oscillations
 becomes much smaller, allowing
  weaker oscillatory terms, such as
  the $E_{\text{MCA}}^{\uparrow\uparrow}$ term
  with the 3 ML and 4 ML periods,
  to  become also significant
  in  the $\overline{\Gamma}$~region contribution
  to $E_{\text{MCA}}^{\text{PT}}$.
  These two longer periods
  come from the QW $sp$ states which originate from
  the majority-spin band in bulk Co and cross the Fermi level
  $\epsilon_{\text{F0}}$
  regularly as the Co layer thickness increases.
  Specifically, these states cross $\epsilon_{\text{F0}}$
  with the period of 4.12 ML at ${\bf k}=(0,0)$
  (the $\overline{\Gamma}$ point) \cite{MC22},
  but the period reduces to  approximately 3~ML at ${\bf k}$ distant
  by $0.1 k_{\text{D}}$ from the  $\overline{\Gamma}$ point
  (where $k_{\text{D}}$ is  $|\overline{\Gamma}$--$\overline{M}|$).
 The diagonal spin-pair terms $E_{\text{MCA}}^{\downarrow\downarrow}$
and $E_{\text{MCA}}^{\uparrow\uparrow}$ include,
apart from interband contributions, also intraband
contributions [Eq. (\ref{eq-mca-PT-spin-pairs-intra-inter})],
which partly cancel out oscillations of the former
and arise due to the lack of the inversion symmetry in the Co/Pt bilayer.

The total intraband term of the MCA energy
 $E_{\text{MCA}}^{\text{PT}}=E_{\text{MCA,intra}}+E_{\text{MCA,inter}}$
is of comparable magnitude as its interband term and
the two terms largely cancel out
(see Fig. \ref{fig-ma_FT_PT_CoPt8_var_temp}), as it was previously found for
the Co/Pd and Co/Cu bilayers \cite{MC22,MC24}.
The oscillations of $E_{\text{MCA,intra}}$
are smaller than those of $E_{\text{MCA,inter}}$
in most of the Co thickness range, especially at lower temperatures.
However, the intraband term can substantially increase
for specific Co/Pt films where one of nearly flat energy bands
$\epsilon_{n\sigma}({\bf k})$
in the vicinity of the $\overline{\Gamma}$ point
lies almost exactly at the  Fermi level,
as for the Co(10 ML)/Pt(8 ML) bilayer;
see Fig. \ref{fig-CoPt-bilayer-energy-bands-zero-soc}(a).
The intraband contribution
 $\frac{1}{2}f'_0(\epsilon_{n\sigma})[\epsilon_{n\sigma}^{(1)}]^2$
 from such a band, proportional to
 $-\delta(\epsilon_{n\sigma}-\epsilon_{\text{F0}}$) at $T\rightarrow 0$,
 then leads  to a strong peak in the
 $E_{\text{MCA}}^{\text{PT}}(N_{\text{Co}})$ variation
 (at $N_{\text{Co}}=10$~ML in Fig. \ref{fig-ma_FT_PT_CoPt8_var_temp}),
 in a similar way as peaks of the DOS arise.

These findings show that the origin of the MCA energy
oscillations in the Co/Pt bilayer, as described by the PT,
is more complex than in the Co film.
Unlike the Co film, where only minority-spin state pairs
contribute to these oscillations,
the bilayer's oscillations arise from pairs of states
with both same and opposite spins,
as well as individual states which give intraband contributions.
Furthermore, although these oscillations are induced by changes
of the Co layer thickness, they do not arise primarily
from the SOC within the Co part of the bilayer.
In fact, the MCA energy and its oscillations in the Co/Pt bilayer
arise almost entirely from the large SOC of the Pt layer,
as it is demonstrated in Ref. \onlinecite{MC22}
by decomposing the PT expression for the MCA energy into four terms
$E_{\text{MCA}}^{\text{XY}}$,
proportional to the products $\xi_{\text{X}}\xi_{\text{Y}}$ of the SOC constants,
where X and Y are Co or Pt.

This scenario is possible because
states in the  Co layer near the Fermi energy
hybridize with the $d$ states in Pt layer,
resulting in states that span over the whole bilayer,
as previously found for similar delocalized states in the Pd/Co/Pd
trilayer (Fig. 7 in Ref. \onlinecite{MC01}) and
the Co/Cu/Pt trilayer (Fig. 4 in Ref. \onlinecite{EBMC15}]).
Such a hybridization occurs for both minority-spin $d$ states
and majority-spin $sp$ states in Co,
because $d$ states of both spins are available near the Fermi level
in the nonmagnetic Pt layer.
Since the resultant states $|n{\bf k}\sigma\ket$
have a finite probability in each layer
[defined with their wavefunctions $\psi_{n{\bf k}\sigma}({\bf r})$
as the spatial integral of
$|\psi_{n{\bf k}\sigma}({\bf r})|^2$ over the Co or Pt layer]
the energies $\epsilon_{n\sigma}({\bf k})$
of these states change
as the thickness of the Co layer increases,
providing a necessary condition for oscillations of the MCA energy
enhanced by the SOC of Pt.
The states with a significant probability  in the Pt layer
are subject to the large spin-orbit interaction in this layer
and, thus, can strongly couple to other states of both spins
that also have a substantial probability  in the Pt layer.
In particular, such coupling occurs within pairs
of minority-spin states that span both Co and Pt layers,
and are degenerate at the $\overline{\Gamma}$ point [${\bf k}=(0,0)$],
as in the Co film.
These states are predominantly composed of the $yz$ and $zx$ orbitals
 [or, more specifically,  their two combinations,
 $\phi_1$ and $\phi_2$, defined in Eq. (\ref{eq-phi12})],
but also contain minor components from other orbitals
 for ${\bf k}\neq (0,0)$.
Pairs of minority-spin states with
such a spatial distribution and orbital composition
are indeed present in the Co/Pt(8 ML) bilayer
and have the probability of around  $Q=0.5$ in the Pt layer for
the Co thicknesses of 10-12 ML
(see Figs. \ref{fig-CoPt-bilayer-energy-bands-zero-soc}
and \ref{fig-CoPt-bilayer-energy-bands-zero-soc-majority-spin}).
They contribute to the 2~ML period oscillations of the $E_{\text{MCA}}^{\downarrow\downarrow}$ term
in the PT approach, with largest contributions coming from
pairs with energies in the immediate vicinity
of the Fermi level; see Fig. \ref{fig-CoPt-bilayer-ma-n-np}(a).
Similar pairs of $d$ states,
predominately composed of the $yz$ and $zx$ orbitals
and degenerate at the $\overline{\Gamma}$ point,
also exist for majority spin
and they are almost entirely confined within the Pt layer ($Q\approx 1$).
Consequently, such majority-spin states
near the Fermi level, as those seen in
Figs. \ref{fig-CoPt-bilayer-energy-bands-zero-soc-majority-spin},
can also strongly couple to the minority-spin QW states
near the Fermi level
[Fig. \ref{fig-CoPt-bilayer-ma-n-np}(b)],
leading to the 2 ML period oscillations  of the
 $E_{\text{MCA}}^{\downarrow\uparrow}+
 E_{\text{MCA}}^{\uparrow\downarrow}=
 2 E_{\text{MCA}}^{\downarrow\uparrow}$
 term in the PT calculations
 (Fig. \ref{fig-CoPt-ma-spin-pair-Gamma-region}).
 Note that the oscillation amplitude of
this term can be significantly reduced by substantial contributions from
some other majority-spin states  coupled
to the minority-spin QW states,
as shown in Fig. \ref{fig-CoPt-bilayer-ma-n-np}(b).
Furthermore,  the majority-spin states
that are mostly localized in the Pt layer
and arise from $d$ states in Pt hybridized with $sp$ states in Co
contribute to oscillations of
the MCA energy with longer periods of 3-4 ML.
As already discussed, this contribution dominate in
the $E_{\text{MCA}}^{\uparrow\uparrow}$ term,
but it is also present, as a minor component, in
the
$E_{\text{MCA}}^{\downarrow\uparrow}+
 E_{\text{MCA}}^{\uparrow\downarrow}$
term.
The described scenario
is similar to the mechanism that
underlies the experimentally observed MCA oscillations
in the Co/Cu bilayer versus the Cu thickness,
which are  substantial
despite a low DOS at the Fermi energy in Cu
and primarily arise from the SOC in the Co layer \cite{MCMP13}.
In that system, the $sp$ states in Cu
hybridize with  minority-spin $d$ states in the Co film and 
the energies of the resultant states, mostly localized in the Co layer,
shift and cross the Fermi level periodically as the Cu thickness increases.

States composed solely of $yz$, $zx$
(which is strictly valid
at the $\overline{\Gamma}$ point)
are coupled only by the $L_z S_z$ terms of $H_{\text{so}}$,
as discussed under Eq. (\ref{eq-mat_element-Hso}).
Thus, states with such an orbital composition and  same spins
can couple to each other only  for out-plane magnetization
(with spins along the easy axis $\zeta=z$)
while the coupling between them vanishes
for in-plane direction of magnetization ($\zeta=x$).
The reverse is true for pairs of states
with opposite spins where the finite coupling
is present only for in-plane magnetization
since the spin operator $S_z$ is
equivalent to $S_{\eta}=(S'_{+}-S'_{-})/2i$
in the rotated frame of reference $O\xi\eta\zeta$,
with the ladder operators $S'_{+}$ and $S'_{-}$ acting on the spin states
$|\downarrow\ket$ and $|\uparrow\ket$ quantized along the easy axis $\zeta=x$.
Similar distinction, though less strict,
with strong coupling
for one  magnetization direction and weak
for the other, depending on the spins of the involved
states,
also holds in the vicinity of the $\overline{\Gamma}$ point
where the considered QW states include minor components from other orbitals,
in addition to $yz$ and $zx$.
As a result, contributions from  pairs  of such states
with same and opposite spins
come with opposite signs in the MCA energy
$E_{\text{MCA}}^{\text{PT}}=
\Omega^{(2)} ({\hat{\bf M}}_{\perp})-
\Omega^{(2)} ({\hat{\bf M}}_{||})$.
This explains why the 2 ML period oscillations of
the spin-pair terms
$E_{\text{MCA,inter}}^{\downarrow\downarrow}$
and
$E_{\text{MCA}}^{\downarrow\uparrow}+
E_{\text{MCA}}^{\uparrow\downarrow}$
are in antiphase and largely cancel out,
which reduces the amplitude of
the  2 ML period oscillations in the total MCA energy,
as seen in  Fig. \ref{fig-CoPt-ma-spin-pair-Gamma-region}.
An additional reduction of its oscillation amplitude arises due to
intraband contributions in $\Omega^{(2)}$
[Eq. (\ref{E2-corr})]
which are non-zero only for in-plane magnetization
so that the term $E_{\text{MCA,intra}}^{\downarrow\downarrow}$
also oscillates with the Co thickness in antiphase to the interband term
$E_{\text{MCA,inter}}^{\downarrow\downarrow}$.

The cancellations of the oscillatory spin-pair terms
in the PT approach
lead to around three-fold reduction in
the oscillation amplitude of the net MCA energy,
compared to its $E_{\text{MCA,inter}}^{\downarrow\downarrow}$ term.
However, this only partly explains
why the amplitude of the MCA energy obtained
with the PT for the CoPt bilayer
is much smaller than that for the Co film with the SOC of Pt,
where only the term
$E_{\text{MCA}}^{\downarrow\downarrow}=
E_{\text{MCA,inter}}^{\downarrow\downarrow}$
due to minority-spin state pairs
contributes to the oscillations of
 $E_{\text{MCA}}^{\text{PT}}$
(see Fig. 3 in Ref. \cite{MC22}).
Another reason for this difference is that
the effective strength of the SOC in the CoPt bilayer
is smaller than that of Pt.
Indeed, minority-spin QW states
 $\psi_{n{\bf k} \downarrow}({\bf r})$
that span over a Co/Pt bilayer
can be formally represented,
as sums of two states
$\psi_{n{\bf k}\downarrow}^{\text{X}}({\bf r})$,
one confined in the X=Co layer and the other in the X=Pt layer,
\begin{equation}
\psi_{n{\bf k}\downarrow}({\bf r})=
a\psi_{n{\bf k}\downarrow}^{\text{Co}}({\bf r})
 +  b\psi_{n{\bf k}\downarrow}^{\text{Pt}}({\bf r}) \, ,
\end{equation}
with respective probability amplitudes
$a$ and $b$  (where $|a|^2+|b|^2=1$;
note that $a$ does not denote the lattice constant here).
The effective SOC constant in the matrix elements
\mbox{
$\bra n{\bf k}\downarrow | H_{\text{so}} |n'{\bf k}\downarrow \ket$
}
entering the $E_{\text{MCA}}^{\downarrow\downarrow}$ term
is then approximated as
 \begin{equation}
\xi_{\text{eff}}^{\downarrow\downarrow}=|a|^2\xi_{\text{Co}} + |b|^2\xi_{\text{Pt}} \approx |b|^2\xi_{\text{Pt}} \, ,
 \label{eq-bilayer-effective-SOC-down-spin}
 \end{equation}
 given that $\xi_{\text{Pt}}$ is nearly eight times
 larger than $\xi_{\text{Co}}$.
 With the probability in  Pt of $Q=|b|^2\approx 0.5$
 for the QW states near the $\overline{\Gamma}$ point
(Fig. \ref{fig-CoPt-bilayer-energy-bands-zero-soc}),
the value of  $\xi_{\text{eff}}^{\downarrow\downarrow}$
is close to $\xi_{\text{Pt}}/2$ and their contribution  to
$E_{\text{MCA}}^{\downarrow\downarrow}$ is proportional to
$(\xi_{\text{eff}}^{\downarrow\downarrow})^2\approx 0.25\xi_{\text{Pt}}^2$.
Majority-spin $d$ states  near the Fermi level
$\psi_{n{\bf k}\uparrow}({\bf r})=
\psi_{n{\bf k}\uparrow}^{\text{Pt}}({\bf r})$,
which are entirely confined within the Pt layer
(Fig. \ref{fig-CoPt-bilayer-energy-bands-zero-soc-majority-spin}),
 are subject to the SOC of Pt only.
Accordingly, the contribution from pairs of QW states with opposite spins
to $E_{\text{MCA}}^{\uparrow\downarrow}$
and  $E_{\text{MCA}}^{\downarrow\uparrow}$
involves the factor
$(\xi_{\text{eff}}^{\downarrow\uparrow})^2=|b|^2\xi_{\text{Pt}}^2$
which defines the respective effective constant
 \begin{equation}
\xi_{\text{eff}}^{\downarrow\uparrow}=|b|\xi_{\text{Pt}}  \, ,
 \label{eq-bilayer-effective-SOC-down-up-spin}
 \end{equation}
close to $0.7\xi_{\text{Pt}}$ for $|b|^2\approx 0.5$.

 Thus, the reduced effective SOC in the Co/Pt bilayer
 helps to explain further
 why the discrepancy between the PT and FT predictions is smaller
  in this system compared to the standalone Co film with the full SOC of Pt
 (see Figs. \ref{fig-Co_film_eso_per_var_xi_300K}(d)
 and \ref{fig-ma_FT_PT_CoPt8_var_temp}).
Another example that follows a similar rule
is the Co/Pd bilayer
for which the PT  describes the MCA energy and
its oscillations with good accuracy (at $T=300$~K),
while a larger discrepancy between the PT and FT  is found
for a Co film with the  SOC of Pd;
see Fig. \ref{fig-Co_film_eso_per_var_xi_300K}(b)
in Ref. \onlinecite{MC22} and Fig. 2(b) in this paper, respectively.

In the FT approach,
the QW state contribution to the grand potential
for  out-of-plane  magnetization
can be approximately described using  a simplified model
of equidistant pairs of minority-spin QW states
(strongly coupled for this magnetization  direction),
with the Hamiltonian decomposed into 2x2 blocks,
as for  the Co film.
An extension of this model to the case of  in-plane
magnetization is not straightforward
since it would also require including a subset of majority-spin states
and diagonal (intraband) matrix elements
 $\langle  n{\bf k}\sigma|H_{\text{so}}| n{\bf k}\sigma\rangle$
 of the SOC operator.
Although these  elements vanish
for the states $|n{\bf k}\sigma\rangle$
composed solely of
$yz$ and $zx$ orbitals,
for both magnetization directions,
they are finite for in-plane magnetization
in the vicinity of the $\overline{\Gamma}$ point,
due to minor orbital components,
the share of which  cannot be easily estimated.
Nevertheless, we can still expect that
since, for each spin,
the subset of unperturbed QW states can be
approximately represented with a sequence of equidistant energies,
the oscillation amplitude of the MCA energy
determined within the FT approach is limited by the finite energy spacings
between the consecutive QW states pairs,
in a similar way as  for the Co film with enhanced SOC where
QW $d$ states of minority-spin are present only;
see  Sec. \ref{sec-ma-QW-state-FT-PT}.
Such a plausible limitation may well explain
why the oscillation amplitudes have  similar  magnitudes
for the Co/Pd \cite{MC22,MC24} and Co/Pt bilayers
 in the FT approach, despite a large difference in the SOC strength
 between  the two systems,
 while the oscillation amplitude is a few times larger for the Co/Pt bilayer
 in the PT calculations.

\section{Conclusions}

The present work examines limitations of the PT
in calculations of the MCA energy for layered systems with strong SOC
and aims to reveal the origin of these limitations.
The differences between the exact FT and
approximate PT results are most evident
in the oscillations of the MCA energy,
with the PT calculations  predicting a considerably larger oscillation amplitude
while the average value is less affected.
This  discrepancy
between the two approaches
grows with increasing the SOC strength, whether it is the SOC of
the nonmagnetic layer in the Co/Cu \cite{MC24}, Co/Pd \cite{MC24}
and Co/Pt  bilayers
or the enhanced SOC in the Co film.
Substantial deviations of the PT predictions
from the FT values of the MCA energy
are also found for the Co film with the nominal SOC
at low temperatures (100~K and lower),
similar to previous findings on MCA energy oscillations
in the Co/Pd bilayer \cite{MC24}.

The MCA energy calculated for the Co/Pt bilayer
in the FT approach shows a moderate dependence on temperature,
which defines smearing of energy levels.
In contrast, this energy strongly depends on temperature in the PT calculations,
with oscillation amplitude rapidly growing with decreasing temperature,
down to $T=0$.
Consequently, the discrepancy between the FT and PT results
for the Co/Pt bilayer becomes particularly pronounced at low temperatures.
Such a temperature dependence of the MCA energy
in systems with strong SOC  can explain
the largely differing values of this energy
for the Fe(1ML)/Pt(1 ML) bilayers
in the two separate PT calculations reported in Ref. \onlinecite{BlancoRey19}.
Namely,  this difference can be attributed mainly not to
the different {\em ab initio} codes applied,
 SIESTA \cite{Soler02-SIESTA} and
 VASP \cite{Kresse93-94,Kresse96b},
 but rather the different values of
 the Fermi-Dirac smearing $k_{\text{B}}T$ assumed
 in these codes,
 0.01~eV and 0.05~eV, respectively,
 which correspond to $T\approx 115$~K and $T\approx 580$~K.
This also explains why the PT well reproduces the FT results
using VASP, where the higher temperature is applied, while the discrepancy
between the PT and FT values of the MCA energy
is large for the lower temperature, used in SIESTA.

The PT contributions to the MCA energy deviate from
its FT distribution
mostly
near the high symmetry points
where the oscillations due to respective QW states arise, with
the largest deviations around the
$\overline{\Gamma}$ point, and smaller, though still significant,
deviations near the $\overline{M}$ point.
Accordingly,
the results for the Co film with enhanced SOC are analyzed in depth
using a simple model of minority-spin QW $d$ states
(composed of  $yz$ and $zx$ orbitals) near the $\overline{\Gamma}$ point.
The model provides an approximate description of
the dominating 2~ML period oscillations of the MCA energy,
showing good agreement with the numerical
calculations for the full film system.
It is shown that the discrepancy
between the FT and PT predictions arises because
the second-order PT formula (\ref{E2-corr})
can fail to adequately describe
contributions to the MCA energy from
pairs of states with
energy spacing smaller or comparable to the SOC strength.
This is specifically the case
for the considered pairs of QW states in the central region of the BZ
which are doubly degenerate at the $\overline{\Gamma}$ point,
and whose energies lie very close to the Fermi level
at this point.
For specific film thicknesses for which such a  pair exists,
the difference between the MCA energies obtained in the PT and FT calculations
significantly grows with decreasing temperature,
and the PT result even diverges at $T=0$
when a pair energy lies exactly at the Fermi level at the $\overline{\Gamma}$ point.
However, even in this case,
the PT still  reproduces
the QW state contribution to
the MCA energy  with good accuracy
when the SOC strength $\xi$ is a few times smaller than
$10k_{\text{B}}T$, as for the Co film with the nominal SOC  at $T=300$~K.
Furthermore, the discrepancy between the FT and PT predictions
for the MCA energy is much smaller and
its temperature dependence is much weaker
for film thicknesses for which no QW state pair is present
in the immediate vicinity of the Fermi level, particularly
when this level lies midway
between  two consecutive pairs at the $\overline{\Gamma}$ point.
The determined dependence
of the MCA energy oscillation amplitude
on the energy of the central QW state pair
(the one closest to the Fermi level)
well explains how local oscillation amplitude changes
with the Co film thickness.

Further,
it is shown that, while the PT oscillation amplitude
grows quadratically with increasing
the SOC strength in the film, the FT amplitude follows this
trend only for weak and moderate SOC (e.g., that of Co)
and  is limited  by the finite energy spacing
between the neighbouring QW state pairs
when the SOC strength becomes comparable to this spacing.
Consequently, the PT predicts a larger amplitude of the MCA energy oscillations
than the exact FT approach, with particularly large difference
for the Co film with the strong SOC of Pt.
A similar limitation in the FT oscillation amplitude due to
a sequence of QW state pairs
is also expected for the Co/Pt bilayer
where the amplitude
predicted by the PT is a few times larger.

In the Co/Pt bilayer, the oscillations of the MCA energy
with increasing thickness of the Co layer
are mediated by the strong SOC in the Pt layer.
Near the Fermi level, states of both spins in Co
(minority-spin $d$ states and majority-spin $sp$ states)
hybridize with $d$ states in Pt and
the MCA energy oscillations arise as
the energies of the hybridized states
change with the Co thickness.
These states are coupled by the SOC to other states of both spins,
particularly strongly to those mostly localized in the Pt layer.
As a result,
the oscillations of the MCA energy calculated with
the PT come from spin-pair terms with same (mainly minority)
and opposite spins,
which leads to large cancelations of  2~ML  period oscillations
and revealing weaker oscillations of longer periods in the thickness dependence
of the MCA energy.
In addition, the intraband terms which emerge due to the lack of the inversion symmetry in the Co/Pt bilayer also contribute to the MCA oscillations.
In contrast, in the FT calculations,
the oscillatory terms with longer periods  play only a minor role
in the oscillations of the MCA energy, which are dominated by the 2 ML period.

The states, which span over the whole Co/Pt bilayer,
are subject to an effective SOC defined
with a weighted average of the SOC constants of Co and Pt.
Although the effective SOC constant is substantially reduced
compared to that of Pt,
it is still high enough to make
the PT predictions for the MCA energy largely
inaccurate, with  oscillation amplitude 2-3 times greater
than in the exact FT calculations.
Since the states calculated in  the presence of the SOC
that are responsible for the MCA oscillations within the FT approach
are also delocalized across the bilayer,
their contributions to the MCA energy are also expected to reflect an effective, reduced SOC.

Although this work focuses on the effect of SOC
on the MCA energy oscillations, it is also relevant
to the total MCA energy.
While average MCA energies calculated with the FT and PT show only moderate discrepancies at strong SOC,
the MCA energy for a system with a specific thickness can exhibit significant discrepancies due to large differences between the oscillations of this energy
 predicted by the two approaches.
Thus, the PT approach should be used with caution
when examining the MCA energy
(e.g., its spatial distribution
\cite{Miura13,Miura18-prb,Okabayashi18})
in layered systems comprising elements with strong SOC.
Accordingly, prior verification of agreement between
FT and PT total MCA energies
is crucial for a reliable application of PT results in further analysis.

\appendix

\section{Symmetrization of MCA energy distribution in Brillouin zone}
\label{app-symmetrization}

\begin{figure}[t]
    \includegraphics*[width=8.6cm]{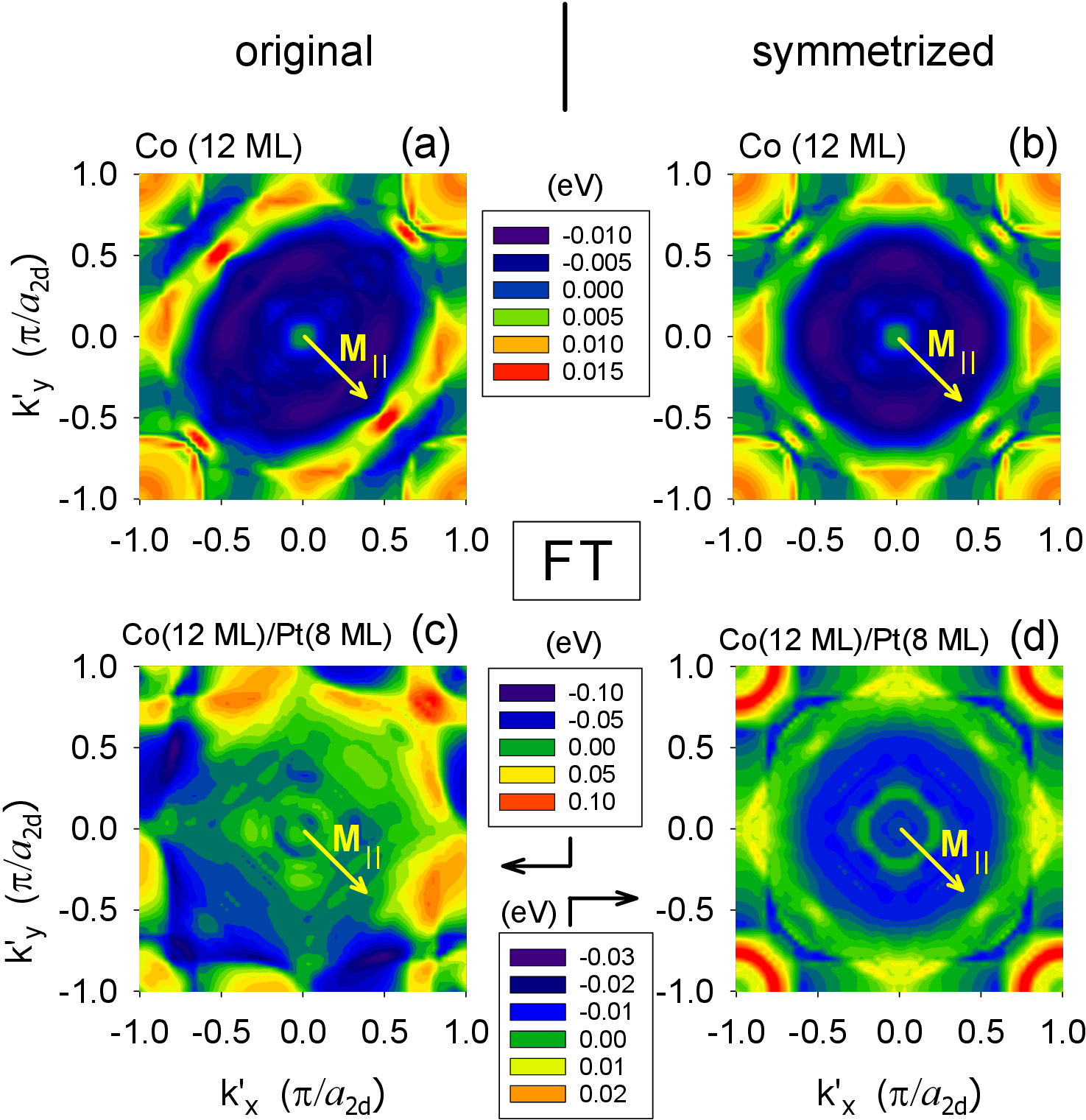}
      \caption{
      (a,c) Original and (b,d) symmetrized  distributions
      of the MCA energy in  the BZ  obtained with the FT  for the (001) fcc
      (a,b) Co(12~ML) film and (c,d) Co(12~ML)/Pt(8~ML) bilayer at $T=300$~K;
      see Appendix~\ref{app-symmetrization}. The assumed direction of in-plane magnetization ${\bf M}_{||}$ is marked with the yellow arrow.
      The respective plots for the Co(12~ML) film obtained with the PT are nearly identical within the figure's scale. Note the different scales of the original and symmetrized  distributions for the  Co(12~ML)/Pt(8~ML) bilayer
      due to the cancelation of the first-order terms from ${\bf k}$ and $-{\bf k}$.  }
  \label{fig-ma-bz-FT-symm4-Co12-Co12Pt8}
\end{figure}

The distribution of  the MCA energy in  the BZ,
as defined with Eqs. (\ref{eq-mca-omega-FT}) and
(\ref{eq-mca-omega-PT}),
lacks the full symmetry of the system without the SOC,
as shown in Figs. \ref{fig-ma-bz-FT-symm4-Co12-Co12Pt8} (a) and (c).
This is due to the reduced symmetry of
the grand potential distribution in the presence of  the SOC
for in-plane magnetization,
in particular, $\bf M$  along the $x$ axis.
However,
the symmetric distribution can  be recovered
by averaging (summing and dividing by eight)
the contributions to
$E_{\text{MCA}}=E_{\text{MCA}}^{\text{Y}}$  (Y=FT and PT)
over the star of each point ${\bf k}=(k_{x},k_{y})$, which
comprises the eight equivalent points
${\bf k}_i$ ($i=1,2,\dots, 8$) defined as
$(\pm k_{x},\pm k_{y})$ or  $(\pm k_{y},\pm k_{x})$
and  generated by the respective
symmetry operations ${\cal R}_i$.

Since the unperturbed energies $\epsilon_{n\sigma}$
are equal at all points ${\bf k}_i$,
symmetrizing the distribution of the MCA energy
defined in Eq. (\ref{E2-corr}) within the PT framework
amounts to averaging of the squared matrix elements
$\left| \langle  n'{\bf k}_i\sigma'|H_{\text{so}}|
n{\bf k}_i\sigma\rangle \right|^2 $
over the different ${\bf k}_i$  points.
For a general ${\bf k}$ point, the wavefunctions
$\psi_{n{\bf k}_i}^{\sigma}({\bf r})$  at the points
${\bf k}_i={\cal R}_i{\bf k}$
are equal to
$\psi_{n{\bf k}_i}^{\sigma}({\cal R}_i^{-1}{\bf r})$
up to a constant phase factor $e^{i\beta}$.
Once the substitution ${\bf r}'={\cal R}_i^{-1}{\bf r}$ is made,
the matrix element $\left| \langle  n'{\bf k}_i\sigma'|H_{\text{so}}|
n{\bf k}_i\sigma\rangle \right|^2 $
can be expressed as $\left| \langle  n'{\bf k}\sigma'|H_{\text{so}}|
n{\bf k}\sigma\rangle \right|^2 $ where the spin states
are quantized along a rotated magnetization direction.
This
is either ${\cal R}_i \hat{{\bf M}}$ or one of other directions ${\cal R}_j \hat{{\bf M}}$ ($j\neq i$) if the operation ${\cal R}_i$ involves
only one mirror symmetry
($x\rightarrow -x$ or $y\rightarrow -y$).
In either case, the polar angle $\theta$ of the magnetization vector $\bf M$
remains unchanged  while its azimuthal angle $\phi$
is transformed to the respective angle $\phi'=\phi_i$.
When the spin  quantization axis $\zeta$ is oriented
along the $(\theta,\phi')$ direction,
the matrix element in question may include
five distinct angular-dependent terms,
proportional
to $\sin^2\theta$,  $\sin^2\theta \cos 2\phi'$,  $\sin^2\theta \sin 2\phi'$,
$\sin 2\theta \cos \phi'$ and $\sin 2\theta \sin \phi'$,
respectively.
The terms originate from the angular dependence
of the SOC operator when the spin
${\bf S}=(S_{\xi},S_{\eta},S_{\zeta})$ is represented
in the rotated frame of reference $O\xi\eta\zeta$
\cite{Li90,MC22}.
However, upon summing over all eight
${\bf k}_i$ points, only the $\sin^2\theta$ term remains
while all other  terms, dependent on
$\phi'=\phi_i$, completely cancel out.
As a result, at any $\bf k$ point,
the symmetrized contribution to the second-order correction
$\Omega^{(2)}(\hat{\bf M})$ to the grand potential
 depends solely on the polar angle $\theta$
 (as $\sin^2\theta$ or, equivalently, as $\cos^2\theta$),
 and is independent of  the azimuthal angle $\phi$.
Consequently, the symmetrized distribution of the MCA energy in the BZ  defined with Eq. (\ref{eq-mca-omega-PT}) in the  PT approach is independent of the choice of the in-plane magnetization direction.

The MCA energy distributions obtained with the FT and PT, both original and symmetrized,
are very similar for systems with weak and moderate SOC, like
the Co film with the nominal SOC strength; Fig. \ref{fig-ma-bz-FT-symm4-Co12-Co12Pt8} (a) and (b).
For a system without the inversion
symmetry, like the Co/Pt bilayer, the original distribution
includes finite first-order terms at each $\bf k$ point.
In the PT approach, such terms are excluded from the outset
 because the total first-order correction  $\Omega^{(1)}$ vanishes
 and the  MCA energy  is then given by the difference
$E_{\text{MCA}}^{\text{PT}}=
\Omega^{(2)} ({\hat{\bf M}}_{\perp})-
\Omega^{(2)} ({\hat{\bf M}}_{||})$.
However, these terms persist  in the FT calculations.
Symmetrization of the
MCA energy distribution  effectively
cancels out these potentially large first-order terms
by summing the contributions to $E_{\text{MCA}}^{\text{FT}}$
from the $\bf k$ and $-\bf k$ points.
In consequence,
the resulting symmetrized FT distribution (its leading term)
is of the second order.
This cancellation effect of the symmetrization is evident in
Figs. \ref{fig-ma-bz-FT-symm4-Co12-Co12Pt8} (c) and (d)
(note the different scales in these plots).


\begin{thebibliography}{100}

\bibitem{Gay-Richter86}
J. G. Gay and R. Richter, Phys. Rev. Lett. {\bf 56}, 2728 (1986).

\bibitem{Gay-Richter87}
J. G. Gay and R. Richter, J. Appl. Phys. {\bf 61}, 3362 ( 1987).

\bibitem{Li90}
Chun Li, A. J. Freeman, H. J. F. Jansen, and C. L. Fu
Phys. Rev. B {\bf 42}, 5433 (1990).

\bibitem{Bruno89}
P. Bruno, Phys. Rev. B {\bf 39}, 865 (1989).

\bibitem{MC94}
M. Cinal, D. M. Edwards, and J. Mathon,
Phys. Rev. B {\bf 50}, 3754 (1994).

\bibitem{MCMP16}
M. D\k{a}browski, M. Cinal, M. Przybylski, G. Chen, A. T. N’Diaye,
A. K. Schmid, and J. Kirschner,
Phys. Rev. B {\bf 93}, 064414 (2016).


\bibitem{MCMP19}
M. D\k{a}browski, M. Cinal, A. K. Schmid, M. Przybylski,
and J. Kirschner,
Phys. Rev. B {\bf 99}, 184420 (2019).

\bibitem{Buhrman12}
L. Liu, C.-F. Pai, Y. Li, H. W. Tseng, D. C. Ralph, R. A. Buhrman,
Science {\bf 336}, 555 (2012).

\bibitem{Buhrman12b}
L.Liu, O. J. Lee, T. J. Gudmundsen, D. C. Ralph, and R. A. Buhrman,
Phys. Rev. Lett. {\bf 109}, 096602 (2012).


\bibitem{Lee13}
K. -S. Lee, S. -W. Lee, B. -C. Choi, J. -W. Moon, S. -H. Oh, W. Kim, J. Ryu, P. Fischer, and S. -K. Kim, Appl. Phys. Lett.  {\bf 102}, 112410 (2013).


\bibitem{Suto25}
P. D. Bentley, Y. Sasaki, I. Suzuki, S. Isogami,
Y. K. Takahashi, and H. Suto,
Appl. Phys. Lett. {\bf 126}, 022404 (2025).

\bibitem{Nakamura25}
Y.-N. Apriati, K. Nawa, and K. Nakamura,
Appl. Phys. Lett. {\bf 126}, 082403 (2025)

\bibitem{Kelly24}
Y. Liu, P,Yang, and P.J. Kelly,
Phys. Rev. B {\bf 109}, 014416 (2024).


\bibitem{Ham21}
W. S. Ham, A.-M. Pradipto, K. Yakushiji, K. Kim, S. H. Rhim, K. Nakamura,
Y. Shiota, S. Kim,and T. Ono,
npj Computational Materials {\bf 7}, 129 (2021).


\bibitem{Maziewski21}
A.K. Dhiman, M. Matczak, R. Gieniusz, I. Sveklo, Z. Kurant, U. Guzowska,
F. Stobiecki, and A. Maziewski,
J. Magn. Magn. Mater. {\bf 519}, 167485 (2021).


\bibitem{Maziewski25}
A.K. Dhiman, A. Fakhredine, R. Gieniusz, Z. Kurant,
I. Sveklo,  P. Dłużewski, W. Dobrogowski, S. K. Jena,
A. Pietruczik, C. Autieri, A. Wawro, A. Maziewski,
Appl. Surf. Sci. {\bf 679}, 161151 (2025).

\bibitem{Zhang22}
Q. Zhang, J. Liang, K. Bi, L. Zhao, H. Bai, Q. Cui, H.-A. Zhou,
H. Bai, H. Feng, et al.,
Phys. Rev. Lett. {\bf 128}, 167202 (2022).

\bibitem{Zhu24}
Q. Liu, L. Liu, G. Xing, and L. Zhu,
Nat. Commun. {\bf 15}, 2978 (2024).


\bibitem{Weinert95}
M. Weinert, R. E. Watson, and J. W. Davenport, Phys. Rev. B
{\bf 32}, 2115 (1985).

\bibitem{Wang96-jmmm}
X. Wang, D.-S Wang, R. Wu, and A. J. Freeman,
J. Magn. Magn. Mater. {\bf 159}, 337 (1996).


\bibitem{MC97}
M. Cinal and D. M. Edwards,
Phys. Rev. B {\bf 55}, 3636 (1997).

\bibitem{MC22}
M. Cinal, Phys. Rev. B {\bf 105}, 104403 (2022).

\bibitem{MC24}
M. Cinal, Phys. Rev. B {\bf 109}, 024424 (2024).



\bibitem{Wang93}
D.-S Wang, R. Wu, and A. J. Freeman, Phys. Rev. B {\bf 47}, 14932
(1993).

\bibitem{Qiao18}
J. Qiao, S. Peng, Y. Zhang, H. Yang, and W. Zhao,
Phys. Rev. B {\bf 97}, 054420 (2018).
%

\bibitem{Miura13}
Y. Miura, S. Ozaki, Y. Kuwahara, M.Tsujikawa,
K. Abe, and M. Shirai,
J. Phys.: Condens. Matter {\bf 25}, 106005  (2013).

\bibitem{Miura18-prb}
K. Masuda and Y. Miura,
Phys. Rev. B {\bf 8}, 224421 (2018).

\bibitem{Miura22-jphys-review}
Y. Miura and J. Okabayashi,
J. Phys.: Condens. Matter {\bf 34} 473001 (2022).


\bibitem{Ong16}
P. V. Ong, N. Kioussis, P. K. Amiri, and K. L. Wang,
Phys. Rev. B {\bf 94}, 174404 (2016).


\bibitem{Sun19}
X. Chen, S.Zhang, B. Liu, F. Hu, B. Shen, and J. Sun,
Phys. Rev. B {\bf 100}, 144413 (2019)

\bibitem{Ke19}
L. Ke,
Phys. Rev. B {\bf 99}, 054418 (2019).


\bibitem{Okabayashi18}
J. Okabayashi, Y. Miura, and H. Munekata,
Sci. Rep. {\bf 8}, 8303 (2018).


\bibitem{Li13}
D. Li, A.  Smogunov, C. Barreteau, F. Ducastelle, and D. Spanjaard,
Phys. Rev. B {\bf 88}, 214413 (2013).

\bibitem{Li14}
D. Li,  C. Barreteau, M. R. Castell, F. Silly and A.  Smogunov,
Phys. Rev. B {\bf 90}, 205409 (2014).

\bibitem{BlancoRey19}
M. Blanco-Rey, J.I. Cerd\'{a} and A. Arnau,
New J. Phys. {\bf 21},  073054 (2019).

\bibitem{Szunyogh97}
L. Szunyogh, B. \'{U}jfalussy, C. Blaas, U. Pustogowa, C. Sommers,
and P. Weinberger
Phys. Rev. B {\bf 56}, 14036 (1997).


\bibitem{Guo99}
G. Y. Guo, J. Phys.: Condens. Matter {\bf 11}, 4329 (1999).

\bibitem{MC03}
M. Cinal,
J. Phys.: Condens. Matter {\bf 15}  29 (2003).

\bibitem{Sandratskii15}
L. M. Sandratskii, Phys. Rev. B {\bf 92}, 134414 (2015).


\bibitem{MCMP13}
S. Manna, P. L. Gastelois, M. D\k{a}browski, P. Ku\'{s}wik, M. Cinal,
M. Przybylski, and J. Kirschner,
Phys. Rev. B {\bf 87}, 134401 (2013).


\bibitem{MC98}
M. Cinal and D.M. Edwards,
Phys. Rev. B {\bf 57},  100 (1998).

\bibitem{MC01}
M. Cinal,
J. Phys.: Condens. Matter {\bf 13},  901 (2001).


\bibitem{Chang17}
C.-H. Chang, K.-P. Dou, G.-Y. Guo, and C.-C. Kaun,
NPG Asia Materials {\bf 9}, e424 (2017).

\bibitem{Werwinski24}
J. Marciniak, M. Werwi{\'n}ski,
J. Magn. Magn. Mater. {\bf 609}, 172455 (2024).


\bibitem{DMEJM91}
D. M. Edwards, J. Mathon, R. B. Muniz, and M. S. Phan,
Phys. Rev. Lett. {\bf 67}, 493 (1991);
Erratum Phys. Rev. Lett. {\bf 67}, 1476 (1991).


\bibitem{MP11}
U. Bauer, M. D\k{a}browski, M. Przybylski, and J. Kirschner,
Phys. Rev. B {\bf 84}, 144433 (2011).



\bibitem{MCMP12}
M. Przybylski, M. D\k{a}browski, U. Bauer, M. Cinal, and J. Kirschner,
J. Appl. Phys. {\bf 111}, 07C102 (2012).


\bibitem{MP16}
S. Manna, M. Przybylski, D Sander, J. Kirschner,
J. Phys. Condens. Matter {\bf 28}, 456001 (2016).

%
\bibitem{MP09}
J. Li, M. Przybylski, F. Yildiz, X. D. Ma, and Y. Z. Wu,
Phys. Rev. Lett. {\bf 102}, 207206 (2009).
%
%
\bibitem{Slezak20}
M. \'{S}l\k{e}zak, P. Dr\'{o}\.{z}d\.{z}, K. Matlak, A. Kozio\l{}-Rachwa\l{},
J. Korecki, and T. \'{S}l\k{e}zak,
 J. Magn. Magn. Mater. {\bf 497}, 165963 (2020).



\bibitem{Hubner96}
T.H Moos, W. H\"{u}bner, K.H. Bennemann,   
Solid State Comm. {\bf 98}, 639 (1996).

\bibitem{Hubner97}
A. Lessard, T. H. Moos, and W. H\"{u}bner,  
Phys. Rev. B {\bf 56}, 2594 (1997).

\bibitem{Balcerzak06}
T. Balcerzak,
Thin Solid Films {\bf 500}, 341 (2006).

\bibitem{Abate-Asdente65}
E. Abate and M. Asdente, Phys. Rev. {\bf 140}, A1303 (1965).


\bibitem{Papaconstantopoulos86}
D.A. Papaconstantopoulos,
{\em Handbook of the Band Structure of Elemental Solids}
(Plenum, New York, 1986).


\bibitem{Freeman79}
C. S. Wang and A. J. Freeman, Phys. Rev. B {\bf 19}, 793 (1979).

\bibitem{Jepsen71}
O. Jepsen and O. K. Andersen, Solid State Commun. {\bf 9}, 1763 (1971).

\bibitem{EBMC14}
E. Barati, M. Cinal, D. M. Edwards, and A. Umerski,
Phys. Rev. B {\bf 90}, 014420 (2014).

\bibitem{EBMC15}
E. Barati and M. Cinal,
Phys. Rev. B {\bf 91}, 214435 (2015).

\bibitem{Soler02-SIESTA}
J. M. Soler, E. Artacho, J. D. Gale, A. Garc{\'i}a, J. Junquera, P. Ordej{\'o}n,
and D. S{\'a}nchez-Portal,
J. Phys.: Condens. Matter {\bf 14}, 2745 (2002).

\bibitem{Kresse93-94}
G. Kresse and J. Hafner, Phys. Rev. B {\bf  47} , 558 (1993);
ibid. {\bf 49} , 14 251 (1994)]

\bibitem{Kresse96b}
G. Kresse  and J. Furthm\"{u}ller, Phys. Rev. B {\bf 54}, 11169 (1996).


\end{thebibliography}
\end{document}